\documentclass[pre,a4paper,superscriptaddress,twocolumn,amsmath,amssymb,floatfix,amstex]{revtex4-2}%
\usepackage{graphics}
\usepackage{rotating}	
\usepackage{dcolumn}

\usepackage{bm}
\usepackage{color}
\usepackage[utf8]{inputenc} 
 \usepackage{array}
\usepackage{multirow}
\usepackage{hyperref}
\usepackage{float}
\usepackage{bm}
\definecolor{mygray}{gray}{0.4}
 \usepackage{soul}   

\newcommand{\red}[1]{{\color{red}#1}}

\usepackage{aurical}
\usepackage[T1]{fontenc}
\newcommand{\raur}{\text{\Fontauri r}}

\def\Cav{{\cal C}_{\rm av}}
\def\Ctyp{{\cal C}_{\rm typ}}
\def\Ctav{\widetilde{{\cal C}}_{\rm av}}

\def\Cttyp{\widetilde{{\cal C}}_{\rm typ}}
\def\chideux{\chi^2}

\begin{document}

\title{Critical properties of the Anderson transition in random graphs: two-parameter scaling theory, Kosterlitz-Thouless type flow and many-body localization}

\author{I.~Garc\'ia-Mata}
\affiliation{Instituto de Investigaciones F\'isicas de Mar del Plata (IFIMAR), CONICET--UNMdP, Funes 3350, B7602AYL Mar del Plata, Argentina}
\affiliation{Consejo Nacional de Investigaciones Cient\'ificas y
Tecnol\'ogicas (CONICET), Argentina}
\author{J.~Martin} 
\affiliation{Institut de Physique Nucl\'eaire, Atomique et de Spectroscopie, CESAM, University of Li\`ege, B-4000 Li\`ege, Belgium}
\author{O.~Giraud}
\affiliation{Universit\'e Paris-Saclay, CNRS, LPTMS, 91405 Orsay, France}
\author{B.~Georgeot}
\affiliation{Laboratoire de Physique Th\'eorique, Universit\'e de Toulouse, CNRS, UPS, France}
\author{R.~Dubertrand}
\affiliation{Department of Mathematics, Physics and Electrical Engineering, Northumbria University, NE1 8ST Newcastle upon Tyne, United Kingdom}
\author{G.~Lemari\'e}
\email[Corresponding author: ]{lemarie@irsamc.ups-tlse.fr}
\affiliation{Laboratoire de Physique Th\'eorique, Universit\'e de Toulouse, CNRS, UPS, France}
\affiliation{MajuLab, CNRS-UCA-SU-NUS-NTU International Joint Research Unit, Singapore}
\affiliation{Centre for Quantum Technologies, National University of Singapore, Singapore}

\date{September 9, 2022} 
\begin{abstract}
The Anderson transition in random graphs has raised great interest, partly out of the hope that its analogy with the many-body localization (MBL) transition might lead to a better understanding of this hotly debated phenomenon. Unlike the latter, many results for random graphs are now well established, in particular the existence and precise value of a critical disorder separating a localized from an ergodic delocalized phase. However, the renormalization group flow and the nature of the transition are not well understood. In turn, recent works on the MBL transition have made the remarkable prediction that the flow is of Kosterlitz-Thouless type. In this paper, we show that the Anderson transition on graphs displays the same type of flow. Our work attests to the importance of rare branches along which wave functions have a much larger localization length $\xi_\parallel$ than the one in the transverse direction, $\xi_\perp$. Importantly, these two lengths have different critical behaviors:  $\xi_\parallel$ diverges with a critical exponent $\nu_\parallel=1$, while $\xi_\perp$ reaches a finite universal value ${\xi_\perp^c}$ at the transition point $W_c$. Indeed, $\xi_\perp^{-1} \approx {\xi_\perp^c}^{-1} + \xi^{-1}$,  with $\xi \sim  (W-W_c)^{-\nu_\perp}$ associated with a new critical exponent $\nu_\perp = 1/2$, where $\exp( \xi)$ controls finite-size effects. The delocalized phase inherits the strongly non-ergodic properties of the critical regime at short scales, but is ergodic at large scales, with a unique critical exponent $\nu=1/2$. 
This shows a very strong analogy with the MBL transition: the behavior of $\xi_\perp$ is identical to that recently predicted for the typical localization length of MBL in a phenomenological renormalization group flow. 
We demonstrate these important properties for a small-world complex network model and show the universality of our results by considering different network parameters and different key observables of Anderson localization.
\end{abstract}
\maketitle
\section{Introduction}

The Anderson transition on random graphs has recently 
generated great interest \cite{abouchacra73, zirnbauer1986localization, fyodorov1991localization, mirlin1994distribution, monthus2011anderson, biroli2012difference, de2014anderson,  kravtsov2015random, altshuler2016nonergodic, facoetti2016non, tikhonov2016anderson, tikhonov2016fractality, sonner2017multifractality,scaling17, biroli2017delocalized, monthus2017multifractality, tarquini2017critical, kravtsov2018non, bogomolny2018eigenfunction, bogomolny2018power, bera2018return, tikhonov2019statistics, tikhonov2019critical, parisi2019anderson, twoloc20, kravtsov2020localization, detomasi2020subdiffusion, roy2020localization, kravtsov2020localization, khaymovich2020fragile, biroli2022critical, biroli2021levy, alt2021delocalization, tikhonov2021anderson, colmenarez2022sub, khaymovich2021dynamical}. There are several reasons for this, but certainly the analogy with the many-body localization (MBL) transition is one of the most important \cite{basko2006metal, gornyi2005interacting, altshuler1997quasiparticle, tikhonov2021anderson}. The MBL transition is a phenomenon which has proven to be extremely subtle, and after over a decade of considerable work, both theoretical and experimental, see \cite{nandkishore2015many, AleLaf18, RevModPhys.91.021001} for recent reviews, it is not even clear whether the MBL phase exists in one dimension, or if so, from what value of the disorder \cite{morningstar2022avalanches, sels2022bath, long2022phenomenology}. In this sense, there is a much better grasp of the Anderson transition on random graphs with disorder. The existence and precise value of a critical disorder is well established \cite{tikhonov2019critical, parisi2019anderson, sierant2022universality}, and a number of critical properties are well understood \cite{scaling17, tikhonov2019statistics, twoloc20, tikhonov2021anderson, khaymovich2020fragile}. 
From the analogy between Anderson and MBL transitions, our understanding of one system can shed light on the other.

In this paper, we investigate the Anderson transition on random graphs by taking advantage of the knowledge on MBL. The numerous studies on the MBL transition, and in particular the difficulty of describing this phenomenon analytically starting from a microscopic model, have led to the development of an approach called the phenomenological renormalization group (RG) \cite{thiery2018many, goremykina2019analytically, dumitrescu19, MorningHuse19, morningstar2020many}. This approach is based on an avalanche mechanism thought to be responsible for an instability of MBL leading to a transition to a thermal phase \cite{de2017stability, morningstar2022avalanches}. The phenomenological RG allows to describe the renormalization flow in the vicinity of the MBL transition. Remarkably, recent studies have shown that this flow is of the Kosterlitz-Thouless type \cite{dumitrescu19} (or at least very similar to it \cite{morningstar2020many}).
Thus, from the MBL side, we have precise analytical predictions of the flow and the nature of the transition. 

This is not the case for the Anderson transition on random graphs: although several critical properties are known, the flow and nature of the transition are not. The purpose of this paper is to remedy this. We will show that the flow of the Anderson transition is of Kosterlitz-Thouless type, extremely similar to the MBL transition, suggesting that they are in the same universality class. 

This result may seem surprising. Indeed, the analogy between the MBL transition and the Anderson transition on random graphs is not consensual with some differences pointed out in the literature \cite{roy2020fock, roy2020localization, pietracaprina2021hilbert, mace2019multifractal, tarzia2020many, tikhonov2021anderson}. 
In the case of the MBL transition, it is known that the phenomenological RG predictions are extremely difficult (some studies even claim impossible \cite{Panda_2020}) to verify numerically \cite{ABANIN2021168415, laflorencie2020chain}. For the Anderson transition on random graphs, characterizing the flow is still a very difficult task. Indeed, recent debates about the nature of the delocalized phase \cite{monthus2011anderson, biroli2012difference, de2014anderson,  altshuler2016nonergodic, tikhonov2016anderson, tarquini2017critical, scaling17, tikhonov2019statistics, khaymovich2020fragile,biroli2022critical} and about the value of the critical exponents \cite{scaling17, kravtsov2018non, tikhonov2019critical, twoloc20, tarzia2022fully, sierant2022universality} have their origin in the subtlety of this critical behavior. A large part of the paper will therefore be devoted to the description of our scaling approach, which enables us to characterize these properties. Once this is done, we will be able to give a coherent physical picture of the transition and show its close analogy with the MBL transition.

In any theoretical approach to complex systems, it is important to have simpler models in which to test predictions.
This analogy between the flow in random graphs and MBL opens many interesting questions.
Could a phenomenological RG approach be established in this case as well? Can we test the relevance of the avalanche mechanism on this simpler model?
We believe that the interplay between the two models will provide considerable insight into the physics of both systems.

The paper is organized as follows. In Sec.~\ref{introbis} we present the context, objectives and main results of the paper.  In Sec.~\ref{Sec:Model}, we present the Anderson model on a small-world network and the exact diagonalization method we used. In Sec.~\ref{firstresults}, we define the different observables we used to characterize the Anderson transition (generalized inverse participation ratios, correlation functions, and spectral statistics), and we describe first results about their behavior. In Sec.~\ref{secFSS}, we describe in detail our finite-size scaling  analysis of the transition. We test systematically different scaling hypotheses and present the quite rich outcomes of this analysis: different types of critical behaviors for different observables and phases. In Sec.~\ref{sec:physint}, we give a global interpretation of the observed behaviors in terms of a two-parameter scaling theory. In Sec.~\ref{sec:comparison}, we show that our results can in fact be understood within the framework of existing theories of the transition. In Sec.~\ref{sec:MBL}, we show that the Anderson transition on random graphs has a Kosterlitz-Thouless type flow similar to the MBL transition, suggesting that these transitions are in the same universality class. The last Sec.~\ref{mainccl} summarizes the main conclusions of our study.

\section{Context, objectives and main results}
\label{introbis}

\subsection{Context}

Anderson localization \cite{PhysRev.109.1492} is a paradigmatic interference phenomenon in the presence of disorder which leads to a total absence of transport instead of a diffusive one. It is now well understood in finite dimension \cite{evers2008anderson, abrahams201050}. In particular,  it induces a metal-insulator transition in dimension three. This is a second-order phase transition, described by a single parameter scaling law \cite{abrahams1979scaling, evers2008anderson}. The analytical understanding has been supplemented by a precise numerical determination of critical properties by finite-size scaling  \cite{pichard1981finite, mackinnon1981one, PhysRevLett.82.382, PhysRevB.84.134209}, confirmed by experiments with cold atoms \cite{PhysRevLett.108.095701}. At the critical point of the transition, the states are multifractal \cite{castellani1986multifractal, cuevas2007two, evers2008anderson}, a non-trivial example of non-ergodic behavior.

In random graphs of infinite effective dimension, most of these properties (critical and non-ergodic) have been much debated recently \cite{abouchacra73, monthus2011anderson, biroli2012difference, de2014anderson,  kravtsov2015random, altshuler2016nonergodic, facoetti2016non, tikhonov2016anderson, tikhonov2016fractality, sonner2017multifractality,scaling17, biroli2017delocalized, monthus2017multifractality, tarquini2017critical, kravtsov2018non, bogomolny2018eigenfunction, bogomolny2018power, bera2018return, tikhonov2019statistics, tikhonov2019critical, parisi2019anderson, twoloc20, kravtsov2020localization, detomasi2020subdiffusion, roy2020localization, kravtsov2020localization, khaymovich2020fragile, biroli2022critical, biroli2021levy, alt2021delocalization, tikhonov2021anderson, colmenarez2022sub}. In fact, they are very different from those in finite dimension, and their understanding is important because of the analogy between the Anderson transition on such graphs and the many-body localization (MBL) phenomenon \cite{basko2006metal, gornyi2005interacting, altshuler1997quasiparticle}. MBL can be defined as the absence of thermalization in a certain type of many-body quantum systems isolated from their environment, in particular in the presence of disorder and in one dimension, see e.g.~\cite{nandkishore2015many, AleLaf18, RevModPhys.91.021001} for recent reviews. 
The analogy between MBL and Anderson localization on random graphs can be understood by the similarity between the structure of the  Hilbert space of many-body systems and that of random graphs of infinite effective dimension (for limitations of the extent of this analogy see e.g. \cite{roy2020fock, pietracaprina2021hilbert,roy2020localization}).

\begin{figure*}
\includegraphics[width=\linewidth]{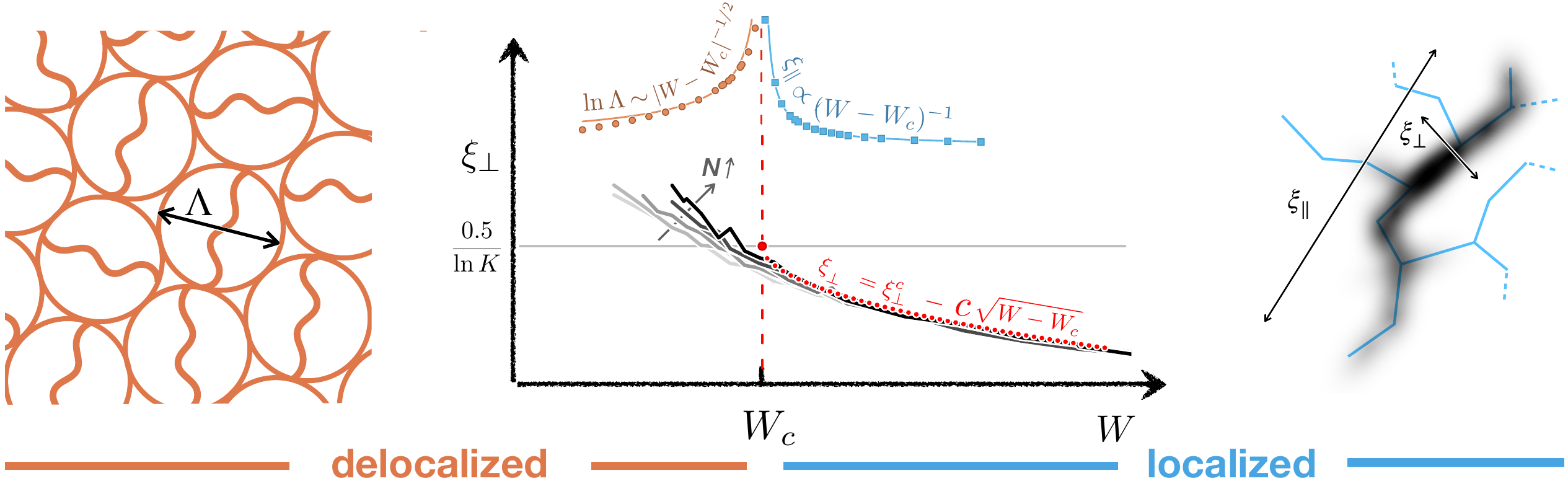}
\caption{Illustration of our main results for the Anderson transition on random graphs. Right panel: in the localized regime, states show a strongly anisotropic localization inherited from glassy non-ergodic properties: they lie mainly on a few rare branches where their localization length is $\xi_\parallel$ while the localization length perpendicularly to these branches $\xi_\perp$ is much smaller.
Middle panel: Close to the transition, $\xi_\parallel \sim (W-W_c)^{-1}$ while $\xi_\perp$ reaches a universal value $\xi_\perp^c = 0.5/\ln K$ with a square root singularity: $\xi_\perp \approx \xi_\perp^c - c \sqrt{W-W_c}$ with $c$ a constant. This is very similar to the predictions for the MBL transition \cite{dumitrescu19} and underlies a two-parameter Kosterlitz-Thouless type flow.
Thus two critical exponents control the finite-size scaling of different observables in the localized phase: $\nu_\parallel = 1$ for average observables and $\nu_\perp = 1/2$ for typical ones. From the delocalized side, the correlation volume $\Lambda$ diverges exponentially at the transition as $\ln \Lambda \sim \vert W - W_c\vert^{-1/2}$. 
Left panel: In the delocalized phase, there exists a characteristic volume $\Lambda$ such that  the behavior is ergodic for $N\gg \Lambda$ (a consequence of the homogeneous structure represented in the figure by the repetition of correlation volumes). For small system sizes $N \ll \Lambda$, the system inherits the strongly non-ergodic properties of the critical behavior, with localization on rare branches.
 }
\label{treecartoon}
\end{figure*}

Anderson localization on random graphs was described quite early, first on the Cayley tree, where one can write an exact self-consistent equation in a certain regime \cite{abouchacra73}, then by means of the supersymmetry method \cite{zirnbauer1986localization, fyodorov1991localization, Mirlin_1991, refId0, tarquini2017critical, tikhonov2019statistics, tikhonov2021anderson, 10.21468/SciPostPhys.12.2.048}. These results predict a transition between a localized phase and an ergodic phase with a number of notable features: the correlation volume diverges exponentially at the transition and the critical eigenstates are localized. Nevertheless, the ergodic character of the delocalized phase has been questioned more recently \cite{monthus2011anderson, biroli2012difference, de2014anderson,  altshuler2016nonergodic}. After a strong debate, it is now understood that the non-ergodic properties of the delocalized phase depend on the system under consideration. Indeed, the case of the finite Cayley tree has a  non-ergodic delocalized phase \cite{facoetti2016non, tikhonov2016fractality, sonner2017multifractality,kravtsov2018non} while in its infinite counterpart, often called the Bethe lattice, and in random graphs with loops and no boundaries, the delocalized phase is ergodic \cite{tikhonov2016anderson, tarquini2017critical, scaling17, tikhonov2019statistics, khaymovich2020fragile,biroli2022critical}. Different random matrix models have been proposed that describe the mechanisms underlying these various behaviors \cite{kravtsov2015random, kravtsov2020localization, khaymovich2020fragile, biroli2021levy, tarzia2022fully}.

\subsection{Main objectives of the paper}

The nature of the Anderson transition on random graphs and the type of its renormalization flow are the two main questions we address in this paper. 
As explained above, we are motivated by the recent predictions of the phenomenological RG on the MBL transition of a Kosterlitz-Thouless type flow \cite{goremykina2019analytically, dumitrescu19, MorningHuse19}.
In this paper, we build a two-parameter scaling theory describing the Anderson transition on random graphs. We published recently two letters \cite{scaling17, twoloc20} discussing some specific aspects of this transition, but a detailed and comprehensive description was still missing.

We address this problem first by exact diagonalization with state-of-the-art algorithms allowing us to reach very large system sizes, up to $N=2^{21}$ sites. 
Second, to overcome the unavoidable finite-size limitations, we use a scaling approach. In finite dimension, where the size constraints are much less stringent than on random graphs, the scaling approach was found necessary to obtain a controlled and precise numerical determination of the critical behavior \cite{pichard1981finite, mackinnon1981one, PhysRevLett.82.382, PhysRevB.84.134209}. The difficulty in the present case is that the scaling laws describing the renormalization flow in the vicinity of the transition are not known analytically, and that they are not controlled by a single scaling parameter, but by two parameters, as we will show. Finally, physical observables are characterized by large distributions, associated with rare events. This has a remarkable effect with regard to the critical properties of the transition, which can be distinct for average or typical observables, as in the MBL transition \cite{dumitrescu19,laflorencie2020chain}.

Our present approach shows that the finite-size scaling  of the transition is quite subtle, but nevertheless allows us to obtain a clear physical image of its properties. It also allows us to get a precise, controlled and unbiased determination of the critical exponents. In this article, we describe this approach in detail. Finally, we make key observations underlining the Kosterlitz-Thouless type of flow and the striking analogy between this transition and the MBL transition.

\subsection{Summary of main results}
\label{mainres}

In the small-world networks considered here, our results indicate the existence of a transition from a localized to a delocalized ergodic phase when the disorder strength $W$ is increased. These results are summarized below and illustrated in Fig.~\ref{treecartoon}.

In the regime where $W$ is larger than a  critical value $W_c$, the system is localized with strongly non-ergodic, glassy-like properties. 
A state is not localized in the same way along the different branches of the graph, but on the contrary explores only a small number of the available branches, analogously to directed polymers \cite{derrida1988polymers, monthus09, monthus2011anderson, monthus2019revisiting}.
Along these rare branches, the localization length $\xi_\parallel$ is much larger than $\xi_\perp$ which controls the extent of eigenstates perpendicularly to these branches, as illustrated in Fig.~\ref{treecartoon} (right). This distinction is all the more marked as one tends towards the transition, where $\xi_\parallel$ diverges, while $\xi_\perp$ reaches a finite universal value. 
In fact we find that the length $\xi_\parallel$ diverges as $(W-W_c)^{-1}$. By contrast,  $\xi_\perp^{-1} \approx {\xi_\perp^c}^{-1} + \xi^{-1}$ with $ {\xi_\perp^c} = 1/(2 \ln K)$ and 
 $\xi \sim  (W-W_c)^{-1/2}$, where $\exp( \xi)$ controls finite-size effects.
We thus see the emergence of a flow with two critical localization lengths, very similar to that predicted for the MBL transition \cite{dumitrescu19}, and of the Kosterlitz-Thouless type \cite{kosterlitz1973ordering}. 
These lengths characterize different properties of the system. Some observables are governed by $\xi_\parallel$: this is the case of the inverse participation ratio and of the average correlation function of eigenstate amplitudes. Other observables are governed by $\xi_\perp$, in particular the spectral statistics or suitably defined typical correlation functions. Another remarkable feature is the strong multifractality: while the large amplitudes of the eigenstates have a localized behavior, their small amplitude fluctuations are multifractal (for more details see Sec.~\ref{def_multif}). This strong multifractality is also controlled by $\xi_\perp$.

At the critical point, the system has an asymptotically localized behavior: the spectral statistics show a slow convergence to Poisson statistics and the multifractality is asymptotically strong.

In the regime $W<W_c$, we find an ergodic delocalized phase with however strongly non-ergodic properties at small scales. The characteristic scale is a  correlation volume $\Lambda$ which diverges exponentially at the transition as $\Lambda \sim \exp[a (W_c - W)^{-1/2}]$ where $a$ is a constant. 
For small system sizes $N\ll \Lambda$, the system inherits the strong multifractal properties of the critical behavior, while at large scales it is found ergodic: it is thus a strongly multifractal metal \cite{cuevas2007two}. This regime is illustrated in Fig.~\ref{treecartoon} (left). These results, which we presented in \cite{scaling17} and that we confirm here, clarified the debate on this important point. Note that there is now a consensus that the delocalized phase on such graphs is ergodic \cite{tikhonov2016anderson, scaling17, biroli2018delocalization, tikhonov2019statistics, tikhonov2019critical, kravtsov2020localization, khaymovich2020fragile, sierant2022universality}.

In this work the above physical picture of the transition is obtained by means of a careful finite size scaling analysis of different observables across the transition. In the random graphs we consider, finite-size scaling  properties are highly non trivial. First, because on such graphs the volume scales exponentially with its length, which implies that scaling laws of the ratio between volumes $N/\Lambda$ or lengths $L/\xi$ are not equivalent. This distinction is crucial as it allows us to deduce whether a phase is ergodic, localized or multifractal. Second, because many observables have large distributions with fat tails which implies the importance of rare events and, as we show, different critical properties for typical or average quantities. In the localized regime, we thus obtain two critical localization lengths $\xi_\parallel$ and $\xi_\perp$.

Our approach gives a completely unexpected description of the localized phase near the transition, where the distinction between two critical localization lengths had never been anticipated. It shows that the flow depends on two parameters, the multifractal dimension $D_2$, which can be seen as a linear density of the equivalent of ``thermal bubbles'' in the MBL problem, and the localization length $\xi_\perp$, and is of Kosterlitz-Thouless type. The localized phase is a ``critical line'' characterized by a vanishing $D_2$ and a finite $\xi_\perp$ up to $W_c$ where $\xi_\perp$ reaches a finite universal value and then has a jump in the ergodic delocalized phase. The results we find for $\xi_\perp$ are identical to what is predicted for the MBL transition by the phenomenological renormalization group \cite{dumitrescu19}. This is a compelling evidence that these two transitions are in the same universality class, despite their important differences.

\begin{figure}
\includegraphics[width=0.99\linewidth]{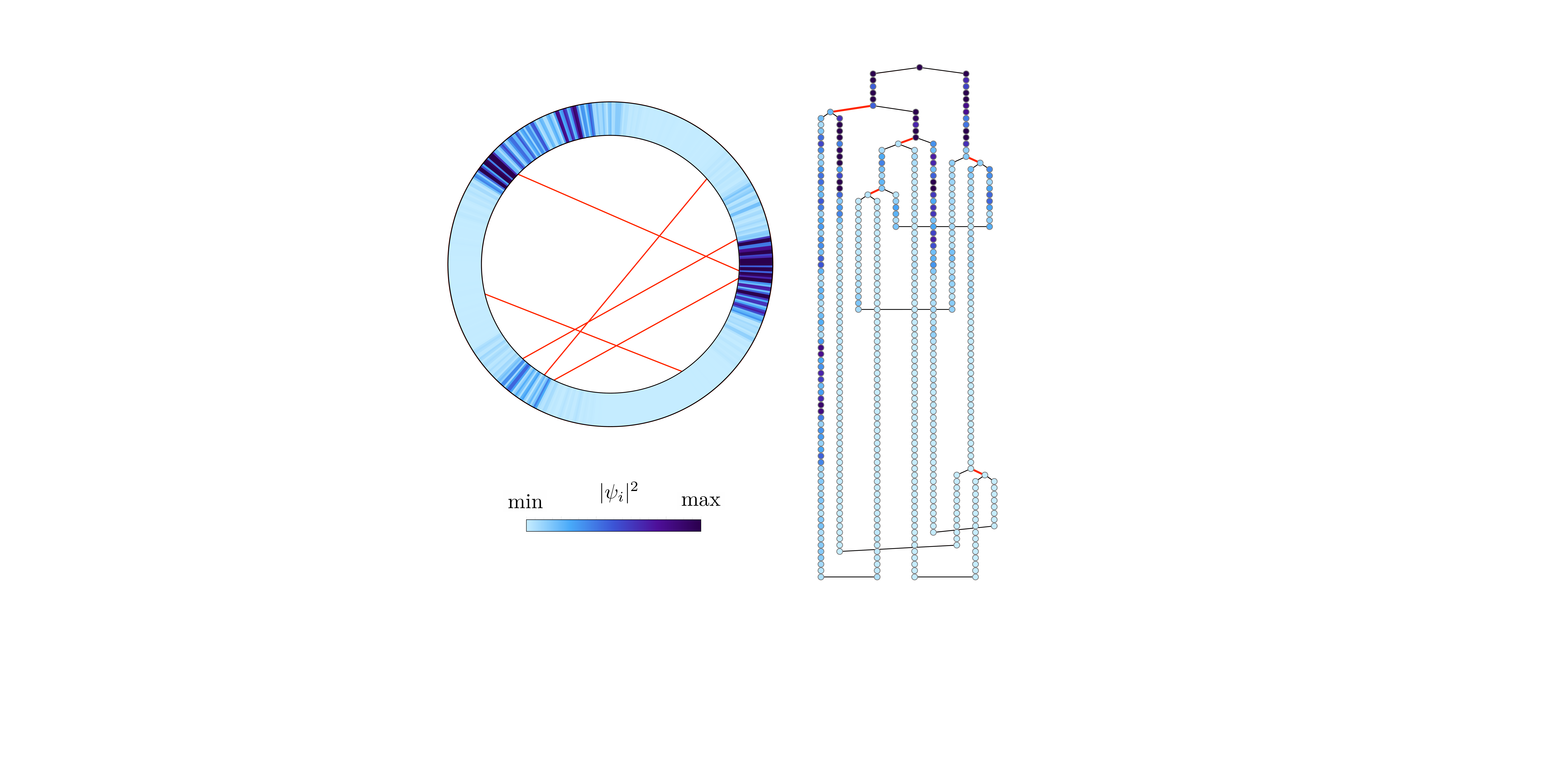}
\caption{Example of smallword model eigenfunction for $N=512$, $p=0.01$ ($\lfloor p N\rfloor =5$ smallworld links) and $W=0.4$. Colors are proportional to $|\psi_i|^2$, from large (dark-blue) to small (white) values. Left : 1D lattice with long-range links; Right: Same graph viewed as a network with vertices arranged according to the underlying tree-like structure. Long-range links (highlighted in red) occur wherever a vertex has three neighbors. 
\label{fig:tree} }
\end{figure}

\section{Model and methods}
\label{Sec:Model}

\subsection{The model}

The Anderson model \cite{PhysRev.109.1492} is the paradigmatic model for studying Anderson localization. It consists of a network where vertices have random energies distributed according to a certain law, and are linked together by hopping amplitudes. The topology of the considered network is of crucial importance. Anderson localization has been widely studied in finite dimensional networks (e.g.~the square lattice in dimension $1$ to $10$, see \cite{garcia2007dimensional, ueoka2014dimensional, mard2017strong}), but also in fractal networks \cite{schreiber1996dimensionality}. We are interested here in networks of infinite effective dimension, which means that the number of nodes $N$ scales exponentially with the diameter $d_N$ (the maximum distance between any two sites) of the graph. Different types of such networks have been considered: random regular graphs (RRG) \cite{tikhonov2016anderson}, critical Erd\"os-Renyi graphs \cite{alt2021delocalization,mard2017strong}, the (finite) Cayley tree and the (infinite) Bethe lattice \cite{derrida1993anderson,monthus09,biroli2010anderson, monthus2011anderson, de2014anderson}.

In this work we consider another type of random network, called smallworld network. The smallworld phenomenon, first introduced by Milgram \cite{milgram1967small}, was subsequently investigated theoretically in the context of network theory \cite{watts1998collective, newman2000mean}.
The model consists in a one-dimensional (1D) lattice with nearest neighbor coupling and additional 
$\lfloor p N\rfloor$ long range links ($\lfloor \, . \, \rfloor$ represents the integer part, and $p\in (0,1/2)$). The corresponding quantum mechanical problem is that of a tight-binding model on the lattice \cite{Chepelianskii,giraud2005quantum}. It can be described by the following Hamiltonian
\begin{equation}
\label{hamil}
H=
\sum_{i=1}^{N} \varepsilon_{i}|i\rangle\langle i|+
\sum_{\langle i, j\rangle}| i\rangle\langle j|+
\sum_{k=1}^{\lfloor p N\rfloor}(|i_{k}\rangle\langle j_{k}|+| j_{k}\rangle\langle i_{k}|).
\end{equation}
The first term is the on-site disorder with $\varepsilon_i$ independent and identically distributed random variables, which follow here, unless otherwise stated, a Gaussian distribution with zero mean and standard deviation $W$. The second term runs over nearest neighbors of the 1D lattice. The third term gives the long-range links that connect pairs $(i_k,j_k)$, randomly chosen with $|i_k - j_k|> 1$. 

In Fig.~\ref{fig:tree} an instance of such a random graph is depicted, together with an example of an eigenfunction of $H$ close to the band center, in two different representations: either as a one-dimensional chain with periodic boundary condition and 5 additional long-range links, or as a tree-like graph. These representations are of course equivalent, since the Hamiltonian \eqref{hamil} only depends on the topology of the graph.

\begin{figure}[!t]
\includegraphics[width=0.95\linewidth]{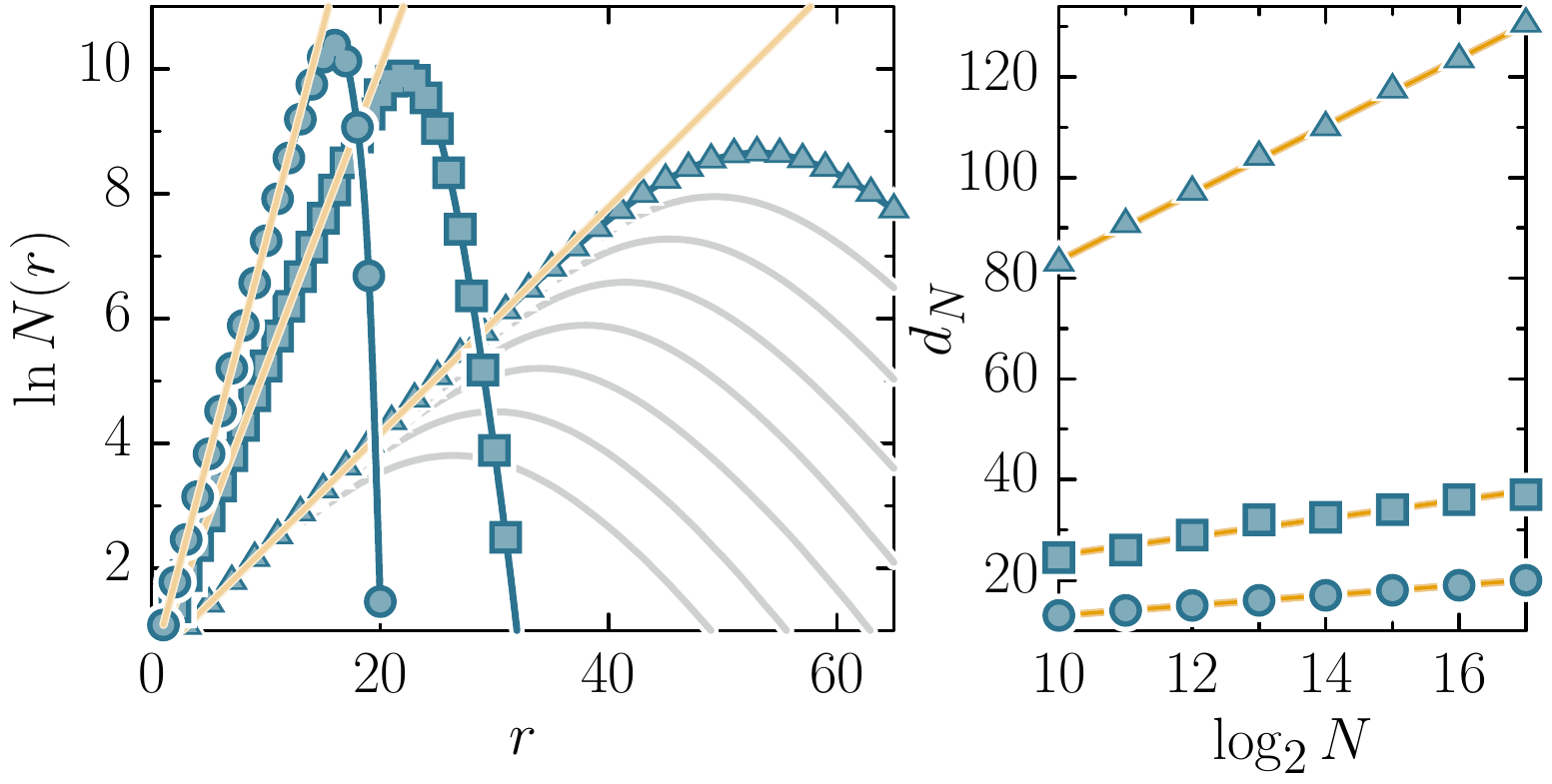}
\caption{\label{fig1Remy}Left:
Number of pairs of nodes at a fixed distance $r$ in the smallworld network  for  $p = 0.06$ (triangles), $p=0.25$ (squares), and  $p=0.49$ (circles). The system size (for the colored symbols) is $N=2^{17}$. The gray lines correspond to smaller system sizes. The straight lines were obtained from fitting $N(r)\sim K^r$ at small $r$. The values of $K$ obtained from these fits are $K=1.199,\, 1.614,\, 1.976$ which differ slightly from the theoretical prediction $K=1+2p$ (see \cite{sierant2022universality}).
Right: Diameter $d_N$ of the graph are well described by a linear function of  $\log_2 N$, as shown by the linear fits (solid lines). The symbol correspondence is the same as in the left panel.
}
\end{figure}

The small-world graph is characterized by the statistical properties of paths that relate pairs of vertices.
For $p = 0$, a 1D lattice (and thus the 1D Anderson model) is recovered. 
For a moderate number of long-range links (small $p$), the graph behaves locally as a tree, with on average $K\approx 1 + 2p$ branches leaving from each vertex. This is best seen by plotting the number $N(r)$ of pairs of nodes at a fixed distance $r$ on the network.
As shown in Fig.~\ref{fig1Remy}, this number grows exponentially up to a distance of the order $\sim d_N /2$, where $d_N$ is the diameter of the graph.
The exponential growth is given by $N(r) \propto K^r$, with $K \approx 1 + 2p$. Thus, at finite $p$, the model is a random graph with mean branching number $K \approx 1 + 2p$, i.e.~the sites have, on average, a number of nearest neighbors $Z=K+1=2+2p$ (see also \cite{sierant2022universality}). Moreover, the average distance between two long-range links is $ 1/(2p)$. While the diameter of random regular graphs of connectivity $K$ possesses similar properties (their diameter scales as the logarithm of their size \cite{bollobas1982diameter}), one of the main interests of our model is that we can continuously control through $p$ the value of the branching number $K$ and thus access the regime $1<K<2$. This is particularly interesting to check the universality of the critical properties (see also \cite{sierant2022universality}) and allows also to reach much larger range of diameters as compared to RRG where $K$ is an integer often taken equal to $K=2$ in numerical studies.

Another important property of this graph is that it has no boundaries and contains loops that typically vary in size like $d_N$. This is analogous to the case of RRG, but distinct from the finite Cayley tree, which has no loops and whose number of sites at the boundaries is proportional to the total number of sites. This distinction turns out to be crucial. In the case of the Cayley tree, it has been shown that the delocalized phase is non-ergodic \cite{tikhonov2016fractality, sonner2017multifractality, kravtsov2018non, biroli2018delocalization}, whereas in the case of RRG, Erd\"os-Renyi graphs and the smallworld network, borderless and with loops, there is now consensus that the delocalized phase is ergodic \cite{tikhonov2016anderson, tarquini2016level, scaling17, biroli2018delocalization, tikhonov2019statistics, tikhonov2019critical, kravtsov2020localization, khaymovich2020fragile, sierant2022universality}.

\subsection{Diagonalization method}
The Anderson model on the swall-world network defined in Eq.~\eqref{hamil} thus results in an ensemble of sparse random matrices that can be diagonalized to characterize the localization properties of this system. We are interested in the statistics of eigenstates and eigenenergies in the middle of the band, i.e., for eigenenergies close to 0. Working in the middle of the band, where the density of states is large, represents a numerical difficulty whose solution is nevertheless well known as the shift-invert method \cite{schenk2008large}. This method, used since a few decades in Anderson localization, has recently been transferred to MBL \cite{luitz2015many, pietracaprina2018shift}, and allows to reach very large Hilbert space sizes (in this work we consider system sizes up to $N =2^{21}$).

We have used the highly optimized library called JADAMILU \cite{JADAMILU} which implements the Jacobi-Davidson method with efficient multilevel incomplete \red{lower-upper} (LU) preconditioning.
The typical number of graph and disorder realizations, for each value of disorder $W$, is $10^4$--$10^5$ for $N\le 2^{16}$ and $10^{3}$--$10^4$  for $2^{16}< N\le 2^{21}$. For each realization and each system size, we compute 16 eigenvalues and eigenfunctions around the center of the band.

\section{Behavior of observables}
\label{firstresults}

The localization properties of the considered model \eqref{hamil} can be characterized from the spatial distributions of the eigenstates and the statistical properties of the energy levels. We describe in this section first the principle and simple results of the multifractal analysis of the eigenstates that we have made. Then we describe the spectral analysis method and its first simple results. Interestingly, we will see that different observables do not necessarily give the same information on the localization transition, and therefore are complementary. This is an aspect that is particularly important in random graphs, contrary to finite dimensional networks where it is much less crucial.

\begin{figure}[!t]
\includegraphics[width=0.85\linewidth]{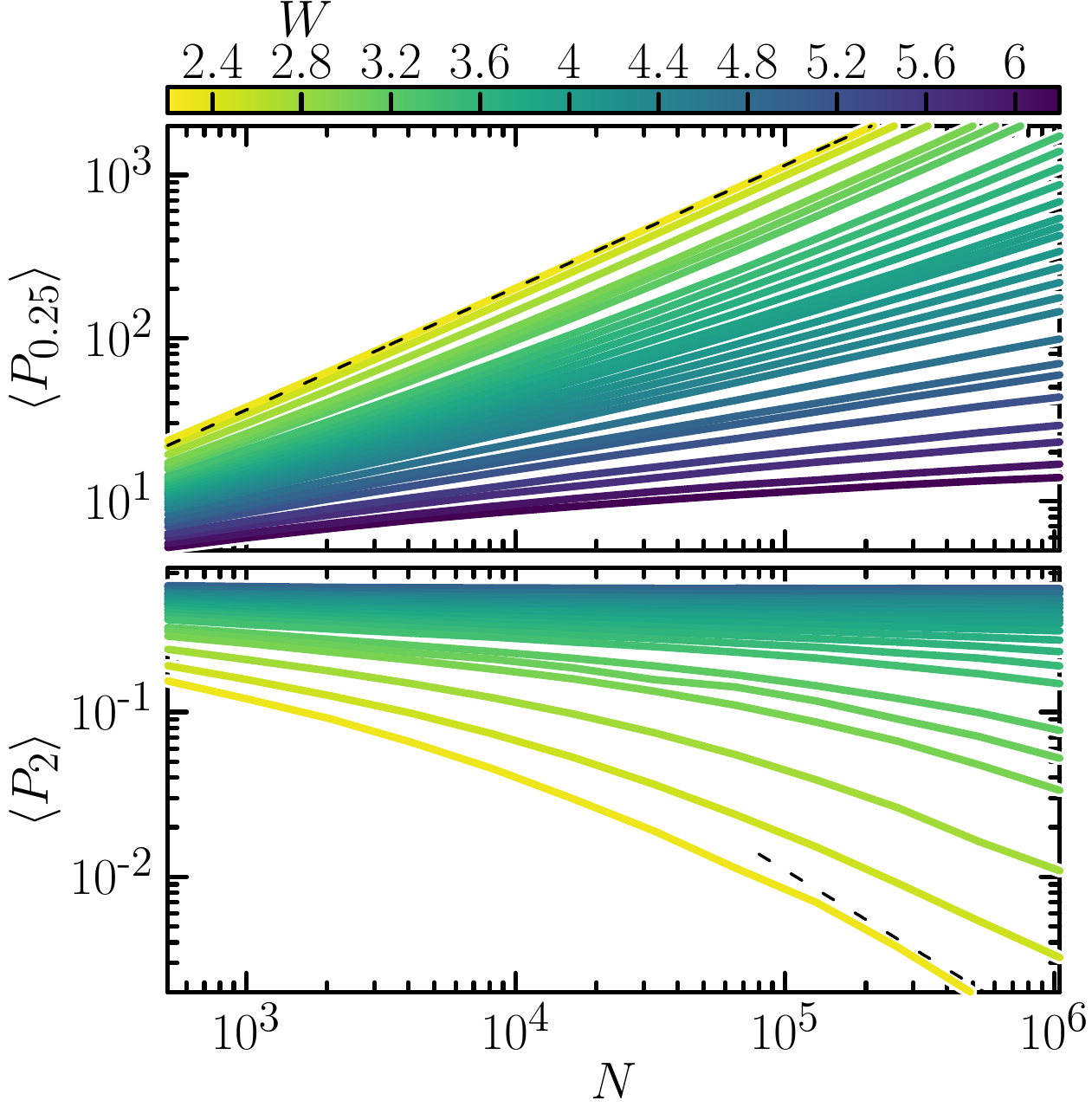}
\caption{\label{figmoments} Average eigenfunction moments $\langle P_q \rangle$ as a function of the system size for $p=0.25$ and $q=0.25$ (top) and $q=2$ (bottom). Dashed lines are the limiting ergodic behavior $\langle P^{({\rm erg})}_q(N)\rangle \sim N^{-(q-1)}$, observed at small $W$. A localized behavior corresponds on the contrary to $\langle P_q\rangle = \text{cst}$.
In contrast to the case of the Anderson transition in finite dimension, there does not exist for $q=2$ a curve with intermediate scale-invariant behavior $\langle P_q\rangle = N^{-\tau_q}$. Therefore, it is not obvious \emph{a priori} to determine the threshold $W_c$ from these data. For $q=0.25$, there exists a finite range of $W$ for which $\langle P_{q}\rangle \sim N^{-\tau_{q}(W)}$ at large $N$, i.e.~a disorder dependent multifractal behavior. This is again very different from the finite dimensional transition and is reminiscent of a delocalized non-ergodic phase. However, as we shall see, this multifractal behavior arises in the localized phase $W>W_c$ for $q<1/2$.
}
\end{figure}

\subsection{Multifractal analysis}
\label{def_multif}

\begin{figure}[!t]
\includegraphics[width=.97\linewidth]{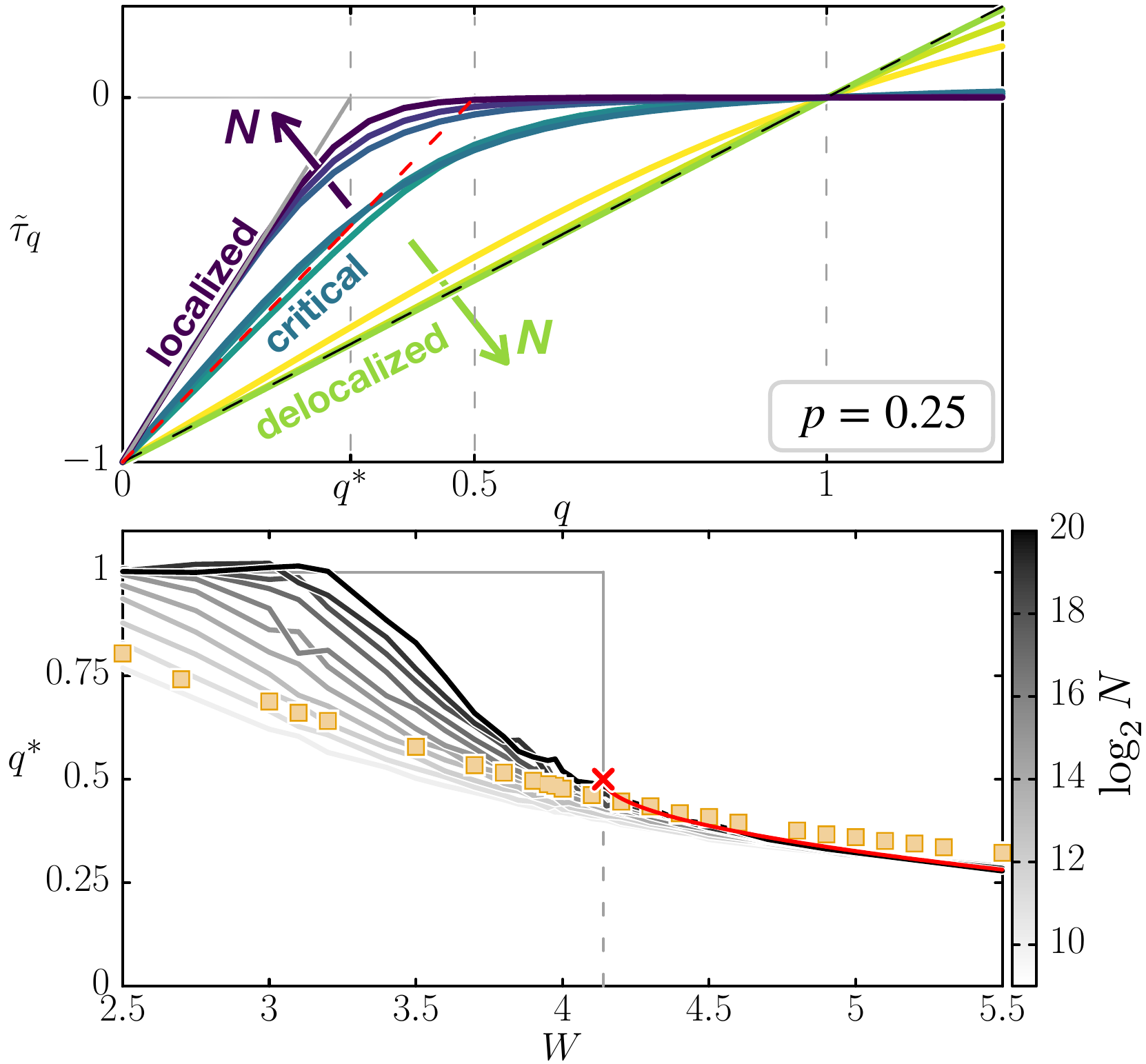} 
\caption{\label{figtauq} Top: Multifractal exponents $\tilde{\tau}_q$ extracted from the moments (see Fig.~\ref{figmoments}) for $p=0.25$ 
and various disorder strengths from small $W$ (delocalized regime) to large $W$ (localized regime). 
In the localized regime, $\tilde{\tau}_q$ tends to the strong multifractal behavior given by Eq.~\eqref{eq:tauqloc}, i.e. $\tau_q=q/q^* -1$ for $q\le q^*$ and $\tau_q=0$ for $q>q^*$. In the critical regime $\tau_q^*$ is given by the same expression, with $q^*=q^*_c=1/2$. In the ergodic phase, $\tau_q=q/q^* -1$ with $q^*=1$.
Bottom: dependence of $q^*$ with $W$ and system size for $p=0.25$. The gray scale goes from  $L=\log_2 N=10$ (light) up to $L=20$ (dark). There is evidence of a discontinuous jump at the transition from 0.5 to 1 in the thermodynamic limit. The red line shows the fit $q^{*}=q_{c}^{*}-C\left(W-W_{c}\right)^{v_{\perp}}$ for the localized phase. The cross marks the $W_c\approx 4.1$ obtained from this fit. The squares show
$\xi_\perp\ln K$ where $\xi_\perp$ was obtained from fitting $\Cttyp(r)\sim e^{-r/\xi_\perp} $. 
}
\end{figure}

\begin{figure}[!t]
\includegraphics[width=0.95\linewidth]{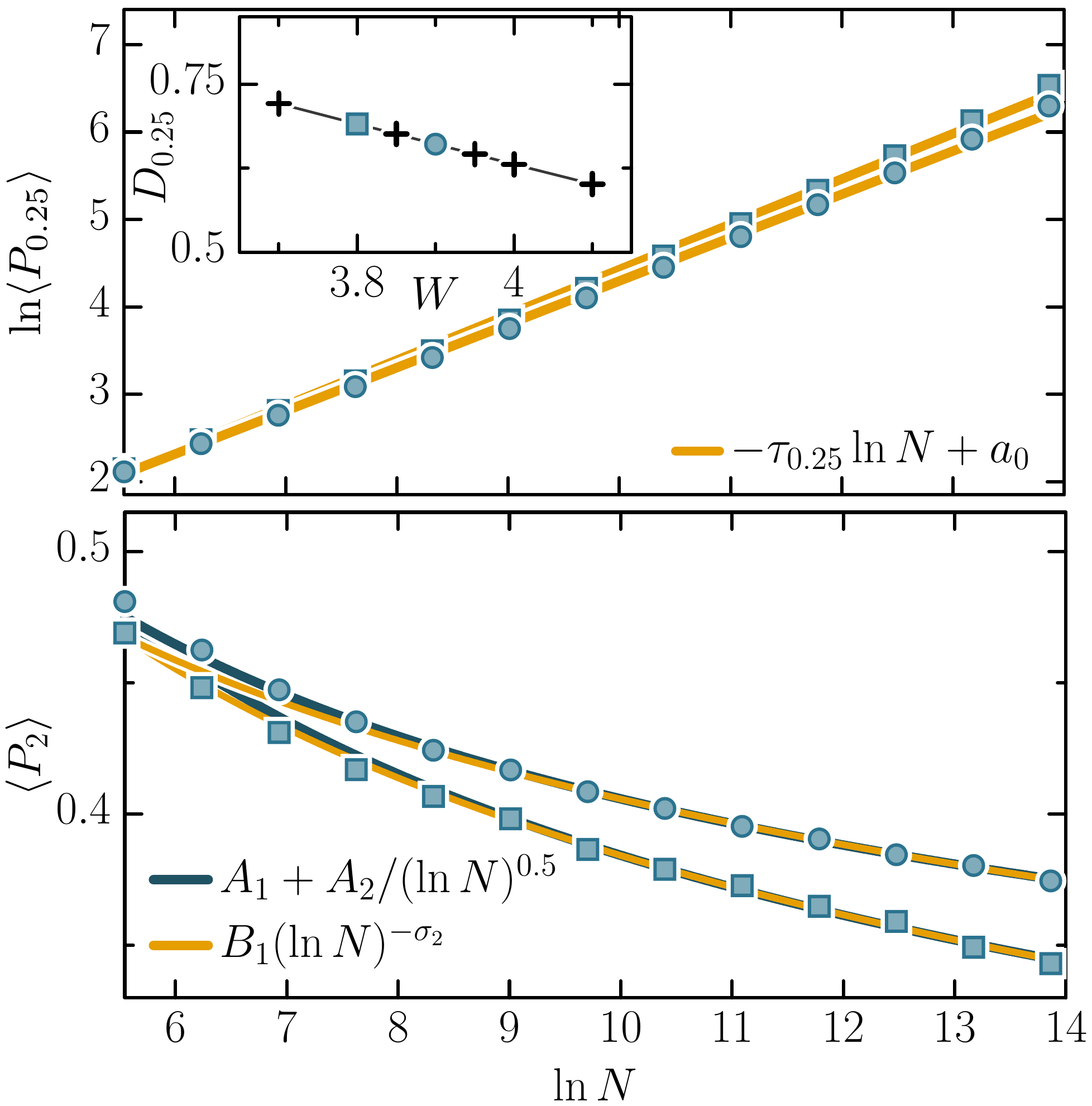}
\caption{\label{figpqcrit} Critical behavior of average moments $\langle P_q\rangle$ as a function of the system size for $W$ near the critical point $W_c$ at $p=0.25$: $W=3.9$ (squares), $W=4.0$ (circles). Top: For $q=0.25<1/2$, the critical behavior is multifractal with $\langle P_q\rangle\sim N^{-\tau_q}$ (solid lines are the corresponding fits); inset shows $D_q=\tau_q/(q-1)$ for various values of $W$ in the vicinity of $W_c$. Bottom: For $q=2>1/2$, the critical behavior is asymptotically localized. We represent two fitting hypotheses: $A_1+A_2/(\ln N)^{0.5}$ \cite{tikhonov2019statistics} (dark) and $B_1 (\ln N)^{-\sigma_2}$ \cite{scaling17} (light) with $\sigma_2=-0.332\pm 0.015$ for $W=3.9$ and $0.24\pm 0.004 $ for $W=4.0$. We are unable to distinguish clearly between these two possibilities.
}
\end{figure}

 Multifractality, introduced in the 1970s \cite{mandelbrot1975objets, hentschel1983infinite}, describes a whole range of phenomena, from turbulence to the economy or the DNA structure  (see references in \cite{falconer2004fractal}). Multifractals are characterized by the existence of a whole range of fractal dimensions. Quantum multifractality is a more recent concept. It was observed, among other systems, at the threshold of the Anderson transition in finite dimension \cite{chalker1990scaling, brandes1996critical, schreiber1991multifractal, janssen1994multifractal}, in quantum Hall systems \cite{chalker1988scaling, brandes1994multifractal}, in various random matrix models such as the power-law random banded matrix model \cite{mirlin1996transition, kravtsov1997new} or quantum maps \cite{meenakshisundaram2005multifractal, martin2008multifractality}, and was very recently analyzed mathematically in singular quantum billiards \cite{keating2022multifractal}. 
 
 One of the manifestations of multifractality is the  algebraic scaling of the moments of eigenfunctions with system size $N$. For an eigenstate $(\psi_i)_{1\leq i\leq N}$, the $q$th moment $P_q = \sum_i |\psi_i|^{2 q}$ scales as $P_q \sim N^{-\tau_q}$. Multifractal dimensions are defined as $0 \le D_q=\tau_q/(q-1) \le 1$. In general, multifractal dimensions depend on $q$ in a nontrivial way. A multifractal state with $0<D_2<1$ thus occupies a volume which increases with $N$ as $N^{D_2}$, but occupies an algebraically small fraction of the total volume $N$, a property which has been called ``non-ergodic extended'' recently \cite{altshuler2016nonergodic}. 
The algebraic behavior of the moments corresponds to the fact that there is no characteristic scale, contrary to the localized and delocalized regimes which are associated with a localization and a correlation length, respectively. Above this characteristic length scale, one expects $\tau_q = 0$ for a localized state (for $q>0$), while $\tau_q=q-1$ for an extended state.

Multifractal analysis allows to extract the above properties from quantum states. It can be done in different ways (see for example \cite{evers2008anderson, PhysRevB.84.134209}). In what follows, we study the average moments $\langle P_q\rangle=\langle\sum_i |\psi_i|^{2 q}\rangle$, where $\langle .\rangle$ is an average over disorder and graph realizations as well as over all the eigenvectors around the band center considered in each realization. 
The behavior of these moments as a function of the size of the system $N$ provides information on the localization of a state $\psi$. In particular, if we consider $q=2$, we obtain the inverse participation ratio, which quantifies the number of sites occupied by a state. Thus, for a localized state, which  occupies $N_\xi$ sites ($N_\xi \sim \xi^d$ in finite dimension, and $N_\xi \sim \exp(a \xi)$ in infinite dimension ), $P_2 \sim 1/N_\xi$ if $N\gg N_\xi$. On the contrary, if the state is delocalized, then $P_2 \sim 1/N$.

Furthermore, different values of $q$ allow to analyze different ranges of the wave function amplitudes. Thus, for large $q$, $P_q$ will be dominated by large amplitudes $|\psi_i|^2$, whereas for small $q$, we will focus on the low values of the amplitudes, close to zero.
In fact, for small $q$, $P_q$ is strongly dependent on rare events where the wave function almost vanishes, and therefore $P_q$ fluctuates strongly despite the average over the states and the disorder. In this case, it is necessary to use a coarse graining approach in order to soften these singularities. We proceed as follows: We partition the system into $N/\ell$ boxes of $\ell$ sites. In the considered Smallworld system, we do this by following the 1D chain. We then define the coarse grained amplitude associated to the box $k=1,\cdots, N/\ell$ as $\mu_k = \sum_{i \in k} \vert \psi_i\vert^2$ where the sum is over the $\ell$ sites $i$ belonging to the box $k$. Note that by normalization of the state $\psi$, $\sum_k \mu_k = 1$. The moments are then expressed as $P_q = \sum_{k} \mu_k^q$. We have considered in this work a coarse graining with $\ell =4$ for all the values of $q$ considered. Note that a multifractal analysis can be done considering the dependence of $P_q$ with $\ell$ instead of $N$. We have presented such a study in \cite{scaling17} which provides interesting information allowing in particular to characterize a non-ergodic volume in the delocalized phase. We will not detail these results in this paper but instead focus on the more standard approach considering $P_q$ as a function of $N$ at fixed $\ell$ \footnote{Note that for the Anderson transition in 3D, it has been found  \cite{PhysRevB.84.134209} that one needs to work at fixed $\ell/N$ ratio to determine accurately the critical exponent of the transition from multifractal data. Whether this is also the case in random graphs of infinite dimension is an open question we leave for further studies.}.

We have obtained data for $\langle{P_q}\rangle$ and their uncertainties for $q \in [0,4]$, $N$ from $N=2^6$ to $N=2^{21}$, and $p$ from $p=0.01$ to $0.49$ for a normally distributed disorder, and for a box-distributed disorder. Numerical data are shown in Fig.~\ref{figmoments} for two values of $q$, one large $q=2$ and one small $q=0.25$ (the distinction between small and large being given by $q<1/2$ for small and $q>1$ for $q$ large, as we will explain in the following), for $p=0.25$.
Numerically we can calculate the {\it local in $N$} multifractal exponents $\tilde{\tau}_q(N)$, defined as 
\begin{equation}
\tilde{\tau}_q(N)=\log_2 \langle P_q(N)\rangle-\log_2 \langle P_q(N/2)\rangle.
\end{equation}
They characterize the flow of the multifractal exponents with system size $N$, which indicates whether we tend to a multifractal, localized, or delocalized behavior.  
 The behavior of such local exponents, obtained from the slopes of the curves in Fig.~\ref{figmoments}, as a function of $q$ and for different system sizes, is shown in Fig.~\ref{figtauq}. In the localized regime, when $N\to\infty$, $\tilde{\tau}_q$  goes to the asymptotic expression
 \begin{equation}\label{eq:tauqloc}
 \tau_q = \begin{cases}
 q/q^*-1 \quad \text{for}\;\; q\le q^* \\
 0 \quad \text{for}\;\; q\ge q^*,
 \end{cases}
 \end{equation} 
 see \cite{de2014anderson, tikhonov2016fractality}. This behavior is called ``strong multifractality'', in the sense that for $q<q^*$ we have a multifractal behavior, while for $q>q^*$ the $D_q$ are 0 and we have a localized behavior.
 In this work, $q^*$ is determined by a linear fit of $\tilde{\tau}_q$ at small $q\gtrsim 0$. It depends crucially on $W$, and the system size close to the transition, as shown in the lower panel of Fig.~\ref{figtauq}. In fact, this dependence encodes, as we will show, a new critical exponent characterizing the transition in the localized regime. 
 The strong multifractality of the localized regime is reminiscent of the multifractality of the MBL phase \cite{mace2019multifractal, tarzia2020many}. Nevertheless, as we will discuss later, it is rather the result of an effective algebraic localization \cite{bilen2021symmetry}.
 
 In the delocalized regime $\tilde{\tau}_q$ goes asymptotically to the function
 \begin{equation}
 \tau_q=q-1
 \end{equation}
 as expected for ergodic delocalized wave functions. Nevertheless, the convergence with system size $N$ to this asymptotic behavior is slow, particularly close to the transition point. This is illustrated in the lower panel of Fig.~\ref{figtauq} where the convergence of $q^*(W,N)$ for $W<W_c$ to its ergodic value $q^* = 1$ is clearly visible only sufficiently far from the transition point. On the contrary, close to the transition, it is unclear at this stage of the analysis whether it would converge to $q^*=1$, thus forming the asymptotic step function depicted by the gray line in the lower panel of Fig.~\ref{figtauq}, or if it converges to a function $1/2<q^*(W)<1$ for a certain range of $W_e<W<W_c$, only reaching $q^*=1$ at $W\le W_e$, similarly to what is found in the Cayley tree \cite{de2014anderson, tikhonov2016fractality, sonner2017multifractality}. 
 In the first scenario, the delocalized phase is ergodic while in the second, there exists a multifractal, i.e.,~a delocalized non-ergodic regime for $W_e < W < W_c$. This question, which has been much debated recently \cite{monthus2011anderson, biroli2012difference, de2014anderson,  kravtsov2015random, altshuler2016nonergodic, facoetti2016non, tikhonov2016anderson, tikhonov2016fractality, sonner2017multifractality,scaling17, biroli2017delocalized, monthus2017multifractality, tarquini2017critical, kravtsov2018non, bogomolny2018eigenfunction, bogomolny2018power, bera2018return, tikhonov2019statistics, tikhonov2019critical, parisi2019anderson, twoloc20, kravtsov2020localization, detomasi2020subdiffusion, roy2020localization, kravtsov2020localization, khaymovich2020fragile, biroli2022critical, biroli2021levy, alt2021delocalization, tikhonov2021anderson, colmenarez2022sub}, motivates the finite-size scaling  analysis that we will describe later. We will see that it is the first scenario that is valid: there exists a characteristic ``correlation'' volume $\Lambda(W)$ such that $q^* \rightarrow 1$ for $N \gg \Lambda(W)$, with $\Lambda(W)$ diverging exponentially at $W=W_c$.     
 
At criticality $W=W_c$ (we will explain how we determine $W_c$ later), local multifractal exponents converge to the same form \eqref{eq:tauqloc} as in the  localized case with $q^*=1/2$. The fact that $q^*=1/2$ at criticality is required by a fundamental symmetry of the multifractal spectrum \cite{mirlin1994distribution, mirlin2006exact, bilen2021symmetry}. This critical behavior implies that $P_q$ scales algebraically with $N$ for $q<1/2$, as shown in the upper panel of Fig.~\ref{figpqcrit} for $q=0.25$, $p=0.25$. On the contrary, $P_q$ at large values of $q>1/2$ tends logarithmically slowly to a localized behavior $P_q \approx A_1+A_2/(\ln N)^{0.5}$ with $A_1$ and $A_2$ two finite constants, in agreement with the theoretical prediction \cite{fyodorov1991localization,tikhonov2019statistics}. Our previous prediction $P_q \sim (\log_2N)^{-\sigma_q}$ \cite{scaling17}, based on the extension to infinite dimension of the multifractal scaling $P_q\sim L^{-\sigma_q}$ at the transition in finite dimension  \cite{evers2008anderson}, with $L \sim \log_2 N$ the linear dimension of the system, fits also very well the data and is almost indistinguishable from the other form predicted analytically \cite{fyodorov1991localization,tikhonov2019statistics}. These two behaviors give, however, two different pictures of the critical behavior: either localized with logarithmic finite size corrections, or multifractal on few branches (because the number of sites on a branch of the graph scales like $\log_2 N$ and $\sigma_2 < 1$). We are unable to distinguish clearly between these two possibilities. There also remains the question of whether the analytical predictions of \cite{fyodorov1991localization,tikhonov2019statistics} are valid for low values of $K<2$ as considered here (see the discussion of correlation functions, in particular). In the following, we will continue to assume $P_q \sim (\log_2N)^{-\sigma_q}$ as it simplifies some aspects of the physical interpretations of our analysis.

These behaviors are valid independently of $p$; only the value of the critical disorder depends on the value of $p$, as illustrated in Fig.~\ref{figpWc} in Appendix \ref{AppendixA}. More refined analysis such as finite-size scaling  will be employed later in Sec. \ref{secFSS} to characterize the transition.

\subsection{Correlation functions}\label{sec:corrfunctions}

\begin{figure}[!t]
	\includegraphics[width=0.95\linewidth]{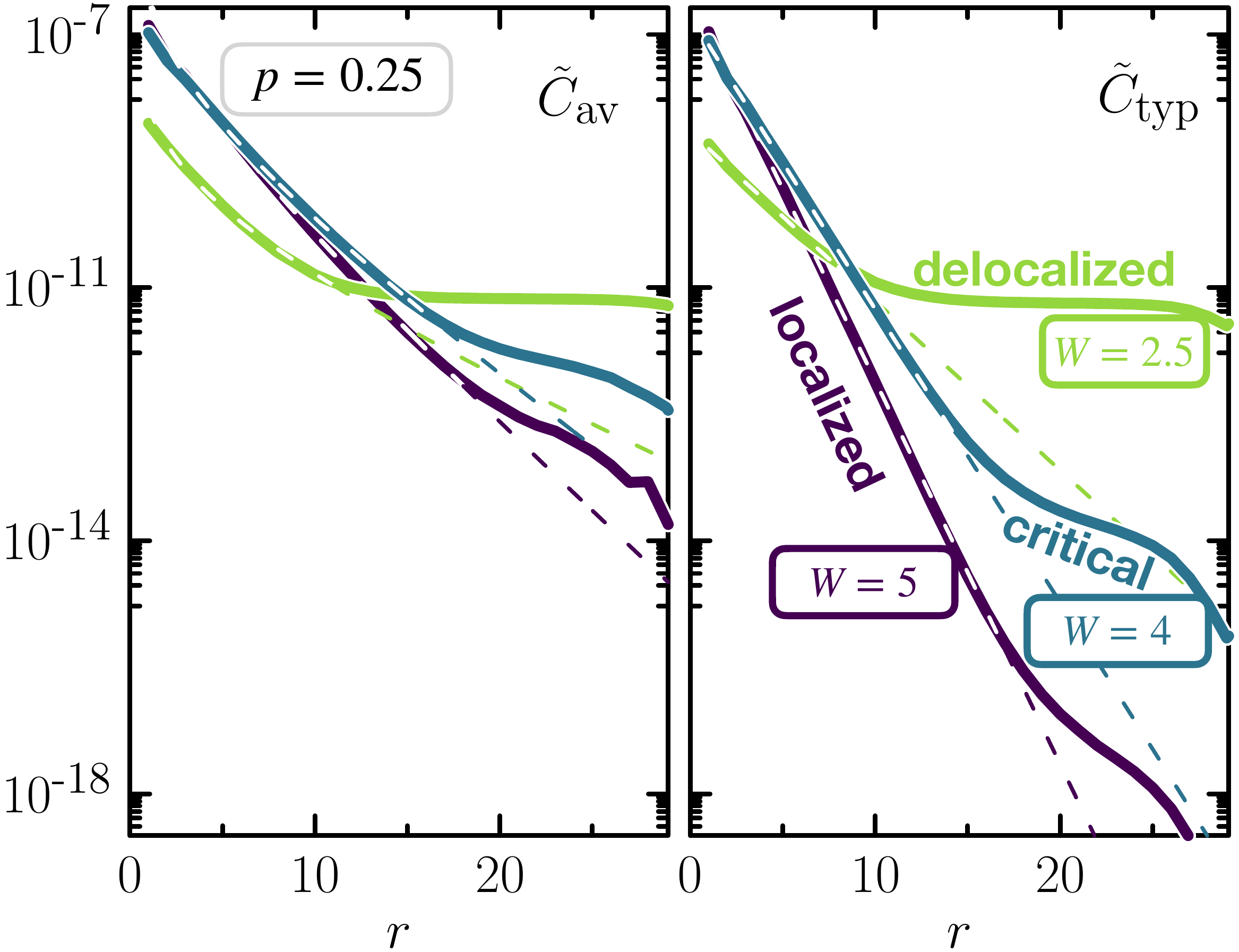}
	\caption{\label{fig:corr1} Average $\Ctav$ (left) and typical $\Cttyp$ (right) correlation functions of eigenvector amplitudes given by Eqs.~\eqref{chav} and \eqref{cttyp}, respectively, for $p=0.25$, $N=2^{17}$, and three values of disorder $W=2.5$ (delocalized),  $4.0\, (\approx W_c),$ and  $5.0$ (localized). The average correlation function is controlled by $\xi_\parallel$ while the typical one is associated with $\xi_\perp$, with $\xi_\parallel \gg \xi_\perp$, resulting in orders of magnitude differences of the two correlation functions. Dashed lines are small-$r$ fits corresponding to Eq.~(\ref{eq:simcav}) for $\Ctav$ and Eq.~(\ref{eq:simctyp}) for $\Cttyp$. $\xi_\parallel$, shown in Fig.~\ref{xi_xipar}, diverges at the transition, while $\xi_\perp$, represented in the lowest panel of Fig.~\ref{figtauq}, reaches a finite universal value at $W_c$. The values obtained for the parameter $\alpha$ of  Eq.~(\ref{eq:simcav}) are $1.48 \pm 0.23$ for $W=2.5$; $1.51\pm 0.18$ for $W=4$ and $1.38\pm 0.15$ for $W=5$, consistent with the theoretical prediction $\alpha=3/2$ \cite{tikhonov2019statistics}.}
\end{figure}

Correlations between wave-function amplitudes are another important characterizion of localization properties \cite{evers2008anderson}. We will see here that they make it possible to highlight in a particularly clear way certain non-ergodic properties of the states. Although many studies have focused on the ergodic or non-ergodic properties of the delocalized phase (see, e.g., \cite{monthus2011anderson, biroli2012difference, de2014anderson,  altshuler2016nonergodic}), the localized phase has quite remarkable properties from this point of view \cite{twoloc20}. Thus, an analogy can be made with the problem of directed polymers \cite{abouchacra73, derrida94, monthus09, somoza2007universal, pietracaprina2016forward, PhysRevLett.122.030401}, an important statistical physics model which, despite its simplicity, has a glassy phase where the replica symmetry is broken \cite{derrida1988polymers, cook1989directed, feigelman10}. This analogy suggests that the quantum states in the localized phase have spatial properties dominated by rare events: they only explore a finite number of rare branches, among the exponential number of branches available. Thus, the states are not isotropically localized, i.e. with the same localization length in all directions, but on the contrary the localization length $\xi_\parallel$ can be abnormally large along certain branches, and the localization length $\xi_\perp$ much smaller in the transverse directions. The rare branches nevertheless dominate the averages \cite{twoloc20}. An important point to emphasize is that these are not rare events in the common sense of a rare configuration of disorder. Here, for each disorder configuration, there exists rare branches. As we will see, this physical image implies several interesting properties, in particular on the critical behavior of the system. It is central to our work.

It is important to try to support this analogy and the physical image we have discussed above. The correlation functions that we are going to introduce here allow this. 
The usual way to define an average correlation function between the amplitudes of a wave function is \cite{evers2008anderson}
\begin{equation}
\label{chav}
\Ctav(r)=\left\langle\frac{1}{N}\sum_{i=1}^N  \frac{1}{N(r)} \sum_{j, d(i,j)=r} \left|\psi_{i}\right|^{2}\left|\psi_{j}\right|^{2}\right\rangle \;,
\end{equation}
where the distance $r$ between two vertices is taken as the minimal path along the small-world network, and $N(r)$ is the average number of sites at a distance $r$ from a given site. 
The supposed presence of rare branches for each state and each configuration implies that the sum over the sites $j$ at distance $r$ from site $i$, $d(i,j)=r$, will be controlled by $\xi_\parallel$. 

In order to compute a meaningful typical correlation function, which escapes the domination of the rare branches, we can replace the sum over sites at distance $r$ in Eq.~\eqref{chav} by the contribution of a single site at distance $r$, taken at random. 
By taking the logarithm, we will end up with a typical estimate of the correlation. Thus, we define:
\begin{equation}
\label{cttyp}
\Cttyp(r)=\exp\left\langle\ln\left( \frac{1}{N}\sum_{i=1}^N |\psi_i|^2|\psi_{j_0}|^2\right)\right\rangle,\quad d(i,j_0)=r,
\end{equation}
where a single site $j_0$ at a distance $r$ from $i$ is picked at random. The position of $j_0$ does not matter, as averaging over disorder is also performed in \eqref{cttyp}.

These definitions, that can be applied for any graph, are similar to the average and typical correlation functions that we defined in \cite{twoloc20}. In that paper, we considered the correlations along the 1D chain at the basis of the smallworld network (see discussion in Appendix \ref{appcorr}). Interpreting in the vicinity of point $i$ this chain as a typical branch, we could then define the average $\Cav(r)=\frac{1}{N}\sum_{i=1}^N \langle |\psi_i|^2|\psi_{i+r}|^2\rangle$ and typical $\Ctyp(r)=\exp\langle\ln( \frac{1}{N}\sum_{i=1}^N |\psi_i|^2|\psi_{i+r}|^2)\rangle$ with the average taken over random realizations. Following the 1D chain is equivalent to choose one arbritrary branch, which is not systematically the rare branch. This is similar to the random choice of $j_0$ in Eq.~\eqref{cttyp}. We have checked the good agreement between these definitions on the 1D chain and the new graph correlation functions \eqref{chav} and \eqref{cttyp} (see Appendix).

The average correlation function was described analytically in \cite{tikhonov2019statistics}. It follows in the localized regime 
\begin{equation}
\label{eq:simcav}
\Ctav(r)\sim K^{-r}\frac{e^{-r/\xi_\parallel}}{r^\alpha},
\end{equation}
with $\alpha = 3/2$, where we have replaced the localization length considered in \cite{tikhonov2019statistics} by $\xi_\parallel$ introduced in our work.
On the contrary, we found in \cite{twoloc20} that 
\begin{equation}
\Cttyp(r)\sim e^{-r/\xi_{\perp}}.
\label{eq:simctyp}
\end{equation}
Figure~\ref{fig:corr1} shows the strong difference between these two correlations and their associated localization lengths. The data for the average correlation function are very well fitted by Eq.~\eqref{eq:simcav} with values of the fitting parameter $\alpha \approx 1.5$, confirming the prediction of \cite{tikhonov2019statistics}. On the other hand, the typical correlations decay exponentially with no power law corrections and a much smaller localization length $\xi_\perp \ll \xi_\parallel$ leading to values of typical correlations orders of magnitude smaller than for the average. We can interpret $\xi_\parallel$ as the average localization length and $\xi_\perp$ as the typical one.
We will go back to $\xi_\parallel$ and $\xi_\perp$ later, by relating these characteristic lengths to multifractal properties and finite-size effects. In particular, we will see that they are associated with different critical exponents.   

\subsection{Spectral statistics}
\label{specstat}

\begin{figure}[!t]
\centering
\includegraphics[width=0.95\linewidth]{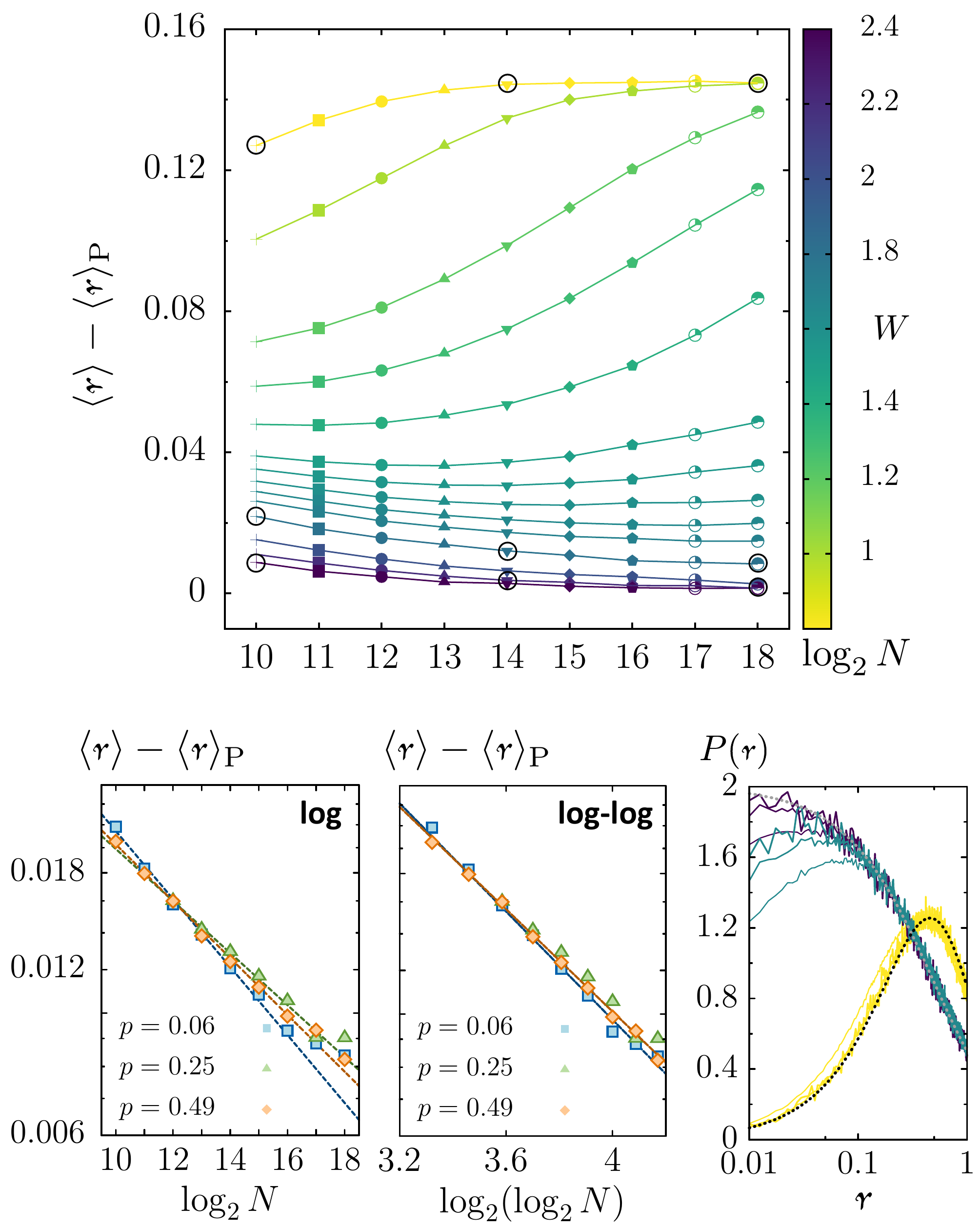} 
\caption{Top: Average spectral gap ratio $\langle\raur\rangle - \langle\raur\rangle_\mathrm{P}$ as a function of system size for $p=0.06$ and various disorder strengths. Bottom left-middle:  $\langle\raur\rangle-\langle\raur\rangle_\mathrm{P}$ slightly above the critical point as a function of system size, for $(p,W)=(0.06,1.8)$ (squares) and $(0.25,4.2)$ (triangles) with Gaussian distributed disorder, and $(p,W)=(0.49,18)$ (rhombus) with box distributed disorder. Straight lines are a fit of the form \eqref{fitlog} (left panel) and \eqref{fitloglog} (middle panel) with $(A,\alpha)\approx (0.90,1.63)$ and $(B,\beta_\raur)\approx (0.09,0.14)$ for $p=0.06$, $(A,\alpha)\approx (0.66, 1.49)$ and $(B,\beta_\raur)\approx (0.06,0.11)$ for $p=0.25$ and $(A,\alpha)\approx (0.73, 1.55)$ and $(B,\beta_\raur)\approx (0.07,0.12)$ for $p=0.49$. Bottom right: Distribution of ratios for $N=2^{10},2^{14},2^{18}$, $p=0.06$ and $W=0.8$ (yellow), $W=1.80$ (blue-green), and  $W=2.4$ (dark-purple). The gray/black dotted curves correspond to the Poisson and Wigner-Dyson distributions (see text).}
\label{pr_regimes}
\end{figure}

We now investigate quantities defined from the spectrum of $H$. Let $E_1,E_2,\ldots$ denote the set of ordered eigenvalues of $H$, and $s_n\equiv E_{n+1}-E_n$ denote the spacing between consecutive energy levels. The distribution of the ratios $\raur_n=\min(s_n/s_{n-1},s_{n-1}/s_n)$ allows us to probe the transition from the localized regime, which should be governed by Poisson level statistics, to the delocalized regime, controlled by random matrix theory (RMT)~\cite{oganesyan2007localization,Kud18}. In the case of ergodic delocalized wave functions, RMT predicts a distribution of ratios close to the Wigner-Dyson surmise $P_{\mathrm{WD}}(\raur)=(27/4)(\raur+\raur^2)/(1+\raur+\raur^2)^{5/2}$, whereas localized states do not show significant level repulsion and thus have randomly distributed energy levels, whose ratios follow the Poisson distribution $P_{\mathrm{P}}(\raur)=2/(1+\raur)^{2}$~\cite{Ata13, atas2013joint}. 

In order to analyze the crossover from Poisson to Wigner-Dyson level statistics, we define the parameter $\eta_\raur$ as
\begin{equation}
\label{defetar}
\eta_\raur=\frac{\displaystyle \langle\raur\rangle-\langle\raur\rangle_{\mathrm{P}}}{\displaystyle \langle\raur\rangle_{\mathrm{WD}}-\langle\raur\rangle_{\mathrm{P}}},
\end{equation}
where $\langle \cdot \rangle$ denotes an ensemble average and $\langle\raur\rangle_{\mathrm{P}}\approx 0.3863$ ($\langle\raur\rangle_{\mathrm{WD}}\approx 0.5359$) is the average of $\raur$ when $\raur$ is distributed according to $P_{\mathrm{P}}(\raur)$ ($P_{\mathrm{WD}}(\raur)$). The parameter $\eta_\raur$ ranges from $0$ for Poisson statistics to $1$ for Wigner-Dyson statistics. The observable $\raur$ has been considered in many studies to characterize the Anderson transition and the MBL transition \cite{biroli2012difference, tikhonov2016anderson, bertrand2016anomalous, fan2020superconductivity, valba2021mobility}. We believe that the quantity $\eta_\raur$ is more appropriate to describe the transition, in particular its critical behavior. We will come back to this statement later.

Figure~\ref{pr_regimes} shows $\langle\raur\rangle - \langle\raur\rangle_\mathrm{P}$ as a function of the system size for large, critical and small disorder strengths. For small disorder ($W=0.8$ for $p=0.06$), $\langle\raur\rangle$ tends to its RMT value associated with states delocalized over the whole network with large overlaps; for large disorder ($W=2.4$ for $p=0.06$), $\langle\raur\rangle$ converges to its Poisson value, reflecting the presence of localized states. The inset of Fig.~\ref{pr_regimes} shows the distribution of ratios. For small disorder, the distribution is close to $P_{\mathrm{WD}}(\raur)$. For large disorder, the distribution is close to $P_{\mathrm{P}}(\raur)$.

Between these two extremes, the behavior of $\eta_\raur$ as a function of $N$ is non-trivial. In the delocalized phase near the transition, $W \lesssim W_c$, we first observe a decrease followed by an increase. This has been described in several studies \cite{tikhonov2016anderson, tarquini2016level, biroli2022critical}, which associate a correlation volume $\Lambda$ with the value of $N$ at the minimum of $\eta_\raur$. In fact, at the transition point (or slightly above the transition point $W_c$ determined by finite-size scaling , see next section) we find, in accordance with~\cite{tikhonov2019statistics, sierant2022universality}, that the spectral statistics converges to Poisson logarithmically slowly with the system size, following
\begin{equation}\label{fitlog}
\langle\raur\rangle \approx \langle\raur\rangle_\mathrm{P}+\frac{A}{(\log_2 N)^{\alpha}}.
\end{equation}
This behavior is shown in Fig.~\ref{pr_regimes} (bottom middle panel). However one can not entirely exclude an algebraic decay as
\begin{equation}\label{fitloglog}
\langle\raur\rangle \approx \langle\raur\rangle_\mathrm{P}+\frac{B}{N^{\beta_\raur}},
\end{equation}
that we will justify later.
The non-monotonic behavior of $\eta_\raur$ as a function of $N$ for $W<W_c$ can therefore be interpreted as first following the critical behavior, until $N$ reaches the correlation volume beyond which the system is of sufficiently large size for delocalization to manifest, i.e. $\eta_\raur$ increases. This is after all a behavior commonly observed in continuous phase transitions. 

What is less so is the behavior of $\eta_\raur$ at the threshold of the transition, which decreases logarithmically to its Poisson value. At the Anderson transition in finite dimension, $\eta_\raur$ takes an intermediate value and the distribution of $\raur$ is intermediate between Poisson and RMT \cite{fan2020superconductivity}. This involves a crossing of the curves of $\eta_\raur$ as a function of $W$ for different values of $N$, thus defining $W_c$. On the contrary, in the present case, a drift of these crossings is observed, which simply means that at the threshold, $\eta_\raur$ is not constant but varies with $N$. The fact that the spectral statistic described by $\raur$ converges slowly towards Poisson indicates a quasi-absence of level repulsion of the states at $W_c$, which is consistent with the strong multifractality of critical states, with $D_q =0$ for
$q>1/2$. Critical states are therefore quasi-localized, thus do not show level repulsion. On the contrary, in finite dimension, critical states display multifractality with $0<D_q<1$ for all $q>0$, and thus show level repulsion and an intermediate statistics for $\raur$ \cite{shklovskii1993statistics, bogomolny1999models}.

Note that such a non-monotonous behavior has been observed for other observables such as $\tilde{\tau}_2$, and analyzed following the same type of reasoning, in the delocalized phase close to the transition \cite{tikhonov2016anderson, biroli2018delocalization}. This has allowed for a determination of the correlation volume $\Lambda(W)$ and its exponential divergence at $W_c$. Nevertheless, we think that this does not allow concluding with certainty as to the ergodic nature of the delocalized phase and does not allow either to describe the nature of the phase transition, for example if it is of the Kosterlitz-Thouless type. This motivates our analysis of the critical properties through finite-size scaling  that we describe in the next section.

\section{Finite-size scaling  theory}
\label{secFSS}

We qualitatively described the behaviors of different observables across the Anderson transition on the smallworld network we consider: moments $P_q$, correlation functions, and spectral statistics. A number of observations strongly motivate the use of a more elaborate method of analysis to characterize the transition:

(i) First, it is not at all clear how to determine the value of $W_c$ from these observations. Indeed, certain observables such as the inverse participation ratio $P_2$ or $\eta_\raur$ tend towards a localized behavior in the vicinity of $W_c$. On the contrary, $P_q$ for $0<q<1/2$ has a more usual intermediate multifractal behavior, similar to what is found in the Anderson transition in finite dimension. Some authors have recently proposed a method to determine $W_c$ indirectly and quite precisely \cite{parisi2019anderson, tikhonov2019critical, sierant2022universality}: the type of graph considered should, in the thermodynamic limit, be equivalent to a Bethe lattice. Indeed, the typical size of the loops present in the smallworld network varies like $\log_2 N$ and therefore diverges in the thermodynamic limit, suggesting, if we neglect the subdominant effects of small-size loops, a mapping between this network and the Bethe lattice. Moreover, it is possible to accurately determine $W_c$ in the Bethe lattice. 

We describe here another approach based on the finite-size scaling  method. This approach that we propose does not presuppose any form for the critical behavior of the considered observable, nor any particular property of the network. It makes it possible to quantitatively determine values of $W_c$ compatible with our numerical data for each observable considered. Of course, the good correspondence between the different values of $W_c$ obtained for different observables and with the predictions based on the mapping to the Bethe lattice \cite{parisi2019anderson, tikhonov2019critical, sierant2022universality} will have to be examined closely.

(ii) Second, finite size effects are particularly marked in the neighborhood of the transition, and do not allow to conclude as to the ergodic nature of the delocalized phase. Several studies \cite{de2014anderson, altshuler2016nonergodic, kravtsov2018non} have described extrapolations of the behaviors of $P_q$ towards the thermodynamic limit suggesting a non-ergodic delocalized behavior. Other studies \cite{tikhonov2016anderson, biroli2018delocalization}, considering the same observables on the same systems, conclude, via an analysis of the minimum of $\tilde{\tau}_2$ in favor of an ergodic 
behavior. They identify a characteristic correlation volume $\Lambda(W)$ from this minimum, which is found to diverge exponentially at $W_c$, $\Lambda(W) \sim \exp[a (W_c - W)^{-1/2}]$, as expected theoretically \cite{fyodorov1991localization, mirlin1994distribution, tikhonov2019critical}. However, the behavior beyond that characteristic scale has not been shown to be ergodic. This illustrates well the subtlety of the analysis of the vicinity of the transition. In a usual phase transition, an extrapolation to the thermodynamic limit is performed by the method of finite-size scaling . It is this approach that we have followed in \cite{scaling17, twoloc20} and that we will describe in detail in this section.

(iii) Third, it is interesting to understand the nature of this phase transition, particularly in view of its supposed link with the MBL transition \cite{altshuler1997quasiparticle, basko2006metal, biroli2017delocalized, roy2020fock, roy2020localization, tarzia2020many, tikhonov2021anderson}. A number of elements suggest that this is not a second-order phase transition, in contrast to the Anderson transition in finite dimension. Is it a Kosterlitz-Thouless type transition as would be the case for the MBL transition in the phenomenological renormalization group approach \cite{}? Note that although many analytic predictions have been made for the Anderson transition on random graphs, see e.g.~\cite{fyodorov1991localization, mirlin1994distribution, tikhonov2019critical, sonner2017multifractality, de2014anderson, kravtsov2018non, aizenman2011absence,aizenman2011extended}, the nature of the transition and its renomalization flow have not been described. A careful finite-size scaling  approach can give some elements allowing progress on this important question.

The scaling theory of localization \cite{abrahams1979scaling, pichard1981finite, mackinnon1981one, evers2008anderson} has played an extremely important role in the field of Anderson localization.
On the theoretical level, it allowed the understanding of the Anderson transition as a second-order phase transition, with the characterization of its lower critical dimension and of its universality classes. Numerically, it allowed the demonstration that dimension two in the orthogonal class is always localized, and the precise and controlled determination of the critical exponent $\nu$ of the transition from different observables, in different dimensions and universality classes.

Implementing the scaling approach to analyze numerical data relies on a number of steps: 

First, a scaling law hypothesis is made. Generally, a single parameter scaling law is sufficient, which takes the form:
\begin{equation}
X(W,L) = X(W_c,L)\, F[L/\xi(W)] \; ,
\end{equation}
where $X$ is the observable considered, $X(W_c,L)$ its critical behavior, and the scaling function $F$ depends only on the ratio of $L$ the linear size of the system (here the diameter of our graph, $L=d_N$) by the characteristic length scale $\xi$. $\xi$ depends only on $W$ and diverges at the transition as $\xi \sim \vert W-W_c\vert^{-\nu}$ with the critical exponent $\nu$. The two asymptotic behaviors of $F$ at large $L\gg \xi(W)$ for $W>W_c$ and $W<W_c$ describe the asymptotic behavior of the observable $X$ in the two phases, localized or delocalized. Multiplying $F$ by the critical behavior allows to describe first a critical behavior for $L \ll \xi(W)$ and then the behavior associated to one of the phases. Irrelevant corrections can be considered in the form of a two-parameter scaling function (see \cite{PhysRevLett.82.382}).

In the second step, we test its compatibility with the numerical data. Different approaches can be used. In the first approach we try to collapse each curve for $X(W,L)/X(W_c,L)$ as a function of $L$ for different values of $W$ onto a single scaling function by plotting the data as a function of $L/\xi(W)$ \cite{mackinnon1981one, PhysRevA.80.043626}. In that procedure, no assumption is made on the form of $F$ or $\xi$: they are determined by the best collapse of the curves. This is a good check of the scaling hypothesis, even far from the transition where non-linear corrections on the scaling function and scaling parameter are usually observed. 

Another approach is to use a Taylor expansion of the scaling function and scaling parameter close to $W_c$ and to fit the data with this ansatz \cite{PhysRevLett.82.382, lemarie2009universality, PhysRevB.84.134209}. This is the most controlled procedure to determine the critical exponent $\nu$, in particular in the presence of irrelevant corrections. Nevertheless, very precise data covering a sufficiently large range of disorder $W$ and system size $N$ are crucial to test conclusively the scaling assumption and arrive at a controlled estimate of the critical parameters $W_c$ and $\nu$. 

In our study, we have used both approaches to describe the critical properties of the Anderson transition on the small-world network we consider. First of all, we will see that there are different scaling assumptions that can be made in this infinite dimension case. In order to determine whether a scaling hypothesis agrees with our data, we have found that one must consider a large range of $W$, the system sizes being limited to $N\le 2^{21}$ in our case. Far from the transition, non-linear corrections are important and make the approach by Taylor expansion difficult. Finally, we found unusual behaviors in the vicinity of the transition, with different scaling functions on either side of the transition and exponentially diverging scaling parameters, properties that are excluded for a second order phase transition, but which could be compatible with a Kosterlitz-Thouless type transition. Nevertheless, in a certain number of cases, we have managed to make an analysis through the Taylor expansion approach of the scaling analysis, and determine the critical exponent in a precise and controlled way.


\subsection{Finite-size scaling  of the moments of the eigenfunctions} 
\label{fssmoments}
\subsubsection{Volumic or linear scaling} 
A specificity of random graphs and tree networks is that the volume scales exponentially with linear system size. As we showed in \cite{scaling17}, this implies the possibility of two types of scaling behavior:
\begin{equation}\label{eq:scahyp}
 \frac{\langle P_q\rangle}{\langle P_q^c\rangle} = \begin{cases}
                       F_\text{lin} \left(d_N/\xi\right) \\[6pt]
                        F_\text{vol} \left(N/\Lambda\right),
                     \end{cases}
\end{equation}
where $P_q^c \equiv P_q(W_c)$ is the critical behavior of the moment at $W=W_c$, $d_N \propto \log_2 N$ denotes the diameter of the random graph of volume $N$ (number of sites), $\xi$ denotes a characteristic length, and $\Lambda$ a characteristic volume. In finite dimension $d$, the two types of scalings $ F_\text{lin}$ and $ F_\text{vol}$ amount to the same behavior, since in that case $\Lambda \sim \xi^d$ and thus $F_\text{lin}(X) = F_\text{vol}\left(X^d\right)$ for $X=d_N/\xi$. In a graph however, $\Lambda = \mathcal{V}(\xi) \sim \exp(a \xi)$ and $N=\mathcal V(d_N)$; as a consequence, $N/\Lambda=\mathcal V(d_N-\xi)$, which is not a function of $d_N/\xi$, thus the two types of scalings are distinct.

In \cite{scaling17} we found that for $q=2$ a linear scaling holds in the localized phase while a volumic scaling better describes the delocalized phase. By contrast, in \cite{twoloc20} we observed a linear scaling in both phases for $q<1/2$. 
Our goal here is to discuss more thoroughly these scalings for different values of $q$.

\begin{figure}[!t]
  \includegraphics[width=0.99\linewidth]{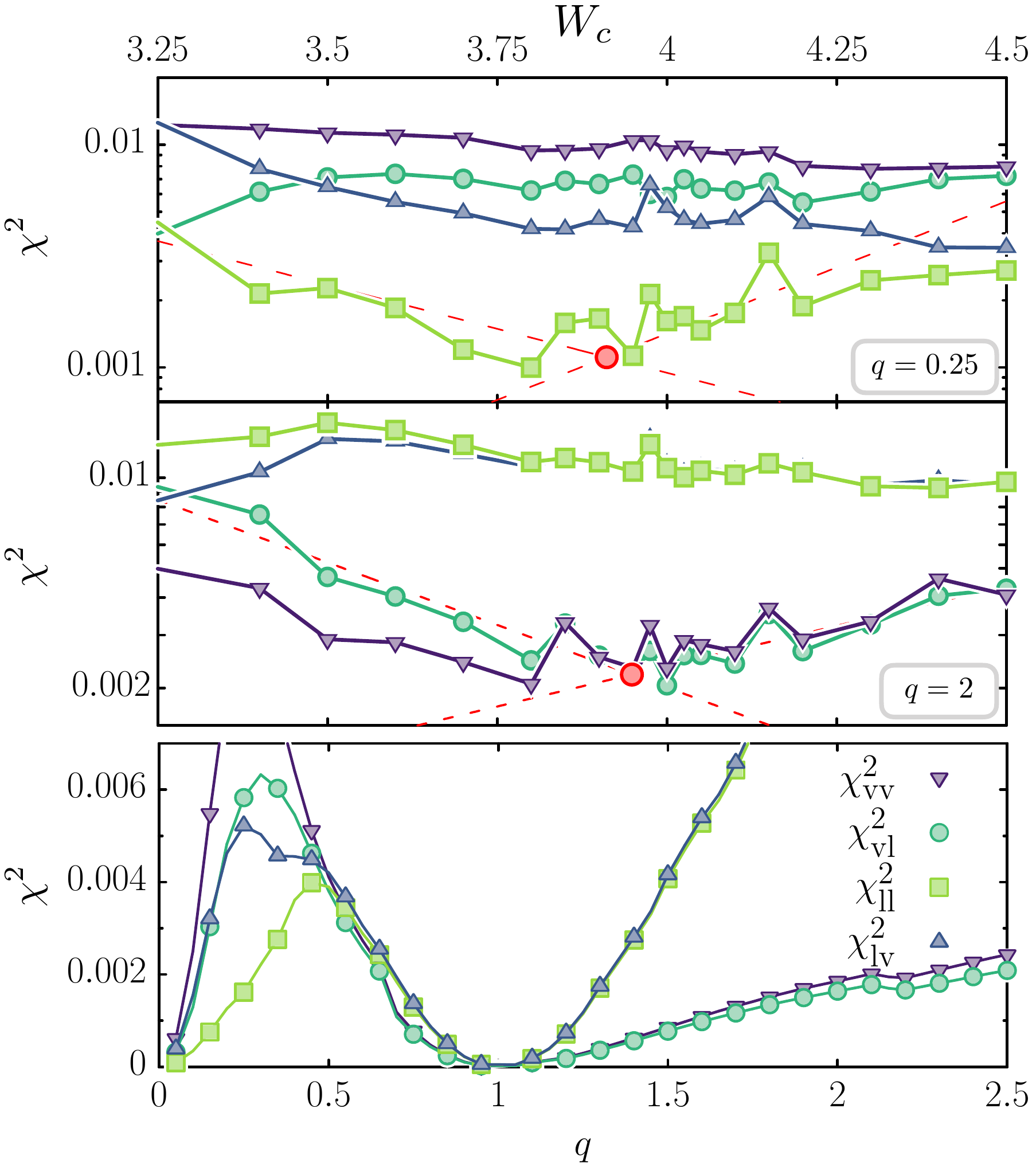} 
\caption{\label{chi2dosp} $\chi^2$ test for different scaling hypotheses. Top and center panel:
$\chideux_{\rm vl},\chideux_{\rm vv},\chideux_{\rm lv},\chideux_{\rm ll}$ as a function of $W_c$, for $q=0.25$ and $2$, with $p=0.25$ (symbol key as in bottom panel). The straight lines correspond to exponential fits near the minima for the lin-lin case (top) and the vol-lin case (bottom). The points marking the line crossings provide an estimate of $W_c$ ($\approx 3.9$ in both cases).
Bottom panel:
$\chideux$ as a function of $q$ for $p=0.25$, $W_c=4$; The finite-size scaling  was done with sizes $2^{10}$--$2^{20}$.
}
\end{figure}

\begin{figure*}[!t]
\includegraphics[width=1\linewidth]{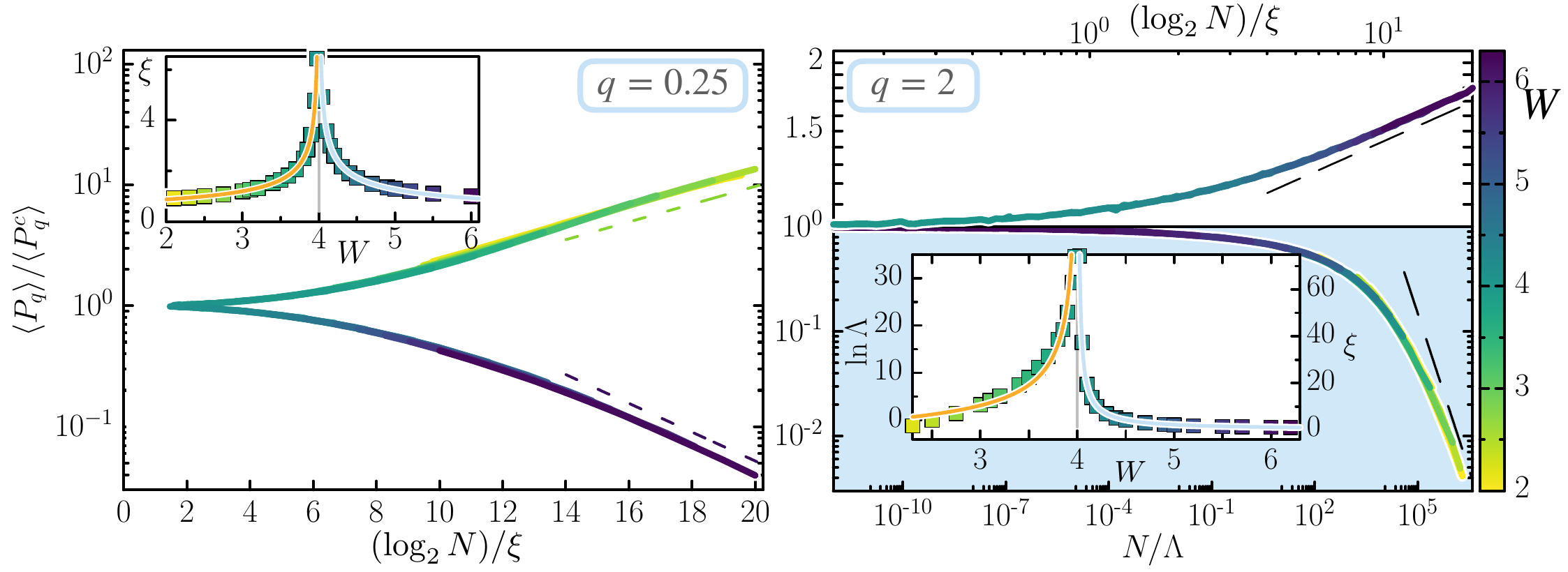}
	\caption{\label{figFSS05p025} Finite-size scaling  of multifractal properties $\langle P_q \rangle (W, N)$ for $p=0.25$, with $W_c$ set to its optimal value $W_c =4$, see Fig.~\ref{chi2dosp}. Left: for $q=0.25<1/2$, the data follow a linear scaling on both sides of the transition. The data shown in Fig.~\ref{figmoments} upper panel, for all system sizes {$2^{9}\le N \le 2^{20}$  and $32$  $W$ values in the large range $2.1\le W \le 6.2$ } have been put onto a single scaling function with two branches. The upper branch corresponds to the delocalized regime $W<W_c$ and the lower branch to the localized phase, $W>W_c$. The straight dashed lines show the asymptotic behavior of the scaling function $F_{\rm lin} (X) \sim \mathcal V(X)^{\beta_{\rm loc/deloc}}$ discussed in Sec.~\ref{sec:physint}.
		The inset shows the scaling length $\xi$ (which is not the localization length for $W>W_c$, as explained in Sec.~\ref{sec:physint}) as a function of $W$. The lines are fits with $\xi\sim |W-W_c|^{-\nu_{\rm loc/deloc}}$. The corresponding critical exponents obtained are: $\nu_{\rm deloc}\approx  0.47$,
		$\nu_{\rm loc} \approx 0.45$. 
		Right: $q=2$, here the upper branch corresponds to the localized phase and has linear scaling, while the lower branch corresponding to the delocalized phase shows volumic scaling, as highlighted by the light blue background. The straight dashed lines correspond to the asymptotic behavior deduced from the scaling hypothesis (see Sec. \ref{sec:physint}). In the localized phase $F_\text{lin}(X)\sim X^{\sigma_2} $ (see Fig.~\ref{figpqcrit} and text), while in the delocalized phase it is given by $F_\text{vol}(X)\sim X^{-1}$ which implies an asymptotic ergodic behavior $\langle P_2 \rangle \sim N^{-1}$ for $N \gg \Lambda(W)$. In the inset we show the logarithm of the correlation volume $\Lambda(W)\sim \exp[\alpha (W_c-W)^{-\nu_{\rm deloc}}]$ and the localization length $\xi(W) \sim (W-W_c)^{-\nu_{\rm loc}}$ as a function of $W$. In the localized phase we get $\nu_{\rm loc} \approx 0.94$. In the delocalized phase the orange line corresponds to a fit of $\ln \Lambda$ with the function $A_1+A_2 (W_c-W)^{-0.5}$, showing that our data are in good agreement with a critical exponent $\nu_{\rm deloc} = 0.5$.}
\end{figure*}

\subsubsection{Determination of $W_c$ and of the type of scaling}
\label{methodFSS}

We consider a finite-size scaling  (FSS) procedure where the curves for different values of $W$ are rescaled so as to collapse onto a single scaling function, with the only assumption that the scaling function is either of the form $F_\text{lin}$ or $F_\text{vol}$, as in Eq.~\eqref{eq:scahyp}. The value of disorder which gives the best collapse of the data is characterized by quantitatively estimating the goodness of the scaling procedure used. This yields a critical value $W_c$, as well as the dependence of the scaling parameter $\xi$ or $\Lambda$ on $W$. 

The value of $W_c$ is determined by testing different possible values of $W_c$ and the two possible scalings (volumic or linear) in a systematic way, and assessing the quality of the resulting scaling via a $\chi^2$ test. This approach is detailed in Appendix \ref{secfssapp}. Depending on the scaling hypothesis we make, we obtain four possible $\chi^2$ values : $\chideux_{\rm vv}$ (delocalized volumic + localized volumic),  $\chideux_{\rm ll}$ (delocalized linear + localized linear), $\chideux_{\rm vl}$ (delocalized volumic + localized linear),  $\chideux_{\rm lv}$ (delocalized linear + localized volumic).
The results are reported in Fig.~\ref{chi2dosp}, where we show $\chideux(W_c)$ for $p=0.25$, and for two values of $q$ (larger or smaller than $1/2$). 
For small $q<1/2$ (here $q=0.25$), the smallest $\chideux$ is attained for the linear-linear hypothesis. By contrast, for $q=2$ the best scaling corresponds to the  delocalized volumic, localized linear case. In both cases the optimal hypothesis has a minimum for $W_c\approx 3.9\pm 0.2$. This estimate of the critical point is compatible with the value $W_c=3.8$ obtained in Sec.~\ref{fss_spec} from finite-size scaling  of the spectrum.
Moreover, using this approach for $p=0.06$ we find $W_c\approx 1.7$ -- $1.75$.
For $p=0.49$ we find $W_c\approx 17$ -- $17.5 $ . 
These values are compatible with the theoretical values recently obtained in \cite{tikhonov2019critical, parisi2019anderson, sierant2022universality}.

As mentioned above, the most relevant scaling hypothesis depends on the value of $q$. In order to investigate this more thoroughly, in Fig.~\ref{chi2dosp} (bottom) we display the values of $\chideux(W_c)$ as a function of $q$ for the four possible scaling hypotheses $\chideux_{\rm vl},\chideux_{\rm vv},\chideux_{\rm lv},\chideux_{\rm ll}$, at the previously determined $W_c=4$. These plots show that  a crossover occurs : while our analysis clearly indicates a linear-linear scaling for small $q\lesssim 0.5$, for larger $q$ ($q>1$) a linear-volumic scaling describes better the data.

\subsubsection{Finite-size scaling  and critical exponents}
Now that the type of scaling and the value of the critical disorder have been determined, we can present the outcome of this procedure which is the behavior of the scaling parameter ($\xi$ or $\Lambda$) and of the scaling function. Close to $W_c$, we expect $\xi \sim \vert W-W_c\vert^{-\nu}$, with $\nu$ the associated critical exponent \cite{scaling17, twoloc20}. In the delocalized phase, the scaling volume $\Lambda$ has been theoretically predicted \cite{refId0} to behave as 
\begin{equation}
\label{comportementlambda}
\Lambda\sim\exp(\alpha \vert W-W_c\vert^{-\nu})
\end{equation}
with $\nu= 1/2$. The results of our finite-size scaling  procedure are displayed in Fig.~\ref{figFSS05p025} for $p=0.25$ and two values of $q$, smaller and larger than $1/2$. Our findings are summarized in Table \ref{tableresume}. 

The left panel of Fig.~\ref{figFSS05p025} corresponds to $q=0.25<1/2$. The data shown in Fig.~\ref{figmoments} upper panel, for all system sizes $2^{9}\le N \le 2^{20}$  and $32$ values of $W$ in the large range $2.1\le W \le 6.2$ , collapse onto a single scaling function with linear scaling on both sides of the transition. The upper branch of the scaling function corresponds to the delocalized regime $W<W_c$ and the lower branch to the localized phase, $W>W_c$. The straight dashed lines show the asymptotic behaviors $F_{\rm lin} (X) \sim X^{-\beta_{\rm loc}}$ and $F_{\rm lin} (X) \sim X^{\beta_{\rm deloc}}$ that we will explain in Sec.~\ref{sec:physint}. The scaling parameter $\xi$ diverges as $\xi\sim |W-W_c|^{-\nu}$ with $\nu\approx 1/2 $ on both sides of the transition. In Appendix \ref{appotherparam} we show the stability and universality of $\nu\approx 1/2$ by obtaining comparable results for a wide range of values of $p$ (see Fig.~\ref{stab_nu}) and in a range of $W_c$ close to the optimal value considered here. 

The right panel of Fig.~\ref{figFSS05p025} corresponds to $q=2$. The scaling is less conventional. In the localized regime, a linear scaling puts the data shown in Fig.~\ref{figmoments} (lower panel) for $W>W_c$ onto the upper branch of the scaling function. On the other hand, volumic scaling better describes the data in the delocalized regime $W<W_c$ corresponding to the lower branch. This volumic scaling implies an asymptotic ergodic behavior $\langle P_2 \rangle \sim N^{-1}$ for $N\gg \Lambda(W)$, shown by the straight dashed line corresponding to $F_\text{vol}(Y)\sim Y^{-1}$. The divergence of the correlation volume $\Lambda(W)$ at the transition is found to be compatible with Eq.~\eqref{comportementlambda} with $\nu_{\rm deloc} = 0.5$.
In the localized regime, the scaling length $\xi(W)$ diverges as $\xi(W) \sim (W-W_c)^{-\nu_\text{loc}}$, with a critical exponent $\nu_\text{loc}\approx 1$. 

We therefore observe a single critical exponent $\nu \approx 0.5$ in the delocalized phase, but two critical exponents $\nu \approx 0.5$ and $\nu_{\rm loc} \approx 1$ in the localized phase. 

{\renewcommand{\arraystretch}{2}
\begin{table}
\begin{tabular}{|c|c|c|}
\hline\hline 
 & Delocalized & $\phantom{{e}}$Localized$\phantom{{e}}$\tabularnewline
\hline 
\hline 
\multirow{4}{*}{$q=0.25$} & linear & linear\tabularnewline
\cline{2-3} 
 & $F_{\textrm{{lin}}}(X)\sim X^{\beta_{\textrm{deloc}}}$ & $F_{\textrm{{lin}}}(X)\sim X^{-\beta_{\textrm{loc}}}$\tabularnewline
\cline{2-3} 
 & $\xi\sim|W-W_{c}|^{-\nu}$ & $\xi\sim|W-W_{c}|^{-\nu}$\tabularnewline
\cline{2-3} 
 & \multicolumn{2}{c|}{$\nu\approx0.5$}\tabularnewline
\hline 
\multirow{4}{*}{$q=2$} & volumic & linear\tabularnewline
\cline{2-3} 
 & $F_{\textrm{{vol}}}(Y)\sim Y^{-1}$ & $F_{\textrm{{lin}}}(X)\sim X^{\sigma_{2}}$\tabularnewline
\cline{2-3} 
 & $\Lambda\sim\exp(\alpha\vert W-W_{c}\vert^{-\nu_{\textrm{{deloc}}}})$ & $\xi\sim|W-W_{c}|^{-\nu_{\textrm{{loc}}}}$\tabularnewline
\cline{2-3} 
 & $\nu_{\textrm{{deloc}}}\approx0.5$ & $\nu_{\textrm{loc}}\approx1$\tabularnewline
\hline 
\end{tabular}
\caption{Summary of the finite-size scaling  properties obtained from Fig.~\ref{figFSS05p025} for $q<q_c^*$ and $q>q_c^*$ where $q_c^*=0.5$. We give the asymptotic behavior of the scaling function \eqref{eq:scahyp} at large $X=d_N/\xi$ or $Y=N/\Lambda$.}
\label{tableresume}
\end{table}}

\subsubsection{Taylor expansion of the scaling function for $q<1/2$}

A more controlled procedure to determine the critical exponent(s) consists in making a Taylor series expansion close to $W=W_c$ of the scaling function and scaling length \cite{PhysRevLett.82.382, lemarie2009universality, PhysRevB.84.134209}. This procedure has the advantage that it allows to assess quantitatively the validity of the scaling hypothesis and the precision of the critical exponent. However, as we explain in Appendix \ref{taylor}, it works only for $q<0.5$ where there is a linear scaling on both regimes sufficiently close to the transition. We detail this approach in Appendix \ref{taylor}. This procedure confirms our previous analysis that $\nu_\text{loc} = \nu_\text{deloc} \approx 0.5$ for multifractal properties $\langle P_q\rangle$ for $q<1/2$.

\subsection{Finite-size scaling  of spectral statistics} 
\label{fss_spec}

\begin{figure}[!t]
\includegraphics[width=0.95\linewidth]{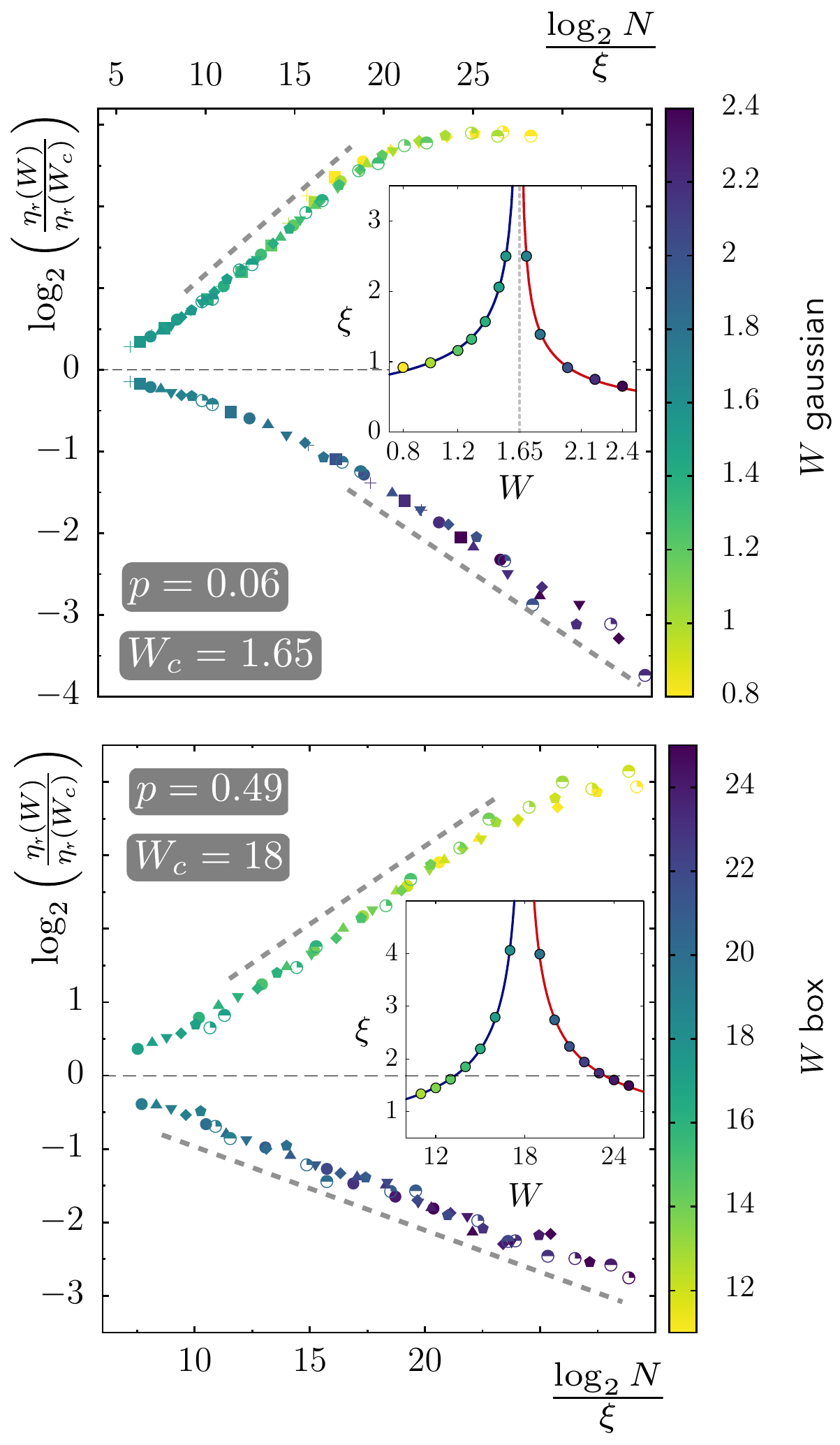}
\caption{Finite-size scaling  of $\eta_r(W)$ (see Eq.~\eqref{defetar}) for different values of $p$ and different disorder distributions (Gaussian for $p=0.06$ and uniform for $p=0.49$). The critical exponents extracted from these finite-size scalings are $\nu_\mathrm{deloc}\approx 0.50$ and $\nu_\mathrm{loc}\approx 0.51$ for $p=0.06$, $\nu_\mathrm{deloc}\approx 0.52$ and $\nu_\mathrm{loc}\approx 0.50$ for $p=0.49$.}
\label{figSS5}
\end{figure}

Let us now perform a finite-size scaling  analysis of $\eta_\raur$ defined in Eq.~\eqref{defetar}. The method is analogous to the one detailed in Sec.~\ref{methodFSS}. Results were presented in Fig.~3 of \cite{twoloc20} for $p=0.06$. Our procedure is the following: $\eta_\raur(W,N)$, plotted as a function of $\log_2 N$, is first rescaled by the critical behavior $\eta_\raur(W_c,N)$ (see Fig.~\ref{pr_regimes}), for system sizes ranging from $N=2^{10}$ to $2^{18}$. A second rescaling of sizes by some factor $\xi(W)$ is then performed along the $x$-axis, yielding a function of the form 
\begin{equation}
\label{eq:scaetar}
\frac{\eta_\raur(W)}{\eta_\raur(W_c)}=F_\mathrm{lin}\left(\frac{\log_2N}{\xi(W)}\right).    
\end{equation}
The index ``lin'' indicates that the rescaling is linear, that is, it is the linear size $\log_2 N$ of the system which is rescaled. In Fig.~\ref{figSS5}, we observe a very good collapse of the data on a single scaling curve in the localized phase and not too far from the transition point $W_c$ in the delocalized phase. However, we do observe small but systematic deviations in the ergodic regime at small disorder strengths. We will discuss the origin of these deviations in Sec.~\ref{sec:physint}. The scaling length $\xi(W)$ diverges in the vicinity of the critical disorder as $\xi(W)=A\,|W-W_c|^{-\nu}$, with a critical exponent $\nu\approx 0.5$. Similar scaling laws where reported in Ref.~\cite{Sad05} for Cayley trees and scale free networks.

In Fig.~\ref{figSS5} we show that such a finite-size scaling  holds for different values of $p$, close to $0$ and $1/2$, and for different disorder distributions. In Appendix \ref{AppendixB} we show that critical exponents are robust to variations in the critical point, for instance if $W_c$ is known only with limited accuracy, and remain close to $1/2$ on both the localized and delocalized sides of the transition. The same exponent is found for different values of $p$ and also for different random distributions of the disorder (uniform or Gaussian distribution). In Appendix \ref{higherorderapp} we present analogous results for higher-order spacing ratio statistics.

\section{Physical interpretation of the behavior of observables using finite-size scaling } \label{sec:physint}

In this section, we give an interpretation of the scaling laws observed previously by relating them to the multifractal properties and to the behavior of correlation functions described in section \ref{firstresults}. What will emerge from this is a simple physical picture of the localized and delocalized phases, and critical regime.

\subsection{Multifractal exponents and scaling functions}
\label{scalinghypotheses}

The different finite-size scaling  laws assumed for $\langle P_q \rangle$ and confirmed in Fig.~\ref{figFSS05p025} can be combined into a single scaling assumption of two variables:
\begin{equation}\label{eq:scahypV2}
\frac{\langle P_q\rangle}{\langle P_q^c\rangle} = F\left(X, Y\right), \ \quad \ X=\frac{d_N}{\xi},\quad Y=\frac{N}{\Lambda}.
\end{equation}
Here $\xi$ is the linear scaling parameter, and $\Lambda=\mathcal V(\xi)$ is the volume associated with the length $\xi$.
We will describe the asymptotic behaviors that $F(X,Y)$ needs to follow in order to recover the localized, multifractal or ergodic properties shown in Figs.~\ref{figtauq} and \ref{figFSS05p025}, and compare these predictions with our numerical results.

\subsubsection{Localized phase $W>W_c$}

\paragraph{Case $0<q< 1/2$:}
In the localized regime $W>W_c$ for $q<1/2$, Fig.~\ref{figtauq} shows that
$\tau_q$ has two distinct regimes, $\tau_q$ finite or $\tau_q \approx 0$, depending on whether $q$ is larger or smaller than a certain value $q^*(W)<1/2$. As a result, eigenstates have two distinct behaviors: multifractal, i.e.~$P_q \sim N^{-\tau_q}$, when $\tau_q$ is finite, or localized, $\tau_q \approx 0$. Note that this multifractality is not transient in $N$. $q^*(W)<1/2$ depends crucially on $W$, as shown in Fig.~\ref{figtauq} bottom. In turn, at fixed $q$, the behavior depends on the value of $W$: it is multifractal in the vicinity of the transition where $q^*(W)>W$, and localized far from the transition where $q^*(W)<q$. 
As we will see later, $q^*(W)$ is one of the most important critical properties of the localized phases.

Another important observation is that in the multifractal regime $q<q^*(W)$, the exponent $\tau_q$ is distinct from the one characterizing the critical behavior. We have thus a renormalized multifractality in that regime, at odds with the finite-dimensional Anderson transition, but similar to the non-ergodic delocalized phase \cite{scaling17}. 

This quite complex behavior can be described by means of the scaling function Eq.~\eqref{eq:scahypV2} provided we assume the following asymptotic behavior:
\begin{equation}\label{eq:asymptlocaql12}
F(X,Y) \sim \mathcal V(X)^{-Aq} + Y^{\tau_q^c},  \ \quad \ \text{for} \quad  X,Y\gg 1 \; ,
\end{equation} 
with $A$ some positive constant. As we explained in \cite{scaling17}, the first linear scaling term of Eq.~\eqref{eq:asymptlocaql12} will renormalize the critical multifractality. On the other hand, the second volumic term will bring a localized behavior.
  
We recall that at the transition $\langle P_q^c\rangle \sim N^{-\tau_q^c}$ with $\tau_q^c = (q/q^*_c -1)$ for $q\le q^*_c\equiv 1/2$ and $\tau_q^c = 0$ for $q> q^*_c\equiv 1/2$ (Eq.~\eqref{eq:tauqloc} with $q^*=q^*_c=1/2$). 
Since $\mathcal V(d_N/\xi)=N^{1/\xi}$, we get from Eqs.~\eqref{eq:scahypV2} and \eqref{eq:asymptlocaql12}: 
\begin{equation}
\label{locqpetit}
\langle P_q\rangle\sim N^{-q/q^*+1}+\Lambda^{-\tau_q^c},\qquad \frac{1}{q^*}=\frac{1}{q_c^{*}} +\frac{A}{\xi} .
\end{equation}
The second term of $ \langle P_q\rangle$ in Eq.~\eqref{locqpetit} is independent of $N$, while the first one goes to 0 or $\infty$ depending on the sign of the exponent. Namely, for $q<q^*$ the first term dominates, while for $q>q^*$ the second one dominates, which allows us to indeed recover the $\tau_q$ observed in Fig.~\ref{figtauq}, corresponding to Eq.~\eqref{eq:tauqloc} with $q^*(W)<q_c^*=1/2$ (since $A<0$).

\begin{figure}
	\includegraphics[width=.99\linewidth]{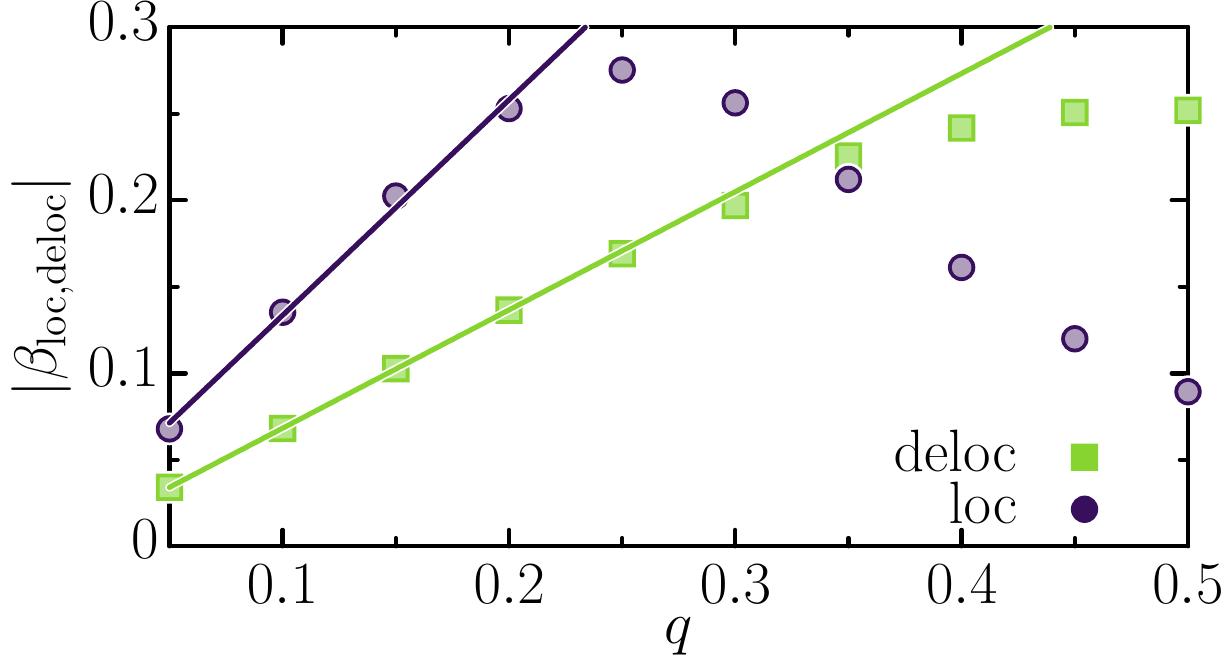}
	\caption{Parameter $|\beta_{\rm loc,deloc}|$ obtained by fitting the asymptotic behavior of the linear scaling function for $q<1/2$ with $\log_2(\langle P_q\rangle/\langle P_q^c\rangle) = \log_2 F_\text{lin}(\log_2 N/\xi) \approx A_0+ \beta_{\rm loc,deloc} (\log_2 N)/\xi\ $ for large $\log_2 N\gg \xi$ on both sides of the transition for $p=0.25$ and $W_c=4$. $A_0$ and $\beta_{\rm loc,deloc}$ are the two fitting parameters. In the localized regime, $\beta_{\rm loc}$ is negative, while $\beta_{\rm deloc}$ is positive in the delocalized regime. Fig.~\ref{figFSS05p025} shows the result of these fits by dashed lines for $q=0.25$. In order to recover from these scaling properties a strong multifractal behavior in the localized regime, Eq.~\eqref{eq:tauqloc} with $q^*(W)$ given by Eq.~\eqref{locqpetit}, we expect that $\beta_{\rm loc} \propto q$. The data for $\beta_{\rm loc}$ as a function of $q$ confirm this expectation for $q \lesssim 0.25$. Beyond that value of $q$, we observe deviations from linear scaling due to the presence of a volumic scaling term, see Eq.~\eqref{eq:asymptlocaql12} and Fig.~\ref{chi2Wstar}, which are consistent with the localized behavior observed far from the transition. This explains the deviations from $\beta_{\rm loc} \propto q$ observed for $q\gtrsim 0.25$ in the localized phase. This behavior is also valid in the delocalized regime but is a transient behavior. Volumic scaling always dominates at large $N$ and gives an asymptotic ergodic behavior, see Eq.~\eqref{delocqpetit}.}
	\label{betadeq}
\end{figure}

The asymptotic behavior Eq.~\eqref{eq:asymptlocaql12} of the scaling function is illustrated in Figs.~\ref{figFSS05p025}, \ref{betadeq} and \ref{chi2Wstar}. 
The dashed line just above the lower localized branch of the scaling function in Fig.~\ref{figFSS05p025} represents a fit by $\log_2(\langle P_q\rangle/\langle P_q^c\rangle) = \log_2 F_\text{lin}(\log_2 N/\xi) \approx A_0+ \beta_{\rm loc} (\log_2 N)/\xi$ for large $\log_2 N\gg \xi$, which corresponds to $F_\text{lin}(X)\sim \mathcal V(X)^{\beta_{\rm loc}}$. We have repeated this analysis for different values of $q$ and plotted the resulting $\beta_{\rm loc}$ as a function of $q$ in Fig.~\ref{betadeq}. As expected from \eqref{eq:asymptlocaql12}, $\beta_{\rm loc} \propto q$ for $q \lesssim 0.25$. 

Moreover, for $W\to W_c$, $\xi$ was found to diverge as $(W-W_c)^{-\frac12}$ (see Fig.~\ref{figFSS05p025}, inset of the left panel), thus $q^*$ is expected to approach $q_c^*$ as $(W-W_c)^{\frac12}$:
\begin{equation}
\label{eq:critq*}
q^* = q^*_c - C (W-W_c)^{\frac12}\;,
\end{equation}
where $C$ is a constant. This prediction, which is one of our most important results, is confirmed by the numerical data shown in Fig.~\ref{figtauq}. It implies in particular a singular behavior of $\xi_\perp = q^*/\ln K$, reminiscent of the predictions for the MBL transition \cite{dumitrescu19}, that we will discuss later.
\\
\quad\\
\paragraph{Case $q>1$:}
As seen in Fig.~\ref{figtauq}, the behavior of $\langle P_q\rangle$ for $q>1$ in the localized phase is much simpler: it is always localized since $\tau_q\simeq 0$ in that regime. As shown in Fig.~\ref{figpqcrit}, the critical behavior is well described by $\langle P_q^c\rangle \sim d_N^{-\sigma_q}$; the localized behavior can thus be obtained by assuming a purely linear-type asymptotic scaling $F(X,Y) \sim X^{\sigma_q}$.
 This yields
\begin{equation}
\label{locqgrand}
\langle P_q\rangle\sim d_N^{-\sigma_q}\left(\frac{d_N}{\xi}\right)^{\sigma_q} \sim \xi^{-\sigma_q},
\end{equation}
which remains constant for $N\to\infty$, and thus $\tau_q=0$, as expected. This asymptotic behavior of the scaling function $F(X,Y) \sim X^{\sigma_q}$ is indicated in the right panel of  Fig.~\ref{figFSS05p025} (corresponding to $q=2$) by the upper dashed line.

Note that this asymptotic behavior corresponds to the usual one in finite dimension where $P_q \sim L^{-\sigma_q} F(L/\xi)$ and $F(X) \sim X^{\sigma_q}$ at large $X$ in the localized phase ($L$ being the equivalent of $d_N\sim \log_2 N$) (see \cite{cuevas2007two, evers2008anderson, PhysRevB.84.134209}).

\subsubsection{Delocalized phase $W<W_c$}

\paragraph{Case $q>1$.}
In the delocalized phase,we showed in Fig.~\ref{figFSS05p025} that the behavior of $\langle P_q \rangle$ is in agreement with a volumic scaling law. Moreover, far from the transition, $\langle P_q \rangle \sim N^{-(q-1)}$. This ergodic limit can be recovered from the following asymptotic behavior of the scaling function ~\eqref{eq:scahypV2}:
\begin{equation}
F(X,Y) \sim X^{\sigma_q} Y^{-(q-1)} \; .
\end{equation} 
Indeed, the behavior in $Y=N/\Lambda$ yields the volumic scaling observed in Fig.~\ref{figFSS05p025}, and leads to the ergodic behavior
\begin{equation}
\label{delocqgrand}
\langle P_q\rangle\sim \xi^{-\sigma_q} (\Lambda/N)^{q-1}.
\end{equation}
This gives $\tau_q=q-1$, as expected from Fig.~\ref{figtauq}.

Our purely volumic scaling analysis, shown in Fig.~\ref{figFSS05p025}, does not take into account the linear-scaling prefactor $X^{\sigma_q}$. It is in fact negligible as compared to the second volumic term $Y^{-(q-1)}$ and thus does not lead to observable deviations in our scaling analysis. The asymptotic behavior $F(X,Y) \sim Y^{-(q-1)}$ is indicated in the right panel of  Fig.~\ref{figFSS05p025} (for $q=2$) by the lower dashed line.

Note that a similar scaling occurs in the finite-dimensional case, where $\langle P_q \rangle \sim L^{-\sigma_q} F(L/\xi)$ and $F(X) \sim X^{\sigma_q}/\mathcal V(X)^{q-1}$ at large $X$ in the delocalized phase ($L$ being the equivalent of $d_N$). However, the main difference with the finite-dimensional situation is that the volume here is not $\mathcal V(X=d_N/\xi)$ but $Y=N/\Lambda$. Indeed, a scaling depending on $\mathcal V(d_N/\xi)$ would mean a linear scaling, which is not compatible with our data; moreover it would imply a non-ergodic delocalized behavior (see \cite{scaling17}):  $\mathcal V(d_N/\xi) = N^{1/\xi}$, thus $\langle P_q \rangle$ would behave as $\langle P_q \rangle \sim \xi^{-\sigma_q} N^{-(q-1)/\xi}$, a (multi)fractal behavior.

The observation of a volumic scaling with the asymptotic behavior~\eqref{delocqgrand} demonstrates that the delocalized phase is ergodic, a question which has been strongly debated \cite{monthus2011anderson, biroli2012difference, de2014anderson,  kravtsov2015random, altshuler2016nonergodic, facoetti2016non, tikhonov2016anderson, tikhonov2016fractality, sonner2017multifractality,scaling17, biroli2017delocalized, monthus2017multifractality, tarquini2017critical, kravtsov2018non, bogomolny2018eigenfunction, bogomolny2018power, bera2018return, tikhonov2019statistics, tikhonov2019critical, parisi2019anderson, twoloc20, kravtsov2020localization, detomasi2020subdiffusion, roy2020localization, kravtsov2020localization, khaymovich2020fragile, biroli2022critical, biroli2021levy, alt2021delocalization, tikhonov2021anderson, colmenarez2022sub}. There is now a consensus on this point \cite{tikhonov2016anderson, scaling17, biroli2018delocalization, tikhonov2019statistics, tikhonov2019critical, kravtsov2020localization, khaymovich2020fragile, sierant2022universality}. 
\\
\quad\\
\paragraph{Case $0<q< 0.5$.}
In the delocalized regime, we have shown that a linear scaling prevails for $q<0.5$, with a scaling function of the form $F_\text{lin} (X) \sim  \mathcal{V}(X)^{Aq}$ (see Table \ref{tableresume}). However, as shown in Fig.~\ref{figtauq}, for $q>1$ and for $q<0.5$ far from the transition the behavior is asymptotically ergodic, which is incompatible with a linear scaling for all $N$. We therefore assume that a volumic scaling dominates at large $N$, with an asymptotic behavior of $F$ in Eq.~\eqref{eq:scahypV2} as:
\begin{equation}
F(X,Y) \sim V(X)^{Aq} + Y^{\tau_q^c-q+1}.
\end{equation}
Under this assumption we get
\begin{equation}
\label{delocqpetit}
\langle P_q\rangle\sim N^{-q/q^*+1}+\Lambda^{-\tau_q^c} (\Lambda/N)^{q-1}
\end{equation}
with the same relation for $q^*$ as in \eqref{locqpetit}. The difference with  \eqref{locqpetit} is that we now have $q_c^*=\frac12<q^*<1$, which is reminiscent of the case of a finite Cayley tree in the non-ergodic delocalized phase \cite{tikhonov2016fractality}.
The second term in \eqref{delocqpetit} yields a contribution $\propto N^{-q+1}$, which always dominates the first one since $0<-q+1<-q/q^*+1$ as soon as $q^*<1$. Thus the asymptotic behavior of \eqref{delocqpetit} is ergodic, $\langle P_q\rangle \sim N^{-(q-1)}$, which reproduces the expected $\tau_q=q-1$ observed in Fig.~\ref{figtauq}. The presence of the first term in \eqref{delocqpetit} accounts for the deviation from the straight line $\tau_q=q-1$, that can be observed in Fig.~\ref{figtauq} at small $N$; this will be discussed in Sec.~\ref{sec:volorlin} below.

Figure~\ref{figFSS05p025} (left panel) confirms that the scaling function $F_\text{lin}(X)$ has an asymptotic behavior $F_\text{lin} (X) \sim  \mathcal{V}(X)^{\beta_{\rm deloc}}$ (upper dashed line). In Fig.~\ref{betadeq}, we plot the exponent $\beta_{\rm deloc}$, extracted from the finite-size scaling  function, as a function of $q$: it is indeed linear in $q$ for $q\lesssim 0.35$; beyond that value, volumic contribution sets in and the linear scaling is not so clear.

Equation \eqref{delocqgrand} and the second term of Eq.~\eqref{delocqpetit} are the signature of a strongly-multifractal metal, with $\langle P_q\rangle\sim \Lambda^{-\tau_q^c} (\Lambda/N)^{q-1}$ for $q<1/2$ and $\langle P_q\rangle\sim \xi^{-\sigma_q} (\Lambda/N)^{q-1}$ for $q>1/2$. This corresponds to the multifractal metal found in the delocalized phase of the Anderson transition in finite dimension, see \cite{cuevas2007two}.

\subsection{Volumic or linear scaling for $0<q<1/2$}\label{sec:volorlin}

\begin{figure}[!t]
\includegraphics[width=0.99\linewidth]{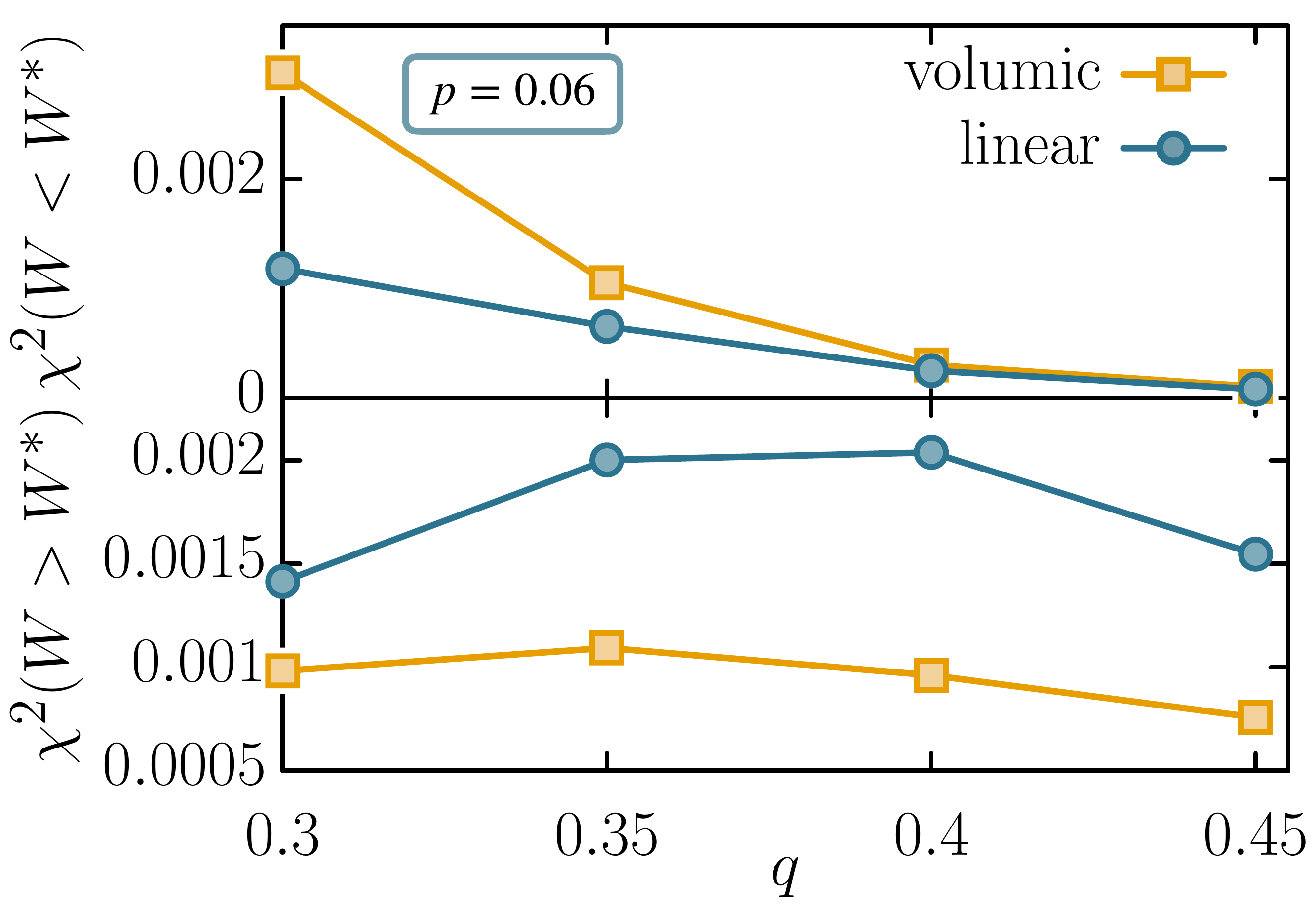} 
\caption{ $\chi^2$ test evaluated as the sum over $W$ (restricted to $W_c<W<W^*$ (top) or $W_c<W^*<W$ (bottom)) of the individual $\chi^2$ calculations (see  Eq.~\eqref{eq:sumW}),  for a volumic (orange squares) or linear (blue circles) scaling, for $p=0.06$. We clearly see that linear scaling is prevalent for $W_c<W<W^*$, leading to a multifractal behavior $\langle P_q\rangle\sim N^{-q/q^*+1}$, while a volumic scaling describes better the data for $W>W^*$, giving a localized behavior $\langle P_q\rangle\sim \Lambda^{-\tau_q^c}$, see Eq.~\eqref{locqpetit}.}
\label{chi2Wstar}
\end{figure}
 
\subsubsection{Localized phase $W>W_c$} 
 
From the considerations below Eq.~\eqref{locqpetit}, in the localized phase the volumic scaling dominates for $q>q^*$ while the linear scaling dominates for $q<q^*$. In the finite-size scaling  plots such as Fig.~\ref{figFSS05p025} left panel, the value of $q$ is fixed, and data correspond to different values of $W$, and thus different values of $q^*$ (recall that $q^*$ is a function of $W$, whose plot is given by Fig.~\ref{figtauq}). 
At the fixed value of $q$ under investigation, one must therefore distinguish between values of $W$, depending on whether the associated $q^*(W)$ is larger or smaller than $q$. To that end, we define $W^*$ such that $q^*(W^*)=q$. Since $q^*$ is a decreasing function, $q^*(W)<q=q^*(W^*)$ is equivalent to $W>W^*$. Therefore, for $W>W^*$ we should have a volumic scaling, associated with localization, while for $W<W^*$ the linear scaling synonym of multifractality should manifest itself. 

This is indeed what we observe, as is confirmed by Fig.~\ref{chi2Wstar}: for each fixed value of $q$ the corresponding $W^*$ is extracted from the plot of $q^*(W)$, and
 the different linear or volumic finite-size scaling  are compared by calculating their respective $\chi^2$ as in Eq.~\eqref{eq:sumW} but with a sum over $W$ restricted to $W_c<W^*<W$ or $W_c<W<W^*$. 
 Figure \ref{chi2Wstar} corresponds to value $p=0.06$; we observe the same effect for $p=0.25$ (data not shown) provided we restrict finite-size scaling  to larger system sizes. 
For $W<W^*$ (top), linear scaling is always better than volumic; the converse is true for $W>W^*$. This is also manifest in Fig.~\ref{chi2dosp} bottom, where the $\chi^2_{ll}$ becomes large close to $q=0.5$. Note that for $q=0.25$ considered in Fig.~\ref{figFSS05p025}, the value $W^*$ such that $q^*(W^*)=q$ is above the largest value of $W$ that we considered, which is why we only see a linear scaling.

\begin{figure}[!t]
\includegraphics[width=0.99\linewidth]{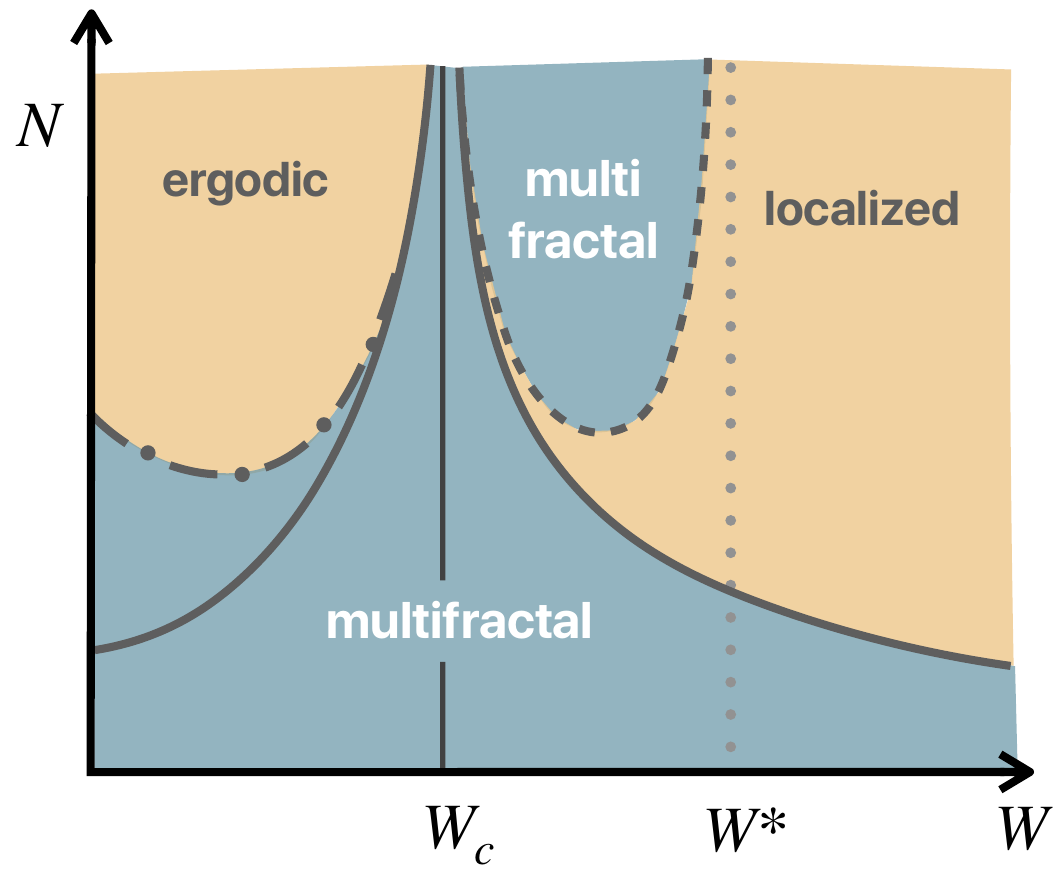}  
\caption{Sketch of the phase diagram for the leading behavior of $\langle P_q\rangle$ for fixed $q<0.5$ extracted from the behavior of the scaling function $F(d_N/\xi,N/\Lambda)$, Eq.~\eqref{eq:scahypV2}. A multifractal behavior is observed when the linear scaling term 
	prevails, while a localized or ergodic behavior is implied by the dominance of the volumic term. In the localized regime, there exists a characteristic strength of disorder $W^*$, defined as $q^*(W^*)=q$, which separates the multifractal and localized regimes. More precisely, a multifractal behavior is observed for either $N\ll \Lambda$ [critical multifractality characterized by $\tau_q^c = (q/q^*_c -1)$] or $N\gg \Lambda^\gamma$ [renormalized multifractality with $\tau_q=(q/q^*-1)$ with $q^*<q^c$, distinct from $\tau_q^c$]. This is indicated by the blue domains of the phase diagram delimited by the solid line for $\Lambda$, and the dashed line for $\Lambda^\gamma$. $\gamma=(1-q/q_c^*)/(1-q/q^*)$, therefore $\Lambda^\gamma$ diverges at $W^*$ where $q = q^*$. For $W>W^*$, only a localized behavior is obtained for $N \gg \Lambda$. In the delocalized regime, the volumic term always dominates for $N\gg \Lambda^{q^*/(1-q^*)}$ which leads to an ergodic behavior. However, $q^*/(1-q^*)>1$ for $q^*>1/2$ as found in the delocalized phase, and diverges when $q^* \rightarrow 1$, i.e., at small $W$. Therefore $\Lambda^{q^*/(1-q^*)}$, indicated by the dashed-dotted line, can be significantly larger than $\Lambda$ (solid line) which explains why linear scaling is observed for $q<q_c^*=1/2$ in the delocalized regime. This leads to a transient multifractal behavior observed for $N\ll \Lambda^{q^*/(1-q^*)}$: critical multifractality for $N \ll \Lambda$ and renormalized multifractality for $\Lambda \ll N\ll \Lambda^{q^*/(1-q^*)}$.  }
\label{phasediag}
\end{figure}

There is an additional effect in the numerical data: finite values of $N$ may lead to transient effects, where a linear scaling can be observed instead of the expected volumic one, or vice-versa.  
For instance, in the localized phase for $q<1/2$, moments are governed by Eq.~\eqref{locqpetit}; while for $q<q^*$ the first term always dominates at large $N$, as $N^{-q/q^*+1}\gg\Lambda^{-\tau_q^c}$, at small $N$ the second term may dominate, in which case one should observe a volumic scaling. 
This latter regime is equivalent to $N\ll \Lambda^\gamma$ with $\gamma=(1-q/q_c^*)/(1-q/q^*)$ (and $\gamma>1$). Close to $W=W_c$ the exponent $\gamma$ is close to 1, and $\Lambda^\gamma\approx\Lambda$ diverges; for $N\ll\Lambda$ the linear scaling dominates, which corresponds to the multifractal insulator regime. For $W\to W^*$ one has $q^*(W)\to q$, and thus $\Lambda^\gamma$ diverges; the linear scaling only dominates at $N\gg \Lambda^\gamma$. This is summarized in the sketch of Fig.~\ref{phasediag}.

\subsubsection{Delocalized phase $W<W_c$} 
\label{VB2}
In the delocalized phase for $q<0.5$, the second term in Eq.~\eqref{delocqpetit} (ergodic behavior) always dominates for large $N$. Nevertheless, using $q_c^*=1/2$, we get that the first linear term dominates iff $N\ll \Lambda^{q^*/(1-q^*)}$. 
The exponent ${q^*/(1-q^*)}>1/2$ in the delocalized phase as $q^*>1/2$, and diverges when $q^* \rightarrow 1$; therefore $\Lambda^{q^*/(1-q^*)}$ can be much larger than $\Lambda$. This explains why we observe a linear scaling for $q<1/2$ in the delocalized regime, as shown in Fig.~\ref{figFSS05p025}. Moreover, this leaves room to observe a transient multifractal behavior in the delocalized regime: a critical multifractality characterized by $\tau_q^c = (q/q_c^*-1)$ for $N \ll \Lambda$, and a renormalized multifractality $\tau_q^* = (q/q^*-1)$ with $q^*>q_c^* = 1/2$ for $\Lambda\ll N < \Lambda^{q^*/(1-q^*)}$.
Nevertheless, close to the transition, $q^*\to 1/2$, and $\Lambda^{q^*/(1-q^*)} \sim \Lambda$. Therefore, we do not obtain another critical exponent associated with the transition to ergodicity, contrary to what has been claimed in certain recent studies \cite{sierant2022universality}. This is summarized in the sketch of Fig.~\ref{phasediag}.

\subsection{Scaling length $\xi$ and localization lengths $\xi_\parallel$ and $\xi_\perp$ from correlation functions}
\label{characscale}

The scaling length $\xi$ that manifests itself in the scaling function \eqref{eq:scahypV2} can be related to the scales $\xi_\parallel$ and $\xi_\perp$ governing the correlation functions and the spectrum.

In \cite{twoloc20} we proposed a simple model of a tree with connectivity $K$ and depth $d$, on which a wavefunction is exponentially localized with localization length $\xi_\perp$ at the root of the tree. This model is a natural consequence of the small-$r$ behavior \eqref{eq:simctyp} of $\Cttyp$. The exponential decrease of the wavefunction along the branches is compensated by the proliferation of sites at a given distance. A simple calculation shows that the moments $P_q$ are given by 
\begin{equation}
\label{toy}
 P_q  = \dfrac{\sum_{r=0}^{d-1} K^r [e^{-r/\xi_\perp}]^q}{\left[\sum_{r=0}^{d-1} K^r e^{-r/\xi_\perp}\right]^q} 
 \sim N^{-\tau_q},
    \end{equation}
with $N=K^d$ and  $\tau_q$ given by
$\tau_q = (q/q^* - 1)$ for $q\le q^*<1/2$ and $\tau_q=0$ for $q^*\le q$, where $q^*$ and $\xi_\perp$ are related by $q^*= \xi_\perp \ln K$. This model suggests to interpret $q^*$ as the localization length $\xi_\perp$ (up to a factor $\ln K$). 
This is corroborated by the plot in Fig.~\ref{figtauq}, where for $W>W_c$ the square symbols given by $\xi_\perp$ extracted from the small-$r$ behavior of $\Cttyp$ follow the values of $q^*$ up to a factor $\ln K$. 

In turn, we saw that $q^*$ is controlled by the scaling length $\xi$ [see Eq.~\eqref{locqpetit}]. Thus, $\xi$ obtained by the scaling behaviors of $\langle P_q \rangle$ for $q<1/2$ controls the typical localization length $\xi_\perp$:
\begin{equation}
\label{eq:xixiperp}
\frac{1}{\xi_\perp} = \frac{1}{\xi_\perp^c} + \frac{D}{\xi} \;,
\end{equation}
where $D$ is a positive constant. In other words, the typical localization length $\xi_\perp$ reaches a universal critical value $\xi_\perp^c = 1/(2 \ln K)$ from below. Since $\xi \sim (W-W_c)^{-\nu_\perp}$ with $\nu_\perp\simeq 0.5$ (see Table \ref{tableresume}), the behavior of $\xi_\perp$ close to $\xi_\perp^c$ has a square-root singularity:
\begin{equation}
\label{eq:xiperp}
\xi_\perp \approx \xi_\perp^c - c (W-W_c)^{1/2} \; .
\end{equation}  
This behavior is exactly what is predicted in \cite{dumitrescu19} for the typical localization length at the MBL transition. Finite-size effects are controlled by the scaling length $\xi \sim (W-W_c)^{-\nu_\perp}$ with the critical exponent $\nu_\perp \equiv 1/2$.  

\begin{figure}[!ht]
	\includegraphics[width=.99\linewidth]{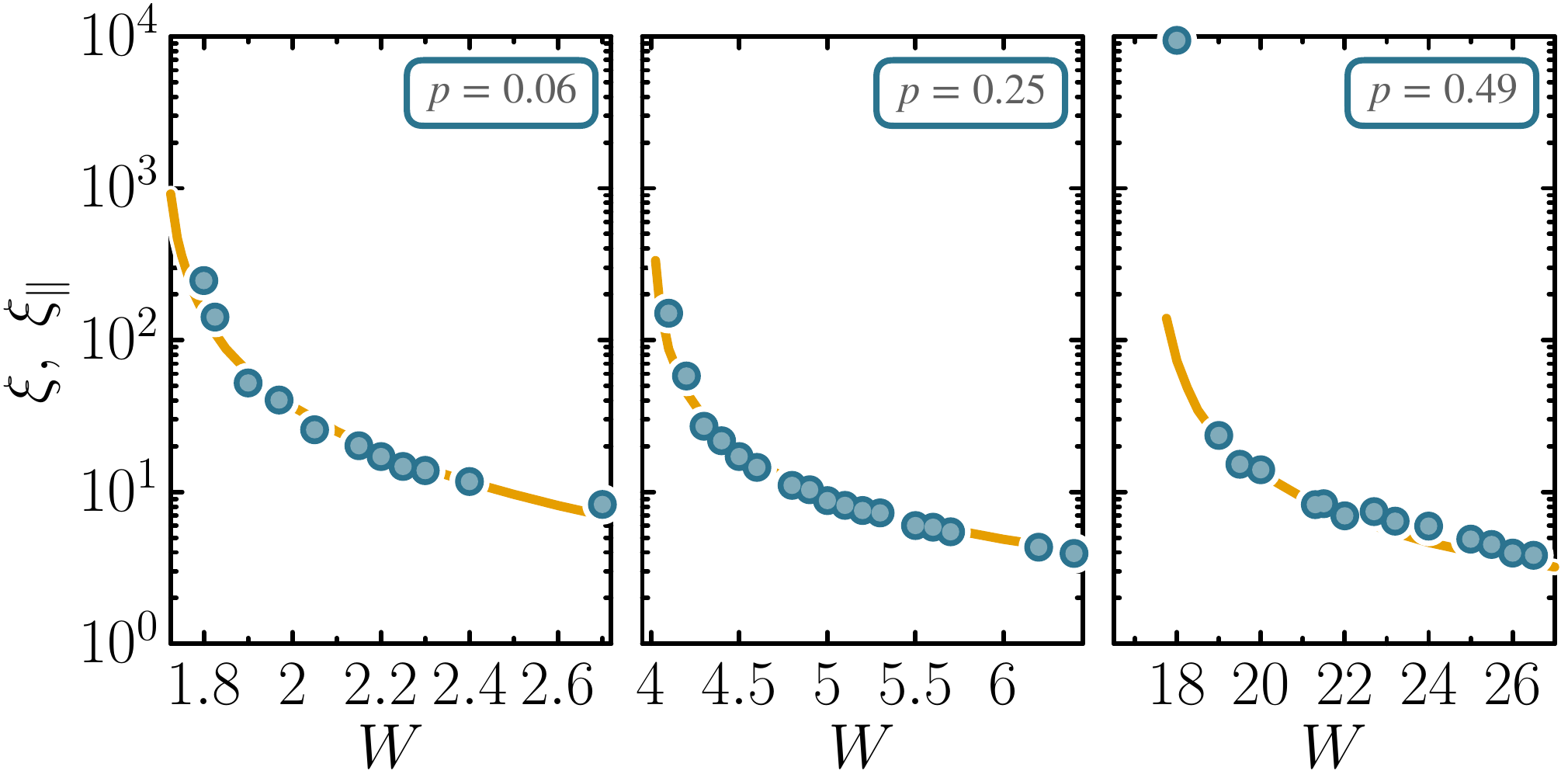}
	\caption{Comparison of scaling lengths obtained from finite-size scaling  and correlation functions. Solid line: $\xi$ obtained from finite-size scaling  in the localized phase for $q=2$ extracted from Figs.~\ref{figFSS05p025} and \ref{figFSS05p006}. Circles: $\xi_\parallel$ obtained from a fit of 
	$\tilde{\Cav}$ by Eq.~\eqref{eq:simcav} with $\alpha$ free. Left: $p=0.06$, $W_c=1.7$. Center: $p=0.25$, $W_c=4$. Right: $p=0.49$ (box disorder), $W_c=17.5$.}
	\label{xi_xipar}
\end{figure}

On the other hand, for large $q>1$, the scaling length diverges as $\xi \sim (W-W_c)^{-1}$ and controls the asymptotic value of $\langle P_q \rangle \approx \xi^{-\sigma_q}$, see Eq.~\eqref{locqgrand}, and thus represents a localization length. Since $\langle P_q \rangle$ for large values of $q$ focuses on large amplitudes of the wavefunctions, following the physical picture drawn from the analogy with the directed polymer problem (see \cite{derrida1988polymers, monthus09, monthus2011anderson, monthus2019revisiting} and Sec.~\ref{sec:corrfunctions}), we expect that they are dominated by the rare branches of the wave functions. We thus expect that $\xi \equiv \xi_\parallel$ extracted from the average correlation function $\Ctav$, Eq.~\eqref{chav}. 

In Fig.~\ref{xi_xipar} we show a comparison between the scaling length $\xi$ for finite-size scaling  in the localized phase for $q=2$ (see Figs.~\ref{figFSS05p025} and \ref{figFSS05p006}) and the localization length $\xi_\parallel$ extracted from a fit of $\Ctav$, Eq.~\eqref{chav}, by Eq.~\eqref{eq:simcav}. The agreement between them is remarkable and confirms our physical picture. Importantly, this implies that there exist two critical localization lengths in the localized phase, the average localization length $\xi_\parallel$ and the typical one $\xi_\perp$ associated with two distinct critical exponents $\nu_\parallel =1$ and $\nu_\perp=1/2$. While a typical and average localization lengths can be defined in finite dimension, they do not have a distinct critical behavior and are both controlled by a single critical exponent $\nu$. The case of random graphs of infinite effective dimension brings the importance of rare events to the point of having two different critical exponents describing the localization along the rare branches and that describing the localization transversely to these rare events.

In the delocalized regime $W<W_c$, a linear behavior is expected for small enough $N$ (see Fig.~\ref{phasediag}). The value of $q^*$ numerically extracted from the plot in Fig.~\ref{figtauq} at small $q$ should therefore also correspond with the $\xi_\perp$ extracted from the small-$r$ behavior of $\Cttyp$, but only for small $N$. This is also manifest in Fig.~\ref{figtauq}, where $q^*$ and $\xi_\perp \ln K$ coincide for the lowest values of $N$. Moreover, as observed in Fig.~\ref{fig:corr1}, $\Cttyp$ reaches a plateau for $r$ larger than a certain characteristic length scale. That scale should be related with $V^{-1}(\Lambda)$, where $\Lambda$ is the correlation volume displayed in the inset of Fig.~\ref{figFSS05p025} (right).

	\subsection{Scaling and level repulsion}\label{sec:levelrep}
	
	In this last subsection, we motivate and interpret the scaling law Eq.~\eqref{eq:scaetar} proposed to describe the spectral statistics. This scaling law was discussed in Sec.~\ref{fss_spec} and tested in Fig.~\ref{figSS5}. As in the case of $\langle P_q\rangle$, we will generalize this scaling assumption to a two-parameter scaling $F\left(X, Y\right)$ with $X=d_N/\xi$ and $Y=N/\Lambda$, with $\xi$ the linear scaling parameter, and $\Lambda$ the scaling volume. We will describe and test the asymptotic behaviors of $F(X,Y)$ required to recover the localized and ergodic behaviors of $\langle \eta_\raur \rangle$. We will start with the localized phase, then we will address the delocalized phase, which has a more complex behavior. 
	
	\subsubsection{Localized phase $W>W_c$}
	
	\paragraph{Level repulsion is governed by $\xi_\perp$.--}
	
	The parameter $\eta_\raur$ defined in Eq.~\eqref{defetar} characterizes the difference between $ \langle\raur\rangle$ and its value in the absence of level repulsion $\langle\raur\rangle_{\mathrm{P}} = 0.3863$.
	As the system size increases, $\eta_\raur$ goes to $\langle\raur\rangle_{\mathrm{P}}$, as shown in Fig.~\ref{pr_regimes}.
	At finite system size $N$, the discrepancy between $\langle\raur\rangle$ and $\langle\raur\rangle_{\mathrm{P}}=0$ is therefore an indicator of the repulsion between nearest energy levels. In random matrix theory, level repulsion is characterized by the small-$s$ behavior of the nearest-neighbor spacing distribution, $P(s)\sim_{s\to 0} s^\beta$, where for Wigner Gaussian ensembles $\beta$ is the Dyson index. For more general systems, following the Brownian motion approach to random matrix spectral statistics proposed by Dyson \cite{dyson1962brownian}, it has been proposed that the analog of the index $\beta$ can be defined as \cite{monthus2016level, chalker1996random, PhysRevB.93.041424}
	\begin{equation}
	\beta = \frac{2 \langle \mathcal{Y} \rangle}{\langle P_2 \rangle - \langle \mathcal{Y}\rangle }\;,
	\label{eq:betaY}
	\end{equation}
	where $P_2$ is the inverse participation ratio and $\mathcal{Y}= \sum_{i=1}^{N} \vert \psi_n(i) \vert^2 \vert \psi_{n+1}(i) \vert^2$ is the density correlation between two eigenstates $n$ and $n+1$ with neighboring energies. The exponent $\beta$ defined by Eq.~\eqref{eq:betaY} goes to $1$ for extended states and to $0$ asymptotically in the localized phase \cite{monthus2016level}.
	
\begin{figure}[!t]
\includegraphics[width=.99\linewidth]{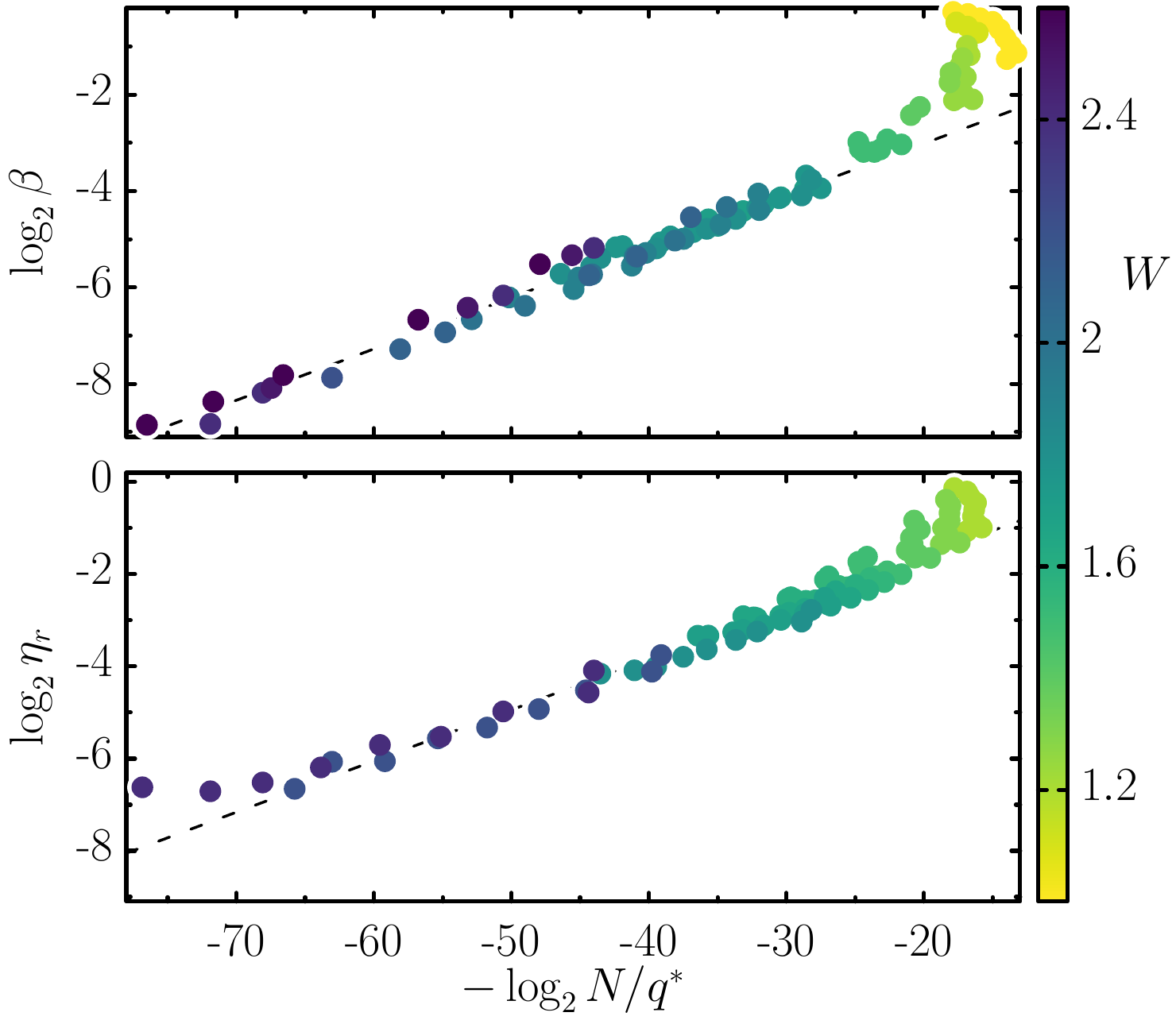}
	\caption{Top: $\log_2\beta$ as  a function of $\log_2N$, for $p=0.06$ and many values of disorder $W$. $\beta$ is the level repulsion exponent of \eqref{eq:betaY}. The dashed black line corresponds to a linear fit with slope $a\sim 0.106$.
	Bottom: $\log_2\eta_r$  [see Eq.~\eqref{defetar}] as  a function of $\log_2N/q^*$, for $p=0.06$ and many values of disorder $W$. The dashed black line corresponds to a linear fit with slope $b\sim 0.111$.}
	\label{beta_repulsion}
\end{figure}

In the present system, eigenstates in the localized phase are localized on rare branches, as discussed in Sec.~\ref{characscale}. If level repulsion is absent or small, then eigenstates with nearby energy $\psi_n$ and $\psi_{n+1}$ should lie on distinct branches, and therefore the finite-size behavior of their correlation $\mathcal{Y}$ should be controlled by $\xi_\perp$ rather than $\xi_\parallel$ (see however 
\cite{Tikhonov20211Eigenstate}). If this holds, we would then expect $\mathcal{Y}$ to decrease exponentially as $\exp(-a d_N/\xi_\perp)=N^{-a/q^*}$, where $a$ is some constant and $d_N$ is the diameter of the graph (that is, the maximum distance that can exist between the localization centers of $\psi_n$ and $\psi_{n+1}$). As a consequence, since in the localized phase the inverse participation ratio is a constant given by Eq.~\eqref{locqgrand}, $\beta$ defined in \eqref{eq:betaY} would decrease as $\beta \sim N^{-a/q^*}$. This is indeed the behavior we observe, as we checked in Fig.~\ref{beta_repulsion} (top panel).

The finite-size behavior of $\langle\raur\rangle$, which directly depends on level repulsion, can be conjectured to also be controlled in the same way by $q^*$, or equivalently $\xi_\perp$, and to decay exponentially in the localized phase, as
	\begin{equation}\label{eq:etaloca}
	\eta_\raur \sim \exp(-b d_N/\xi_\perp) = N^{-b/q^*} 
	\end{equation}
	with $b$ some constant. This is confirmed by the data in
	Fig.~\ref{beta_repulsion} (bottom panel), where it is found $b\approx a$.\\

	\paragraph{Decay of level repulsion at criticality.--}
	Naively, one may conclude from the above arguments that $\eta_\raur$ should decay exponentially with $d_N$ at the transition
	since $q^*_c=1/2=\xi_\perp^c \ln K$ is finite at the transition. However, $\xi_\parallel$, which represents the localization length along the rare branches, diverges at the transition and thus the states $\psi_n$ and $\psi_{n+1}$ cannot be simply thought of as being at a maximal distance $d_N$ from each other.
	Hence, the arguments above need to be revisited at criticality. 
	At the critical point, or slightly above it, we saw in 
	Fig.~\ref{beta_repulsion} (bottom panel) that our data are compatible with the exponential decay \eqref{eq:etaloca} with $q^*=q^*_c =1/2$, but also with a power-law decay with $d_N$ as in Eq.~\eqref{fitlog}, $\eta_\raur \sim d_N^{-\alpha}$. In the following, we analyze the consequences of assuming an exponential decay as Eq.~\eqref{eq:etaloca} at criticality. \\
	
  	\paragraph{Asymptotic behavior of the scaling function.--}
	The asymptotic exponential decay in the localized regime, Eq.~\eqref{eq:etaloca}, can be recovered by the following asymptotic dependence of the scaling function Eq.~\eqref{eq:scaetar} in the localized phase: 
	\begin{equation}\label{eq:scafuncetaloca}
	F_\text{lin}(X) \sim \exp(-D X) \;,
	\end{equation}
	for $X=d_N/\xi \gg 1$, with $D$ a constant. Indeed, if the critical behavior is Eq.~\eqref{eq:etaloca} with $q^*_c =1/2$, i.e.~$\eta_\raur^c \sim \exp ( - b d_N/\xi_\perp^c)$, then Eq.~\eqref{eq:scaetar} gives
	\begin{equation}
	\eta_\raur \sim \exp ( - b d_N/\xi_\perp^c) \exp(-D d_N/\xi) \sim \exp(-b d_N/\xi_\perp) \;.
	\end{equation}
	The coefficient $D$ in \eqref{eq:scafuncetaloca} coincides with the coefficient $D$ appearing in the relation \eqref{eq:xixiperp} between $\xi_\perp$ and $\xi$. 
	The asymptotic behavior \eqref{eq:scafuncetaloca} of the scaling function is shown by the lower dashed line in Fig.~\ref{figSS5}. Since in the localized regime $q^*<q^*_c$ (see Fig.~\ref{figtauq} bottom), $\xi_\perp^c<\xi_\perp$ and thus $D>0$. 

\subsubsection{Delocalized phase $W<W_c$}
	The description of the delocalized phase is more subtle, but it follows from two key observations. \\

	\paragraph{Linear scaling close to the transition.--}
	For system sizes smaller than the correlation volume, $N \lesssim \Lambda$, the system behaves as in the critical regime. Indeed, inside a correlation volume, wave functions lie on a few branches with a transverse localization length $\xi_\perp$. In this transient regime in $N$, we thus expect an exponential decay of level repulsion $\eta_\raur \sim \exp(-b d_N/\xi_\perp) \sim N^{-b/q^*}$ following the critical result \eqref{eq:etaloca}.  As discussed in Sec.~\ref{VB2}, for $q<0.5$ in the delocalized phase, we obtain a linear scaling. We should thus get an asymptotic behavior 
	\begin{equation}\label{eq:scafunclinetadeloc}
	F_\text{lin}(X) \sim \exp(\tilde D X)
	\end{equation}
	for $X\gg 1$. The asymptotic behavior \eqref{eq:scafunclinetadeloc} of the scaling function is shown by the upper dashed line in Fig.~\ref{figSS5}. 
	Since for $W<W_c$ we have  $q^*>q^*_c=1/2$ [see Fig.~\ref{figtauq} (bottom)], we now have $\xi_\perp^c<\xi_\perp$ and thus $\tilde D>0$. The behavior \eqref{eq:scafuncetaloca} for the localized case is very similar to the behavior \eqref{eq:scafunclinetadeloc} for the delocalized case; only the sign in the exponential differs. \\

\begin{figure}[!t]
\centering
\includegraphics[width=0.95\linewidth]{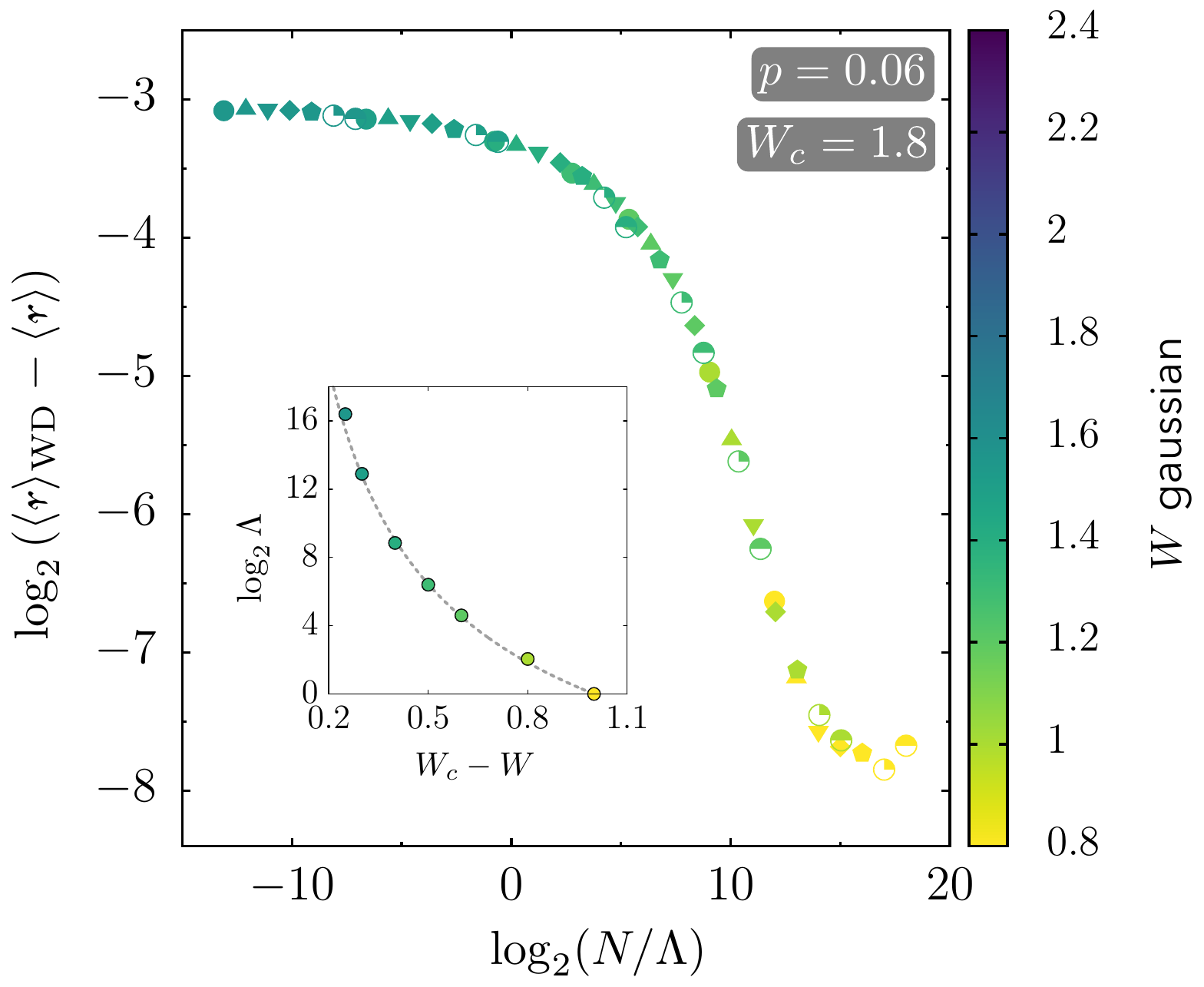} 
\caption{Volumic scaling of the average spectral gap ratio $\langle\raur\rangle_\mathrm{WD}-\langle\raur\rangle$ in the ergodic delocalized regime at small $W$ values, for $p=0.06$. Inset: critical volume $\Lambda$ as a function of $W_c-W$. The gray dashed line is a two-parameter fit of the form
\eqref{lnLambda} (see text) over the interval $W\in[0.8,1.5]$, with $W_c=1.8$ fixed and fit parameters $a_1\approx 15.8$ and $\nu\approx 0.49$.
This confirms that the delocalized phase is governed by a unique critical exponent $\nu=1/2$.
\label{spectralvol}}
\end{figure}

	\paragraph{Volumic scaling far from the transition, in the ergodic regime.--}
	By contrast, as discussed in Sec.~\ref{VB2} for multifractal properties, in the delocalized ergodic regime far from the transition we expect a volumic scaling: 
	\begin{equation}
	\langle\raur\rangle = \langle\raur\rangle_{\mathrm{WD}} (1 - F_\text{vol} (N/\Lambda)) \;.
	\end{equation}
	This is confirmed in Fig.~\ref{spectralvol}. The scaling function has the asymptotic behaviors $F_\text{vol}(Y) \rightarrow 0 $ for $Y\gg 1$ and  $F_\text{vol}(Y)\rightarrow 1$ for $Y\ll 1$.
	
We can fit the correlation volume by
	\begin{equation}\label{lnLambda}
	\ln \Lambda = a_0 + a_1 (W_c - W)^{-\nu}\;
	\end{equation}
with $W_c=1.8$ fixed (the value for $p=0.06$ given in \cite{sierant2022universality}), $a_1$ and $\nu$ two free parameters, and $a_0$ fixed by the fact that $\Lambda=1$ for the smallest disorder value considered. 
The corresponding fit is displayed in the inset of Fig.~\ref{spectralvol}. We find a critical exponent $\nu\approx 0.49$. This confirms that there is a unique critical exponent $\nu\approx 0.5$ in the delocalized phase and that $\Lambda = \mathcal V(\xi)$ is the correlation volume associated with $\xi$ (contrary to what is reported in \cite{sierant2022universality}). These results based on spectral properties corroborate those of Sec.~\ref{VB2} for the eigenstates.\\
	
	\paragraph{Two-parameter scaling in the delocalized phase.--}
	To take into account the presence of the linear scaling close to the transition and the volumic scaling in the ergodic regime one can conjecture the following two-parameter scaling function:
	\begin{eqnarray}
	\langle\raur\rangle = &\left[ \langle\raur\rangle_{\mathrm{P}} + \langle \tilde{\eta}_\raur^c \rangle F_\text{lin}(d_N/\xi) \right] F_\text{vol}(N/\Lambda)  \nonumber \\
	& +  \langle\raur\rangle_{\mathrm{WD}} [1 - F_\text{vol} (N/\Lambda)]\;,
	\end{eqnarray}
	where $\langle \tilde{\eta}_\raur^c \rangle = \langle {\eta}_\raur^c \rangle (\langle\raur\rangle_{\mathrm{WD}} - \langle\raur\rangle_{\mathrm{P}})$.
	The volumic scaling function $F_\text{vol} (N/\Lambda) \rightarrow 1$ for $N\ll \Lambda$ and we recover the linear scaling Eq.~\eqref{eq:scaetar} close to the transition.
	Globally testing the data using this two-parameter scaling law is a difficult challenge due to the exponential divergence of the correlation volume $\Lambda$. This two-parameter scaling is left for future study.
	
	\section{Comparison with previous theoretical results}
	\label{sec:comparison}
	
    \subsection{Analytical calculations}
	
	One of the most important results we have obtained is the equation for $q^*$ in the vicinity of the transition, Eq.~\eqref{eq:critq*}, and the equivalent equation Eq.~\eqref{eq:xiperp} for $\xi_\perp$ governing the exponential decay of the typical correlation function $\Cttyp(r)$, Eq.~\eqref{eq:simctyp}.
	These important formulas, even though they had never been stated explicitly before, can in fact be understood within the framework of the theory developed in \cite{mirlin1994distribution, sonner2017multifractality, de2014anderson, kravtsov2018non}.   

	\begin{figure}
	\includegraphics[width=.9\linewidth]{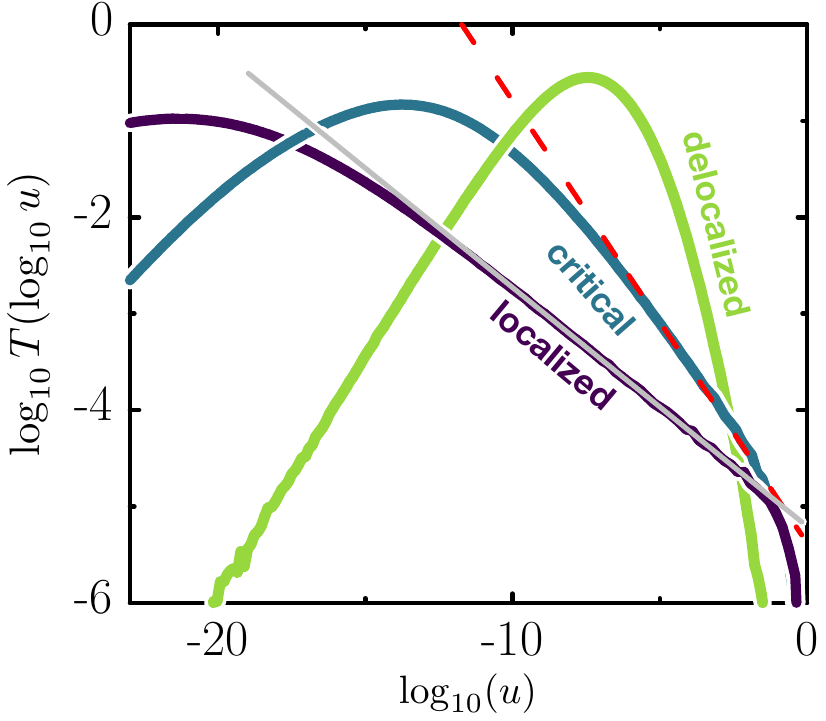}
	\caption{Probability distribution $T(\log_{10}(u))$ with $u=|\psi_i|^2$ for $N=2^{19}$, $p=0.06$ and three disorder values: delocalized $W=1.05$, critical $W=1.725$ and localized $W=2.3$. The lines are fits $T(\log_{10}u)\sim u^{-\gamma(W)}$ for the critical and localized case. The values obtained from the fit are $\gamma(W=1.725)\approx 0.46$ and $\gamma(W=2.3)\approx 0.25$. This confirms the theoretical expectation that $\gamma$ reaches a value $1/2$ at the transition, while $\gamma<1/2$ in the localized regime \cite{mirlin1994distribution, biroli2012difference, de2014anderson, kravtsov2018non}. 
	This exponent controls the typical decay of wavefunction amplitudes, i.e. $\xi_\perp$, see 
	Eq.~\eqref{ur}.\label{prob_psi2}}
    \end{figure}
	
	Indeed, a similar formula concerning the typical decay of localized wave functions can be deduced from \cite{mirlin1994distribution}. In fact, the distribution of amplitudes $u=\vert \psi\vert^2$ of localized eigenfunctions is known to follow $T(u)\sim u^{-(1+\gamma)}$ \cite{mirlin1994distribution, biroli2012difference, de2014anderson, kravtsov2018non}, as illustrated in Fig.~\ref{prob_psi2}. The exponent $\gamma$ reaches a maximum at the transition $\gamma \approx 1/2 - \text{const} (W-W_c)^{1/2}$ [see Fig.~15 of \cite{kravtsov2018non} and Eq.~(21) of \cite{tikhonov2019critical}].
	The typical value of the wave function at the maximal distance $r=d_N= \ln N / \ln K$ from its localization center can then be obtained from normalization [see Eq.~(14) of \cite{mirlin1994distribution}], and it reads	
	\begin{equation}
	\label{ur}
	u(r=d_N) \sim \exp [-r (2 \ln K + \text{const} (W-W_c)^{1/2})] \;.
	\end{equation}
	Using the identities derived earlier, we have $2 \ln K=1/\xi_\perp^c$ and $1/\xi \sim  (W-W_c)^{1/2}$. Equation \eqref{ur} can then be rewritten $u_\text{typ}(r)\sim \exp(-r/\xi_\perp)$. In other words, we recover the typical decay of wave function amplitudes that we derived.
	
	In turn, Eq.~\eqref{eq:critq*} for $q^*$ can be recovered from Eqs.~(26) and (30) of \cite{sonner2017multifractality} (see also \cite{de2014anderson}). In \cite{sonner2017multifractality}, the authors describe the multifractal properties of a finite Cayley tree, which are found to be similar to what we find in the localized phase. In the delocalized phase, the finite Cayley tree has a delocalized non-ergodic phase with multifractal properties, while this is true only for a transient regime of $N$ for $q<1/2$ in the delocalized phase of small-world networks. 
	The appearance of the critical exponent $1/2$ is quite similar to the appearance of the index $1/2$ for the correlation length on the delocalized side [see e.g. Eqs.~(20)-(25) of \cite{tikhonov2019critical}].

	It is quite remarkable that our systematic finite-size scaling  analysis highlighted these critical behaviors, the importance of which had not been understood in the localized phase. Indeed, let us recall that the localized phase was thought to be understood in an exact way, characterized in particular by a single localization length diverging with a critical exponent $1$ [see e.g. \cite{tikhonov2019critical}, Eq.~(30)]. On the contrary, we have shown that there exist two critical localization lengths $\xi_\parallel$ and $\xi_\perp$ and two critical exponents $1$ and $1/2$. Both critical exponents control the finite-size scaling  properties of distinct observables.

\subsection{Comparison with mathematical results} 
On the Cayley tree, there exist rigorous results \cite{aizenman2011absence,aizenman2011extended} which give a criterion for the transition. Consider the two-point Green function $G_{ij} = \langle i \vert (E-H)^{-1} \vert j \rangle$. The exponential decay of this observable is characterized by the typical Lyapunov exponent $\lambda_{\text{typ}} = -\text{lim}_{d \rightarrow \infty} \langle \ln G_{ij}\rangle/d$, and the averaged $\lambda_{\text{av}}=-\text{lim}_{d \rightarrow \infty} \ln (\langle  G_{ij}\rangle)/d$, where $d$ is the distance between $i$ and $j$ (see \cite{kravtsov2018non}). In \cite{aizenman2011absence}, it was shown that a sufficient condition for delocalization is that the typical Lyapunov exponent is $\lambda_{\text{typ}} < \ln K$. A simple heuristic picture can be provided. 
The density of the states whose localization center is at distance $d$ from that of a given state increases exponentially with a rate governed by 
$\ln K$ while $\lambda_{\text{typ}}$ governs the typical exponential decay of wavefunctions; if this exponential decay is slower than the increase in the density, delocalization must set in. This argument is reminiscent of the proliferation of many-body resonances in the MBL problem, leading to the prediction of an ETH-prethermal MBL 
crossover (where ETH denotes eigenstate thermalization hypothesis) discussed recently in \cite{garrat2021local, morningstar2022avalanches, long2022phenomenology} (see also \cite{khaymovich2021dynamical} for a random matrix perspective).

However, in \cite{aizenman2011extended} the authors showed that the localization transition is in fact controlled by the other averaged Lyapunov exponent, $\lambda_{\text{av}} = \ln K$ at the delocalization-localization transition. This is a consequence of a large deviation analysis which takes into account slower wave function decays (along rare branches in our physical picture). 

We stress that the Green's function $G_{ij}$ is distinct from the correlators $\Ctav$ and $\Cttyp$ that we have defined in Eqs.~\eqref{chav} and \eqref{cttyp}. In some sense, $G_{ij}$ is a correlator of wave functions $\psi$ while  $\Ctav$ and $\Cttyp$ are correlators of wavefunction amplitudes $\vert \psi \vert^2$. Nevertheless, it is tempting to say that the exponential decay of $\langle  G_{ij}\rangle$ with $\lambda_{\text{av}} = \ln K$ is equivalent to the $K^{-r}$ decay of the correlation function $\Ctav$ at the transition. Our results thus suggest \cite{kravtsovprivate}: $\lambda_{\text{av}} \approx  \ln K + c (W-W_c)$ in the localized phase $W>W_c$, with $c$ a positive constant.
In \cite{aizenman2011absence,aizenman2011extended} the importance of $\lambda_\text{typ}$ as a critical quantity was not discussed.  Our results suggest \cite{kravtsovprivate}: $\lambda_\text{typ} \approx 2 \ln K + c' (W-W_c)^{1/2}$ for $W>W_c$, with $c'$ a positive constant. It would be interesting to assess this important conjecture on the behavior of the Green function.

\section{Evidence that the Anderson transition on random graphs is of Kosterlitz-Thouless type, in the same universality class as the MBL transition}
\label{sec:MBL}
\begin{figure}[!t]
  \includegraphics[width=1.\linewidth]{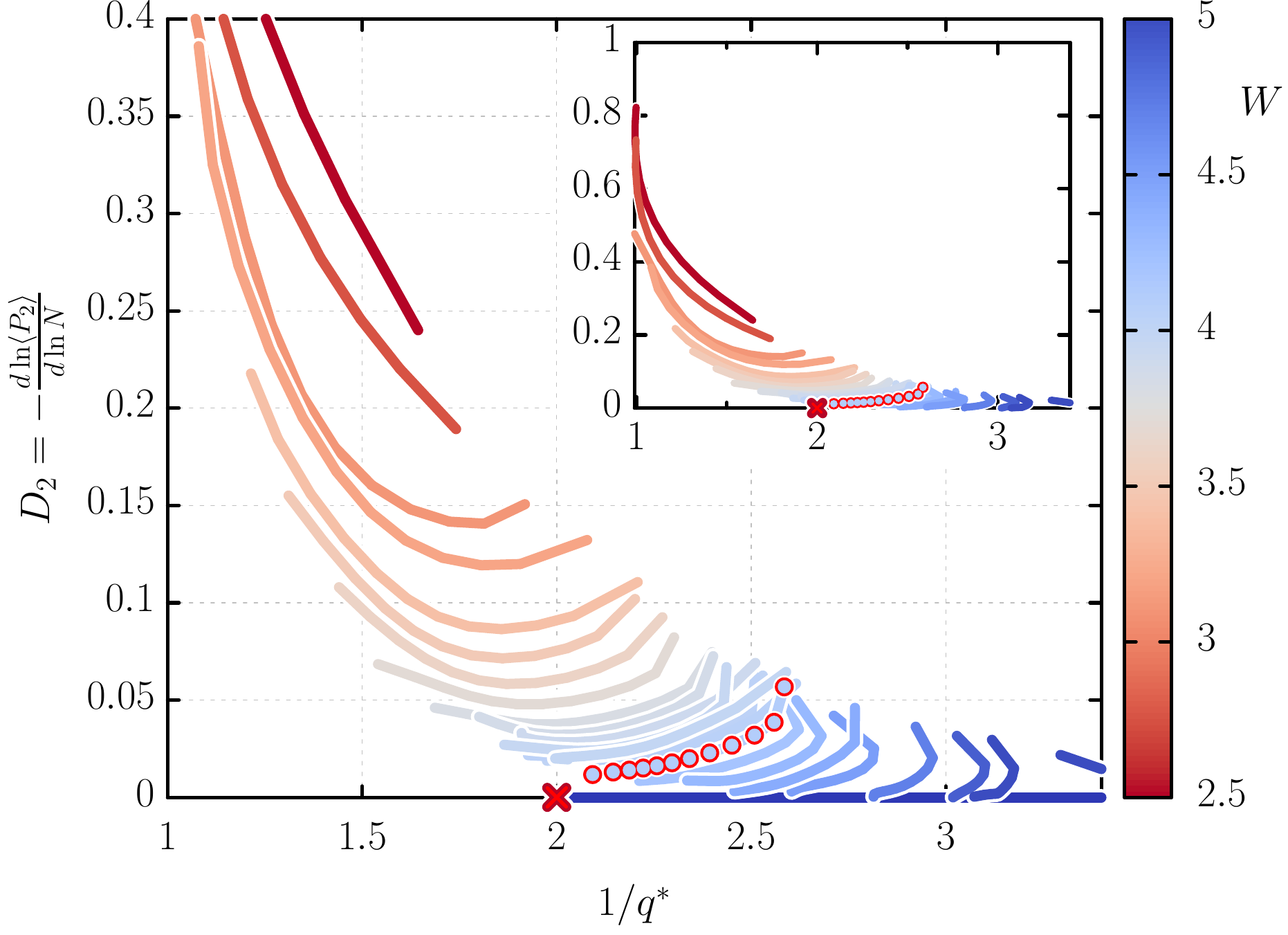}
\caption{Two parameter  RG Kosterlitz-Thouless type flow for $p=0.25$. The $q^*$ is directly obtained from fitting $\tilde{\tau}_q$ with $q/q^*-1$ for $q\sim 0$. The ordinate is the ``flowing'' multifractal dimension $D_2$, i.e. the logarithmic derivative of the moment $\langle P_2 \rangle$ as a function of system size (for $N=2^8$ to $2^{20}$). $D_2$ can be interpreted as the linear density of the equivalent of thermal bubbles in the MBL problem (see text), while $q^*=\xi_\perp \ln K$ controls the value of the typical localization length $\xi_\perp$. In the localized phase $W>W_c$, the flow is towards a ``critical line'' with $D_2=0$ and $q^*$ finite, while at the critical point $q^*$ reaches a universal value $q^*=1/2$. Both $D_2$ and $q^*$ flow to $1$ in the ergodic delocalized regime. The critical behavior is shown by the red open circles and separates these two types of flow. This important figure demonstrates a Kosterlitz-Thouless type flow for the Anderson transition on random graphs and is very similar to what is predicted for the MBL transition \cite{dumitrescu19}.
In the inset we show the whole scale where the limit $q^*\to 1$, $D_{2}\to 1$ is observed.}
\label{figurasRGflow}
\end{figure}

The many-body localization (MBL) phase transition separates a delocalized from a many-body localized phase in quantum systems in the presence of interactions and disorder. In the delocalized phase thermalization occurs and the eigenstate thermalization hypothesis (ETH) is expected to hold \cite{deutsch1991quantum, srednicki1994chaos, rigol2008thermalization}. In the MBL phase thermalization is precluded. 
In \cite{de2017stability}, a theory was proposed to describe how small ergodic regions due to disorder fluctuations trigger an instability leading to the MBL transition. This instability argument is closely related to the 
argument involving the typical Lyapunov exponent for the Anderson transition on the Cayley tree \cite{aizenman2011extended,warzel2013resonant}.
The MBL transition was described via a phenomenological renormalization group  approach involving this quantum avalanche mechanism \cite{thiery2018many, dumitrescu19, MorningHuse19}, where locally thermalized parts of the many-body system gradually extend to the whole system \cite{warzel21}.
In \cite{goremykina2019analytically, dumitrescu19}, it was shown that such an approach leads to a Kosterlitz-Thouless type renormalization group (RG) flow \cite{kosterlitz2016kosterlitz}, described by a system of differential equations coupling the density of thermal parts with the inverse of the characteristic scale. The typical decay length $\zeta$ of the effective interactions, which governs the localized phase, follows these RG equations, and a critical line separates a phase where $\zeta^{-1}$ converges to a finite value $\zeta_\infty^{-1}$ from a phase where it goes to zero under renormalization. In the vicinity of the transition, one predicts \cite{dumitrescu19}:
\begin{equation}\label{eq:zetaMBL}
    \zeta_\infty^{-1} = \zeta_c^{-1} + c_1 \sqrt{W-W_c} \; ,
\end{equation}
for $W>W_c$. This equation is identical to our Eq.~\eqref{eq:xiperp} for $\xi_\perp$ in the vicinity of the transition. In particular, $\zeta_c^{-1} = \ln 2$ in \eqref{eq:zetaMBL} is a universal number which depends only on how the Hilbert space volume grows with system size $L$, similarly to our critical value ${\xi_\perp^c}^{-1} = 2 \ln K$. Finally, finite-size effects are predicted to be controlled by an exponentially diverging scale $\xi_+\sim e^{c_+/\sqrt{W-W_c}}$, which is the analog of our localization volume $\Lambda = \mathcal V(\xi) \sim e^{\text{cst}/\sqrt{W-W_c}}$, where $\xi \sim (W-W_c)^{-1/2}$ controls the scaling properties of typical observables.

Interestingly, a similar Kosterlitz-Thouless type flow can be obtained in our system, with linear system size $d_N$ playing the role of the renormalization parameter.
In the localized regime, multifractal wave functions are essentially supported on a few branches, which can be seen as non-ergodicity bubbles, whose extent is governed by $\xi_\parallel$ along the branch \cite{scaling17}. These wave functions extend from that branch with characteristic length $\xi_\perp$, which from the model described in Sec.~\ref{characscale} can be related to $q^*$ by $q^*= \xi_\perp \ln K$. Either $\xi_\perp$ or $q^*$ itself thus plays the role of the decay length $\zeta$ of interactions.
In the vicinity of the critical point, the volume occupied by a multifractal wavefunction is given by its inverse participation ratio, which scales as $\sim N^{D_2}$; as a consequence, its linear extension scales as $\sim D_2\ln N$. The quantity $P_2^{-1}/N$ therefore gives the volumic density, and $\ln P_2^{-1}/\ln N\sim D_2$ corresponds to the linear density of non-ergodicity bubbles. Since the inverse of the decay length behaves as $\xi_\perp^{-1}\sim(\xi_\perp^c)^{-1}+C |W_c-W|^{-\kappa}$, $\kappa\approx 0.5$ at the transition, $1/q^*$ approaches the termination point $1/q_c^*=2$. The RG flow goes to a density 0 in the localized regime, and to a finite value in the delocalized regime (with $D_2=1)$.

This RG flow is illustrated in Fig.~\ref{figurasRGflow}, where we plot the multifractal dimension $D_2$ as a function of $1/q^*$. Each curve corresponds to a fixed $W$, and $N$ (or the RG flow parameter $d_N$) increases from right to left along a given curve. The critical line (circles) separates a localized phase (blue curves), where the curves have a finite termination point at $D_2=0$, from a delocalized phase (red curves), where $D_2$ goes to 1. At the critical value of disorder, one has $q^*=q_c^*=1/2$. Figure~\ref{figurasRGflow} strikingly resembles the RG flow picture of the MBL setting in \cite{dumitrescu19} or \cite{MorningHuse19}, showing the close relationship between our model and MBL. Analogous results are shown in Fig.~\ref{figurasRGflow2p} in Appendix \ref{twoparamapp} for $p=0.06$ and $0.49$. 

We also note that linear and volumic scalings have been observed in the MBL transition \cite{mace2019multifractal, laflorencie2020chain, roy2021fock, sutradhar2022scaling}. Another analogy with MBL manifests itself in the behavior of the radial probability distribution of an eigenstate around its maximum. In Appendix \ref{radialprob} we show that we obtain qualitatively similar results to those obtained in \cite{warzel21, roy2021fock} for the MBL transition in random quantum Ising models.

The great similarity of these behaviors, in particular the exact correspondence between the equation \eqref{eq:xiperp} for $\xi_\perp$ and \eqref{eq:zetaMBL} for $\zeta$ and the correspondence between the RG flows, are strong evidence that the Anderson transition on random graphs is in the same universality class than the MBL transition. 
However, this should not obliterate important differences between these two problems. First, the comparison between the finite-size effects for the MBL transition \cite{dumitrescu19, laflorencie2020chain} and the Anderson transition on random graphs suggests that the size of the system $L$ for a many-body system would correspond to the volume $N$ of a graph (see however \cite{mace2019multifractal} which suggests that $L$  corresponds to $d_N$). This helps to understand why we can clearly observe the transition in the case of random graphs, where $N$ reaches several million sites, whereas it is much more delicate in the MBL transition (where $L$ reaches 24 sites in exact diagonalization). Second, the latest developments on the phenomenological renormalization group approach on the MBL transition \cite{MorningHuse19, morningstar2020many}, which take into account large fluctuations in $\zeta$, indicate a renormalization flow which is distinct from \cite{dumitrescu19}, although similar to the Kosterlitz-Thouless flow found previously. Third, the MBL phase is not strictly speaking localized in Hilbert space, but on the contrary is multifractal \cite{mace2019multifractal, tarzia2020many}, thus can be seen as a non-ergodic delocalized phase \cite{tikhonov2016fractality, khaymovich2020fragile, kravtsov2015random}.
Fourth, disorder in the Hilbert space of an MBL system is strongly correlated \cite{roy2020fock}. There are indications that this type of strong correlations might change the nature of the Anderson transition on random graphs \cite{roy2020localization}. Fifth, there has been a long debate about the value of the critical disorder at the MBL transition. This debate does not exist for the Anderson transition, where rigorous approaches make it possible to estimate the threshold $W_c$ very precisely \cite{parisi2019anderson, tikhonov2019critical, sierant2022universality}. Finally, the MBL transition is thought to be driven by avalanches \cite{de2017stability, thiery2018many, dumitrescu19, MorningHuse19}. It is not clear at present whether such a mechanism is relevant
for the Anderson transition, although the strong similarity between the RG flows suggests it might be the case.

\section{Conclusion}
\label{mainccl}

In this paper, we have described the critical properties of the Anderson transition on random graphs from a scaling approach. This problem has raised a very strong interest recently \cite{abouchacra73, monthus2011anderson, biroli2012difference, de2014anderson,  kravtsov2015random, altshuler2016nonergodic, facoetti2016non, tikhonov2016anderson, tikhonov2016fractality, sonner2017multifractality,scaling17, biroli2017delocalized, monthus2017multifractality, tarquini2017critical, kravtsov2018non, bogomolny2018eigenfunction, bogomolny2018power, bera2018return, tikhonov2019statistics, tikhonov2019critical, parisi2019anderson, twoloc20, kravtsov2020localization, detomasi2020subdiffusion, roy2020localization, kravtsov2020localization, khaymovich2020fragile, biroli2022critical, biroli2021levy, alt2021delocalization, tikhonov2021anderson, colmenarez2022sub, sierant2022universality}, in particular because of its analogy with the MBL transition \cite{altshuler1997quasiparticle, tikhonov2021anderson, roy2020localization, tarzia2020many, biroli2017delocalized, warzel21}. 
It has generated several debates, on whether the delocalized phase is ergodic or not
\cite{monthus2011anderson, biroli2012difference, de2014anderson,  kravtsov2015random, altshuler2016nonergodic, facoetti2016non, tikhonov2016anderson, tikhonov2016fractality, sonner2017multifractality,scaling17, biroli2017delocalized, monthus2017multifractality, tarquini2017critical, kravtsov2018non, bera2018return, tikhonov2019statistics, tikhonov2019critical, parisi2019anderson, twoloc20, kravtsov2020localization, detomasi2020subdiffusion, kravtsov2020localization, khaymovich2020fragile}, and more recently on the value of the critical exponents \cite{scaling17, kravtsov2018non, tikhonov2019critical,  twoloc20, biroli2022critical, sierant2022universality}. 
The scaling theory that we describe in this paper makes it possible to answer these questions and many others: 
In fact, even if the Anderson transition on random graphs was reputed to be better understood than the MBL transition, in particular by the existence and the precise value of a critical disorder $W_c$ \cite{tikhonov2019critical, parisi2019anderson, sierant2022universality}, the nature of the transition and its renormalization group flow were not known. In contrast, the phenomenological RG approach of the MBL transition allows to describe the corresponding RG flow as being of the Kosterlitz-Thouless type \cite{goremykina2019analytically, dumitrescu19,morningstar2020many}. It is therefore of high interest to clarify the nature and the type of flow of the Anderson transition.
This is what we have done in this paper.

We have found compelling evidence that there exist not one but two critical localization lengths, associated with two distinct critical exponents. In other words the RG flow has two parameters, and we find that it is of Kosterlitz-Thouless type. This had never been anticipated: the localized phase was deemed to be exactly understood and associated with a single critical exponent $\nu=1$ 
(see, e.g., \cite{tikhonov2019critical}). In fact, all recent attention on the Anderson transition on random graphs has been focused on the delocalized phase \cite{monthus2011anderson, biroli2012difference, de2014anderson,  altshuler2016nonergodic}. The picture we present is also very different from what is known for the Anderson transition in finite dimension, which has only one scaling parameter \cite{evers2008anderson}.

 Although there are many subtleties associated to the behavior of the finite size scaling, our approach has allowed us to give a global and consistent physical picture of the Anderson transition on random graphs. We constructed the scaling analysis in the least biased way, by making the fewest possible assumptions and by characterizing quantitatively the different possibilities.
The robustness of the results and their universality have been carefully tested 
by analyzing the behavior of different key observables: 
multifractality,  correlation functions and spectral statistics. 

In the type of random graphs considered here, of infinite effective dimension, with loops and without boundary, the results presented in the paper indicate the existence of a single transition from a localized to a delocalized ergodic phase. Again, the critical behavior is quite different from the Anderson transition in finite dimension \cite{evers2008anderson, PhysRevLett.82.382, PhysRevB.84.134209}.

The localized regime, $W>W_c$, is strongly non-ergodic with distinct localization properties along the different branches of the graph. There are a few rare branches where the localization length $\xi_\parallel$ is much larger than $\xi_\perp$ which controls  how  eigenstates extend perpendicularly to these branches.
As one approaches the transition, the distinction between the two lengths becomes more prominent, since
$\xi_\parallel$ diverges, while $\xi_\perp$ tends to a finite universal value. 

We have been able to show how these lengths control different critical properties. 
On the one hand, the inverse participation ratio  and the average correlation function are governed by $\xi_\parallel$. On the other hand, other key observables, like the spectral statistics or a suitably defined typical correlation function, depend on $\xi_\perp$. Furthermore, we have shown that there are strong multifractal features that depend on the small amplitudes of the eigenstates that are also controlled by $\xi_\perp$.

One additional important result we obtained is that $\xi_\parallel$ and $\xi_\perp$ have different  
critical exponents. 
Indeed, we find that the length $\xi_\parallel$ diverges as $(W-W_c)^{-1}$. By contrast,  $\xi_\perp^{-1} \approx {\xi_\perp^c}^{-1} + \xi^{-1}$ with $\xi \sim (W-W_c)^{-1/2}$ controlling finite-size effects in that case.
We thus see the emergence of a flow with two parameters: the multifractal dimension $D_2$, which can be interpreted as the linear density of the equivalent of ``thermal bubbles'' in the MBL problem, and the localization length $\xi_\perp$. This two-parameter flow is of Kosterlitz-Thouless type. In the localized phase up to the critical point, we reach at large system sizes a ``critical line'' where $D_2=0$ and $\xi_\perp$ reaches a finite value. At the transition point, $\xi_\perp$ reaches a finite universal value. In the ergodic regime it goes to a constant value which presents a discontinuity with the critical values.
This is very similar to the flow predicted for the MBL transition \cite{goremykina2019analytically, dumitrescu19,morningstar2020many}. In fact, the behavior for $\xi_\perp$ is exactly that recently predicted for the typical localization length at the MBL transition using the phenomenological renormalization group approach \cite{dumitrescu19}.
 
The delocalized regime
is ergodic at large scales. However  at small scales, below a  characteristic correlation volume $\Lambda$, it exhibits signs of strong non-ergodicity, inheriting the strong multifractality of the critical behavior. 

The existence of the new critical exponent $\nu_\perp=1/2$ in the localized phase is proved \emph{a posteriori} compatible with previously formulated theories \cite{mirlin1994distribution, sonner2017multifractality, de2014anderson, kravtsov2018non}. Moreover, $\nu_\perp=1/2$ can be understood as the limit of the critical exponent in finite dimension, which tends towards $1/2$ and not $1$ when the dimension increases \cite{garcia2007dimensional, ueoka2014dimensional, tarquini2017critical}. In this sense, it is rather the exponent $\nu_\parallel=1$ that is abnormal, i.e., specific to random or tree graphs of effective infinite dimension, and reflects the importance of rare events, i.e., rare branches. On the other hand, our results clearly indicate the existence of a unique critical exponent $\nu=1/2$ in the delocalized phase, clarifying a recent debate \cite{scaling17, kravtsov2018non, tikhonov2019critical,  twoloc20, biroli2022critical, sierant2022universality}.

Our work shows a very strong analogy between the Anderson transition on random graphs and the MBL transition. This opens interesting perspectives. First of all, the analogy of the critical behaviors suggests that these two transitions could be governed by the same avalanche mechanism. This mechanism is crucial in the MBL problem \cite{de2017stability, thiery2018many}, but remained difficult to characterize numerically and experimentally \cite{luitz2017how, goihl2019exploration, crowley2020avalanche, leonard2020signatures, morningstar2022avalanches, sels2022bath}. It would be interesting to clarify if an analogous mechanism takes place in random graphs. If this is the case, a phenomenological RG approach could be developed for the Anderson transition, complementary to other known analytical approaches \cite{zirnbauer1986localization, fyodorov1991localization, tikhonov2019statistics, kravtsov2018non, khaymovich2020fragile, tarquini2016level}, where, according to several studies \cite{Panda_2020, morningstar2022avalanches, long2022phenomenology}, finite-size effects are too strong to properly describe the transition. 

The methods presented here could also be used to understand the critical properties of transitions to non-ergodic delocalized phases, whose existence has been demonstrated in a large class of systems \cite{kravtsov2015random, tikhonov2016fractality, khaymovich2020fragile, biroli2021levy}. In particular, the MBL phase can be seen as a non-ergodic delocalized phase in the Hilbert space, since it is multifractal there \cite{mace2019multifractal, tarzia2020many}. Moreover, many-body ground states are multifractal \cite{PhysRevB.80.184421, PhysRevE.86.021104, luitz2014universal, PhysRevLett.122.106603}, thus non-ergodic delocalized, and their transitions could be interpreted as transitions between different non-ergodic delocalized phases, where our scaling theory could be very useful.

\begin{acknowledgments}
 The authors thank D.~Huse, V.~Kravtsov, A.~Mirlin, A.~Scardicchio, P.~Sierant, and K.~Tikhonov for interesting discussions. 
 This study has been (partially) supported through the EUR grant NanoX ANR-17-EURE-0009 in the framework of the "Programme des Investissements d'Avenir", the French-Argentinian LIA LICOQ, and also by research funding Grants No.~ANR-17-CE30-0024, ANR-18-CE30-0017 and ANR-19-CE30-0013. We thank Calcul en Midi-Pyr\'en\'ees (CALMIP) and the Consortium des Equipements de Calcul Intensif (CECI), funded by the Fonds de la Recherche Scientifique de Belgique (F.R.S.- FNRS) under Grant No.~2.5020.11, for computational resources and assistance. I.G.-M.~received funding from CONICET (Grant No.~PIP 11220150100493CO) and ANCyPT (Grants No.~PICT-2020-SERIEA-00740 and PICT-2020-SERIEA-01082).
\end{acknowledgments}


\appendix

\section{Dependence of $W_c$ on the parameter $p$}
\label{AppendixA}

\begin{figure}
\includegraphics[width=0.85\linewidth]{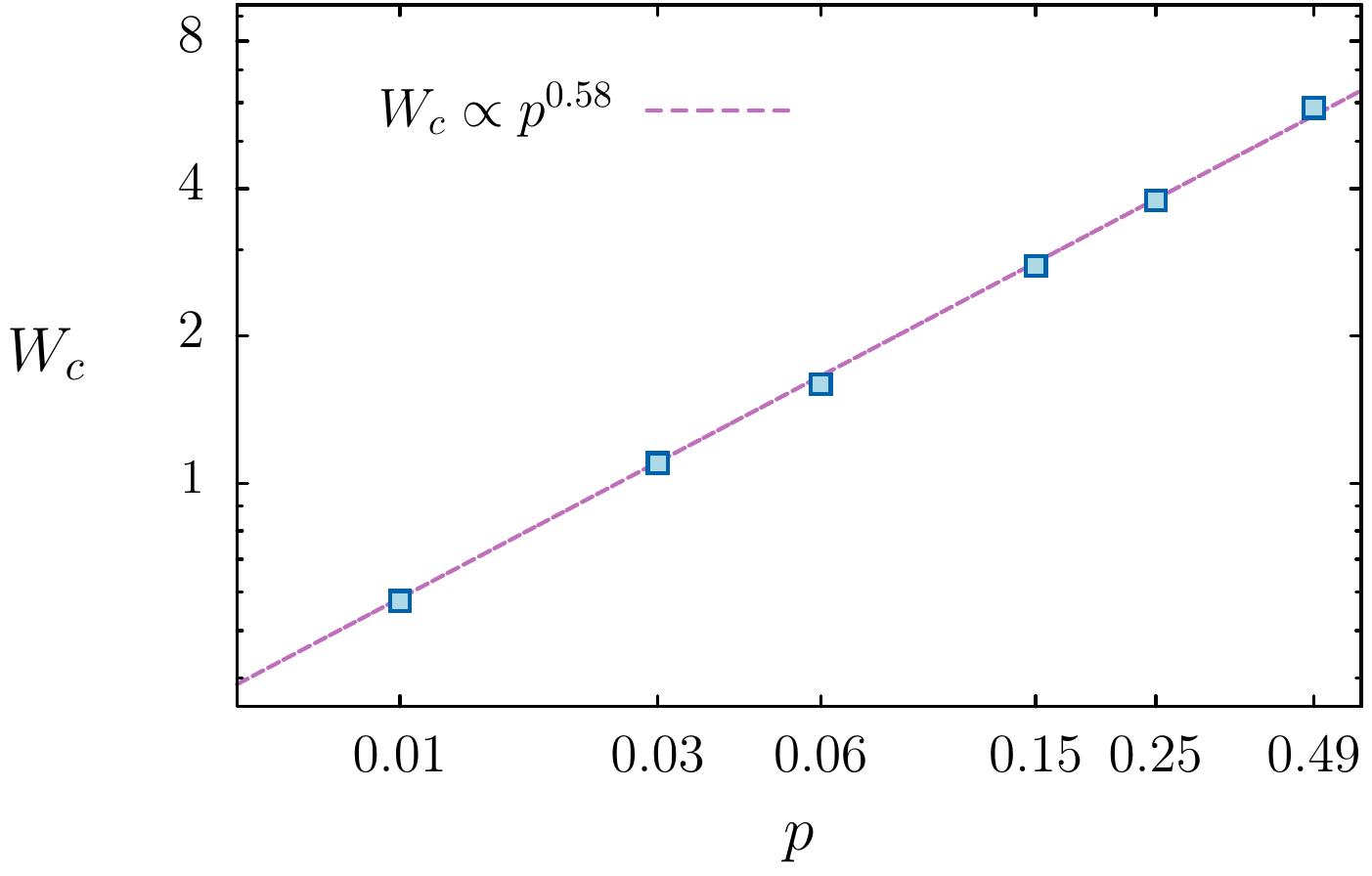}
\caption{Log-log plot of the critical value $W_c$ of the (Gaussian distributed) disorder strength as a function of the long-range link probability $p$. Data points have coordinates $(p,W_c)=(0.01,0.58)$, $(0.03,1.10)$, $(0.06,1.60)$, $(0.15,2.78)$, $(0.25,3.78)$ and $(0.49,5.85)$. The straight dashed line shows the best power-law fit, $W_c^{\mathrm{fit}}=8.55\,p^{0.58}$.
}
\label{figpWc}
\end{figure}

Figure~\ref{figpWc} shows that the critical disorder strength $W_c$ determined on the basis of spectral statistics follows a power law as a function of $p$ with an exponent $0.58$ (purple line). This exponent is close to $1/2$, expected from the following simple physical picture: In the absence of long-range links, the localization length of the 1D Anderson model scales as $\xi_\mathrm{loc}\sim 1/W^2$ at small $W$. When this length is smaller than the mean
distance $1/(2p)$ between two long-range links, the states remain localized. By contrast, when $\xi_\mathrm{loc}$ exceeds $1/(2p)$, delocalization sets in. Hence, the transition should occur at $W_c\sim p^{1/2}$~\cite{Chepelianskii}.\\

\section{Other correlation functions}
\label{appcorr}

In \cite{twoloc20}, we considered the average $\Cav$ and typical $\Ctyp$ correlation functions \textit{along the 1D chain}, defined as
\begin{align}
\label{cav}
\Cav(r)&=\left\langle\frac{1}{N} \sum_{i=1}^N |\psi_i|^2|\psi_{i+r}|^2\right\rangle, \\
\Ctyp(r)&=\exp\left\langle\ln\left( \frac{1}{N}\sum_{i=1}^N |\psi_i|^2|\psi_{i+r}|^2\right)\right\rangle\ .
\label{ctyp}\end{align} 
These correlation functions are also characterized by the two critical localization lengths $\xi_\parallel$ and $\xi_\perp$. It was shown in \cite{twoloc20} that they behave asymptotically as
\begin{equation}
\label{eq:simcavApp}
\Cav(r)\sim \frac{e^{-r/\xi_{\parallel}}}{r^\alpha},
\end{equation} 
\begin{equation}
\Ctyp(r)\sim e^{-r/\xi_{\perp}}.
\label{eq:simctypApp}
\end{equation}
This is similar to the behavior of the correlation functions $\Ctav$ and $\Cttyp$. Note that the chain can be considered a branch only up to a certain length $\approx d_N/2$ (after that length, the local tree-graph structure is lost, as shown in Fig.~\ref{fig1Remy}, where  $N_r $ deviates strongly from its $K^r$ behavior).
Since for each realization only a few branches are significantly populated, in some instances of the random disorder the main chain will be one such branch, which justifies the analogy between $\Cav$, $\Ctyp$ and $\Ctav$, $\Cttyp$.

\section{Numerical implementation of finite-size scaling }
\label{secfssapp}

\begin{figure}
\includegraphics[width=0.99\linewidth]{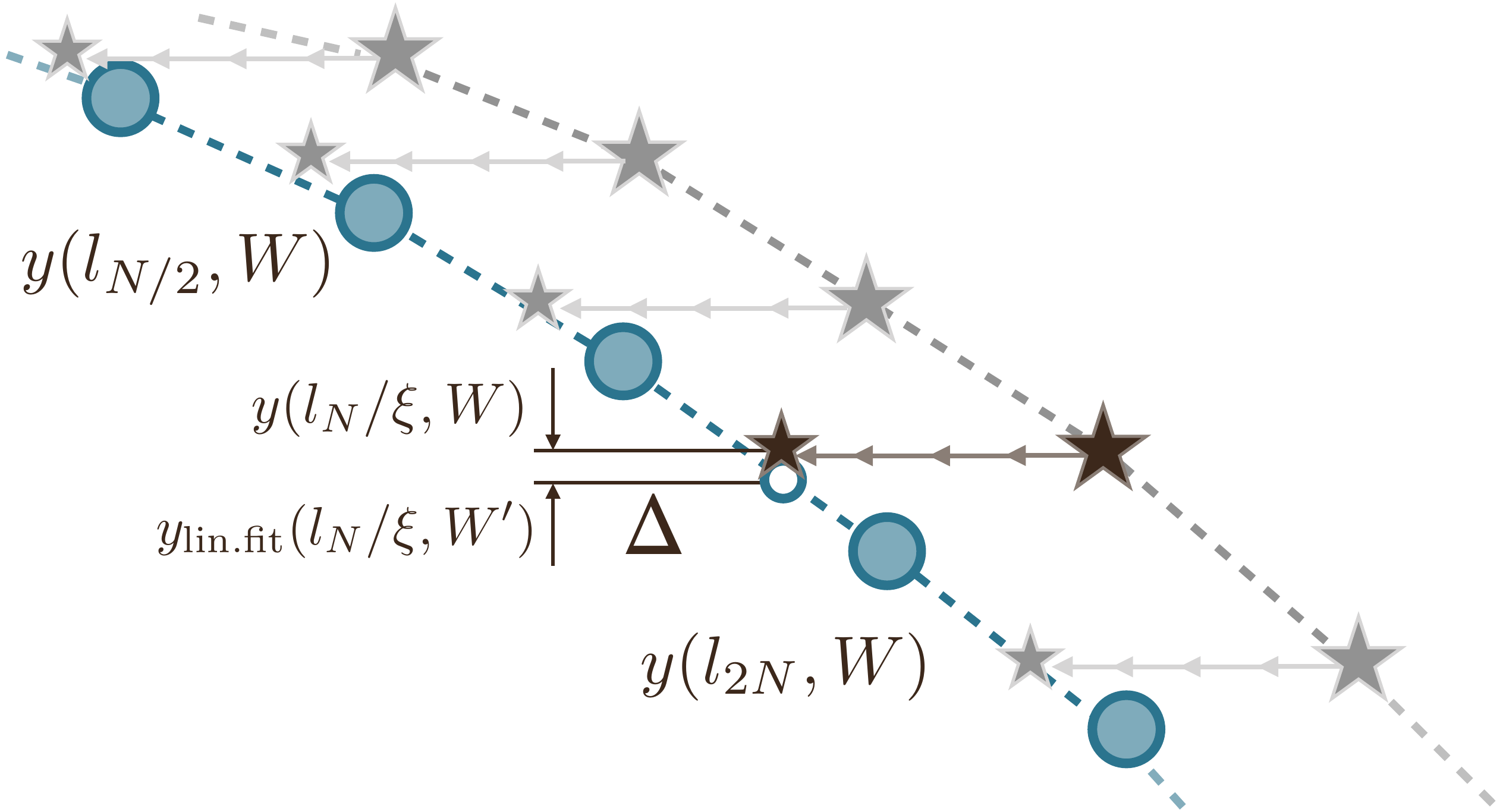}
\caption{\label{schematicFSS}
Schematic representation of the finite-size scaling  procedure in the numerical implementation. Here $\Delta\equiv y(l_N/\xi,W)-y_{\rm lin. fit}(l_N,W')$ is the quantity used to compute $\chideux$ in Eq.~\eqref{eq:chi2}}
\end{figure} 
  
 In practice, the numerical implementation of the finite-size scaling  procedure is done in the following way, schematically depicted in Fig.~\ref{schematicFSS}.  A candidate value for $W_c$ is chosen, and either volumic or linear behavior  (see Eq.~(\ref{eq:scahyp}) is assumed on each side of the transition. All curves $\langle P_q(N,W)\rangle$ are then rescaled by $\langle P_q(N,W_c)\rangle$; namely, we calculate the quantities $y(l_N,W)=\ln [\langle P_q(l_N,W_r)\rangle/\langle P_q(l_N,W_c)\rangle]$, where $l_N=\ln N$ for volumic scaling and $\ln\ln N$ for linear scaling.
We fix the value of $W$ which is farthest from $W_c$ (on the side of the transition we are considering) as the reference value (denoted $W_r$);  our aim is then to find the  parameters $\xi$ (or $\Lambda$) such that the curves $y(l_N,W)$ as a function of $N$ collapse best, for all $W$ onto the curve $y(l_N,W_r)$. This is done by collapsing the curves one after the other, starting from the value of $W$ closest to $W_r$. For instance in the linear scaling case, suppose the curve for some $W'$ has already been collapsed on top of the curve for $W_r$. Then the length $\xi$ associated with $W$ is obtained by finding the rescaling which minimizes the quadratic error between all data points. Since the curve for $W'$ has been shifted by $\xi(W')$, the available point abscissas do not necessarily coincide for different $W$. We therefore introduce $y_{\rm lin. fit}(l_N,W)$, the value of $y$ obtained by a linear fit of consecutive points of $y$.
The parameter $\xi$ is the one minimizing
\begin{equation}
\label{eq:chi2}
\chideux(W; W_c)=\sum_{N} \left[y(l_N/\xi,W)-y_{\rm lin. fit}(l_N,W')\right]^2.
\end{equation}
For each candidate $W_c$ we can then assess the validity of our hypothesis by defining
\begin{equation}
\label{eq:sumW}
\chideux(W_c)=\sum_W\chideux(W; W_c).
\end{equation}

\section{Finite-size scaling  of moments for other parameter values}
\label{appotherparam}
\begin{figure}
  \includegraphics[width=.99\linewidth]{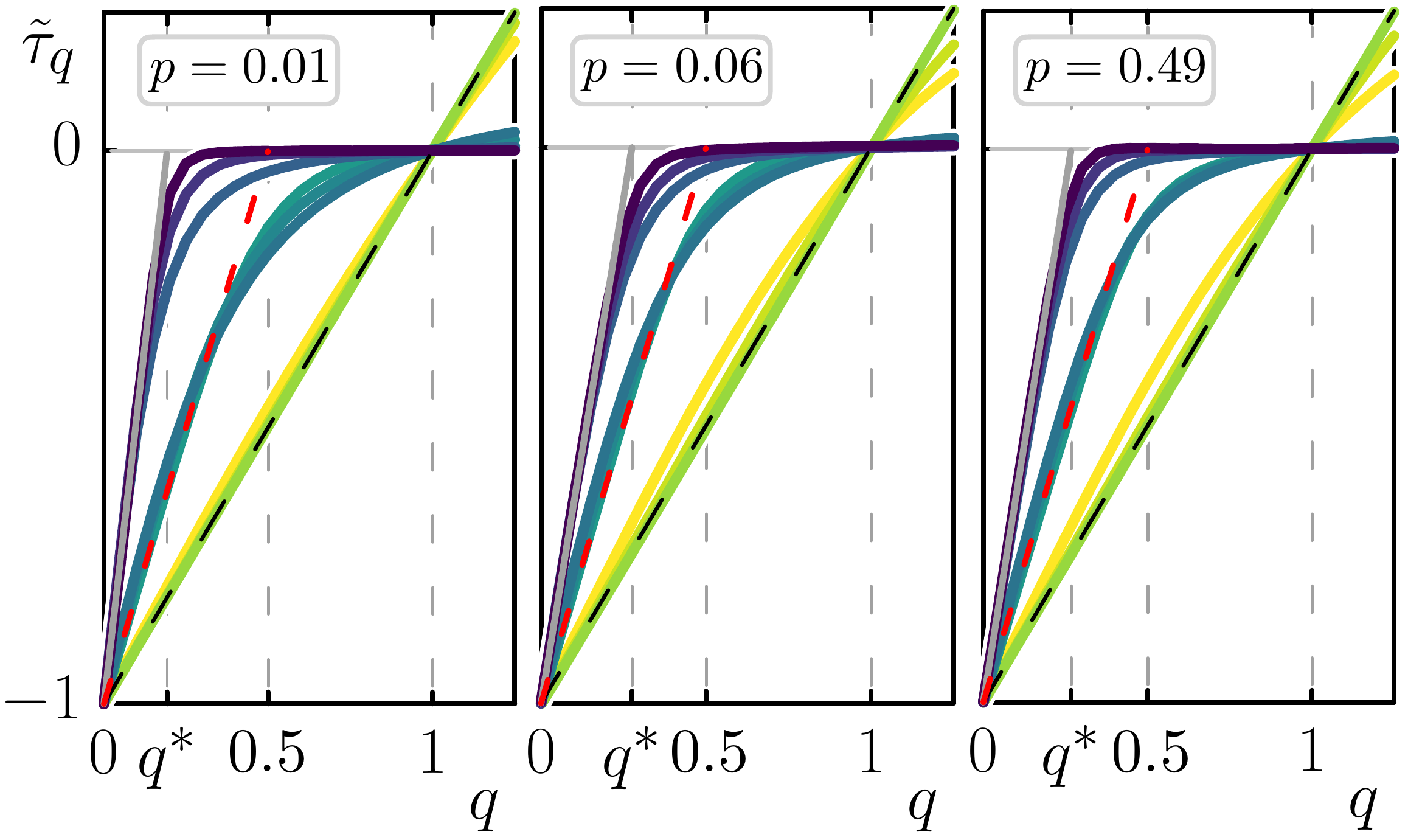}
\caption{Multifractal exponents $\tilde{\tau}_q$ extracted from the averaged moments $\langle P_q\rangle$ for $p=0.1,\ 0.06,\ 0.49$  (for $p=0.49$ we used a box disorder distribution instead of a Gaussian distribution),
for various disorder strengths, from small $W=0.25,\, 1.05,\, 11$ (delocalized regime), critical $W=0.64,\, 1.725,\, 17$, to large $W=1,\, 2.2,\, 26$ (localized regime) for $p=0.01,\, 0.06,\, 0.49$, respectively. 
 We show system sizes $N=2^{12},2^{15},2^{20}$.}
\label{tauqtilde3p}
\end{figure}

\begin{figure*}
\includegraphics[width=0.94\linewidth]{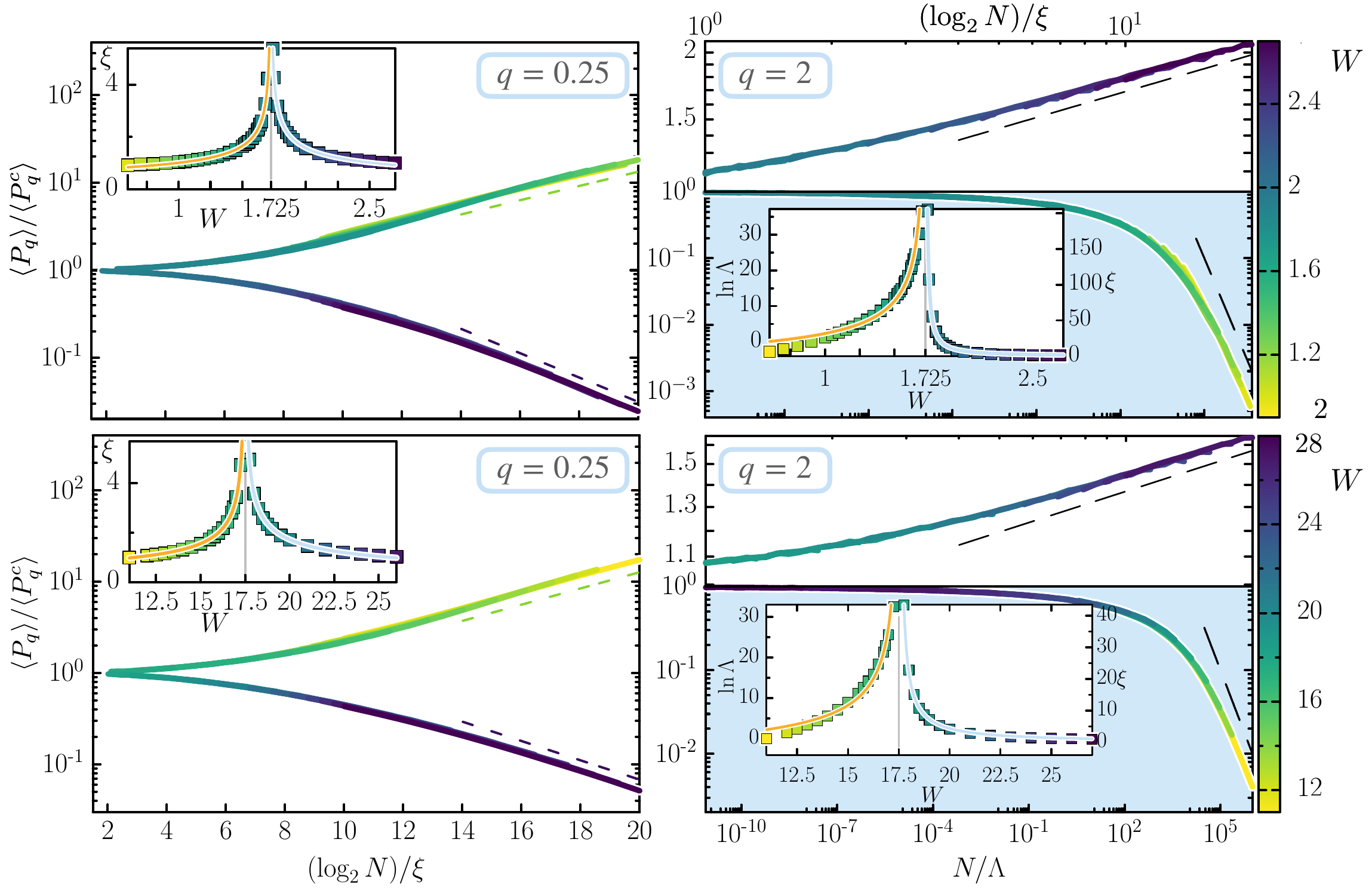} 
\caption{\label{figFSS05p006} Finite-size scaling  of moments for $p=0.06$ with Gaussian disorder (top row) and $p=0.49$ with box disorder (bottom row). In both cases the left column corresponds to $q=0.25$ and the right column to $q=2$. For $p=0.06$ the finite-size scaling  was done assuming $W_c=1.725$ and for $p=0.49$, $W_c=17.5$. For $q=0.25$ we find linear scaling on both sides of the transition. In the insets we show  the length $\xi$ as a function of $W$. The lines are fits $\sim |W-W_c|^{-\nu_{\rm loc/deloc}}$. The corresponding critical exponents obtained are: $\nu_{\rm loc}\approx 0.42$ and  
$\nu_{\rm deloc}\approx 0.42$ for $p=0.06$;  $\nu_{\rm deloc}\approx 0.48$ and
$\nu_{\rm deloc}\approx 0.45$ for $p=0.49$.
The right column corresponds to  $q=2$. Here, the top branch corresponds to the localized phase and has linear scaling, while the bottom branch corresponding to the delocalized phase shows volumic scaling. In the inset we show the logarithm of the scaling volume $\Lambda$ and the scaling length $\xi$ as a function of $W$. In the localized phase we get  $\nu_{\rm loc}\approx 1.14$ for $p=0.06$, and  $\nu_{\rm loc}\approx 0.99$ for $p=0.49$.  Note that in the delocalized regime for $q=2$ (orange curve)  we fixed the critical exponent to $0.5$ and plotted the fit of $\ln \Lambda$ with the function $A_1+A_2 (W_c-W)^{-0.5}$ with $A_1$ and $A_2$ as fitting parameters.}
\end{figure*}
In the main text, we presented results for a parameter $p=0.06$. Here we show that our analysis extends to a range of parameter values (up to $p\approx\frac12$, which corresponds to a 3-regular random graph), and that the critical properties are not sensitive to the choice of disorder distribution (Gaussian or box disorder distributions). 

In Fig.~\ref{tauqtilde3p} we show the analog of Fig.~\ref{figtauq} for different values of $p$. For each case we plot the behavior   $\tilde{\tau}_q=q/q^* - 1$ for three regimes: localized, critical, and delocalized. In the critical case $q^*=1/2$, in the delocalized case $q\to 1$ (as $N$ grows) and in the localized case $q^*<1/5$. We show very stable numerical results in a wide range of connectivity values ($p=0.01,0.06, 0.49$). The critical disorder can be estimated from these curves for the case where $q^*$ is closest to $1/2$. We use these values to compute the finite-size scaling  curves shown in  Fig.~\ref{figFSS05p006} for $p=0.06$ and $p=0.49$
 in analogy with Fig.~\ref{figFSS05p025}. We observe that for $q=0.25$ the finite-size scaling  is compatible with linear scaling on both sides of the transition. For large values of $q$ on the localized side a linear scaling is better suited, while on the delocalized side we see that a change to volumic scaling works best. Straight lines in Fig.~\ref{figFSS05p006} indicate the asymptotic behaviors of the scaling functions. These data confirm the results reported in Table \ref{tableresume}.
 
\begin{figure*}
\includegraphics[width=1\linewidth]{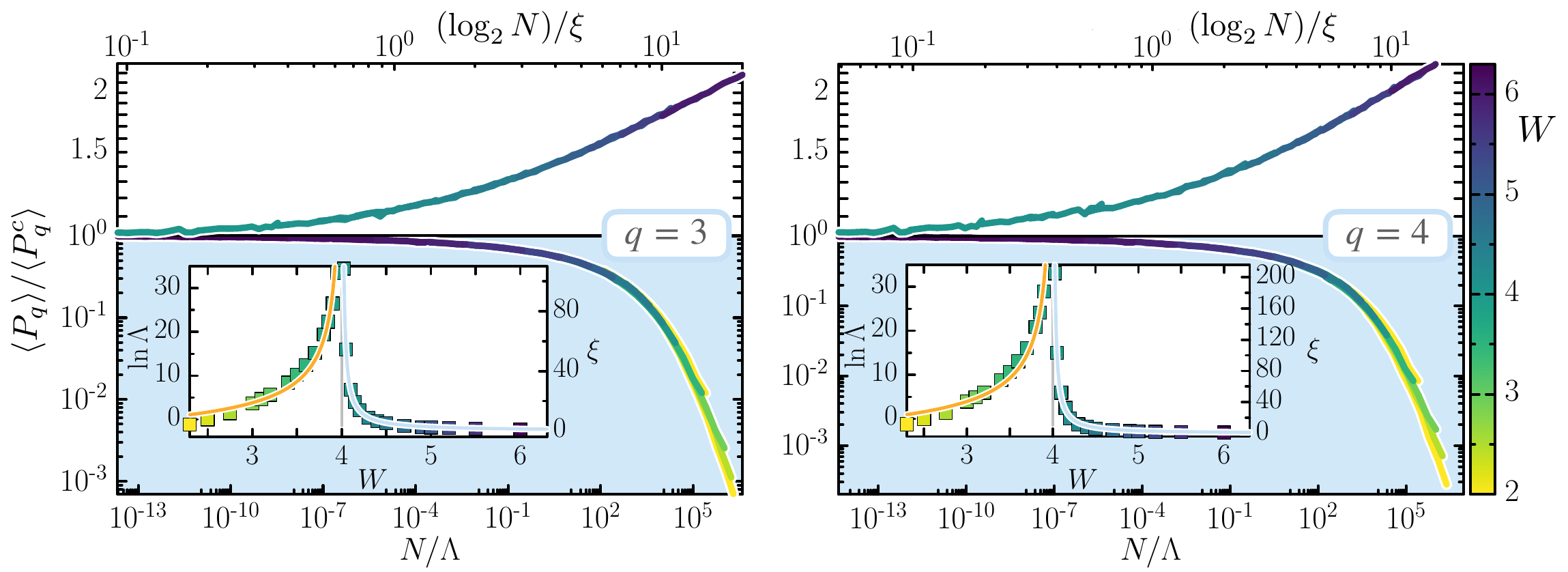}
\caption{\label{figFSS05p025q34} Finite-size scaling of moments for $p=0.25$  and $q=3$ (left) and $q=4$ (right). Finite-size scaling was done assuming $W_c=4$. 
The top branch corresponds to the localized phase and has linear scaling, while the bottom branch corresponding to the delocalized phase shows volumic scaling.
 In the inset we show the logarithm of the volume $\Lambda$ and the length $\xi$ as a function of $W$. In the localized phase we get  $\nu_{\rm loc}\approx 0.97$ for $q=3$, and $\nu_{\rm loc} \approx 1.09$ for $q=4$.  In the delocalized regime, in orange we  fit  $\ln \Lambda$ with the function $A_1+A_2 (W_c-W)^{-0.5}$.}
\end{figure*}

\begin{figure}
\includegraphics[width=0.99\linewidth]{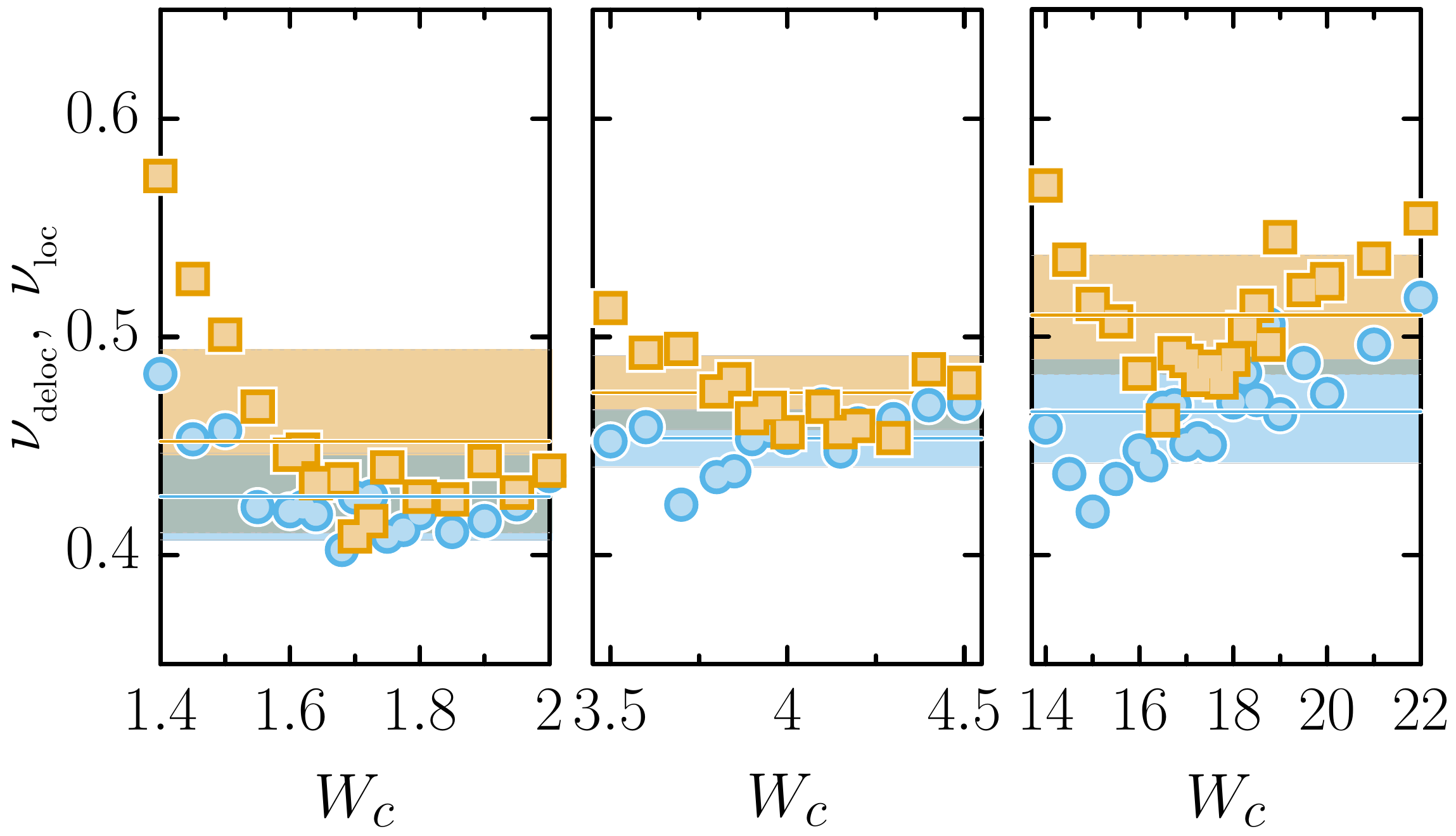}
\caption{\label{stab_nu} Stability of the critical exponent $\nu_\perp$ obtained from the finite-size scaling  analysis of moments (assuming linear scaling on both sides) for $q=0.25$, for different values of $p$. The straight lines correspond to the average value and the shaded regions mark the standard deviation $\sigma$. Left: $p=0.06$ the lines show the average $\langle\nu_\text{loc}\rangle=0.43\pm 0.02$, $\langle\nu_\text{deloc}\rangle=0.45\pm 0.04$. Center: $p=0.25$, $\langle\nu_\text{loc}\rangle=0.45\pm 0.01$, $\langle\nu_\text{deloc}\rangle=0.47\pm 0.02$. Right: $p=0.49$, $\langle\nu_\text{loc}\rangle=0.47\pm 0.02$, $\langle\nu_\text{deloc}\rangle=0.51\pm 0.03$.     }
\end{figure}

  In Fig.~\ref{figFSS05p025q34}, we show that for large moments $q=3$ and $q=4$  we obtain 
  the same critical properties  as for the case $q=2$ shown in Fig.~\ref{figFSS05p025} for $p=0.25$. In Fig.~\ref{stab_nu}, we show the stability of the value of the critical exponent obtained when $q=0.25$ and $p=0.25$ for different choices of $W_c$ close to the optimal value found $W_c \approx 4$.


\section{Details of the Taylor expansion of the scaling function around $W_c$}
\label{taylor}

\begin{figure}
    \includegraphics[width=0.99\linewidth]{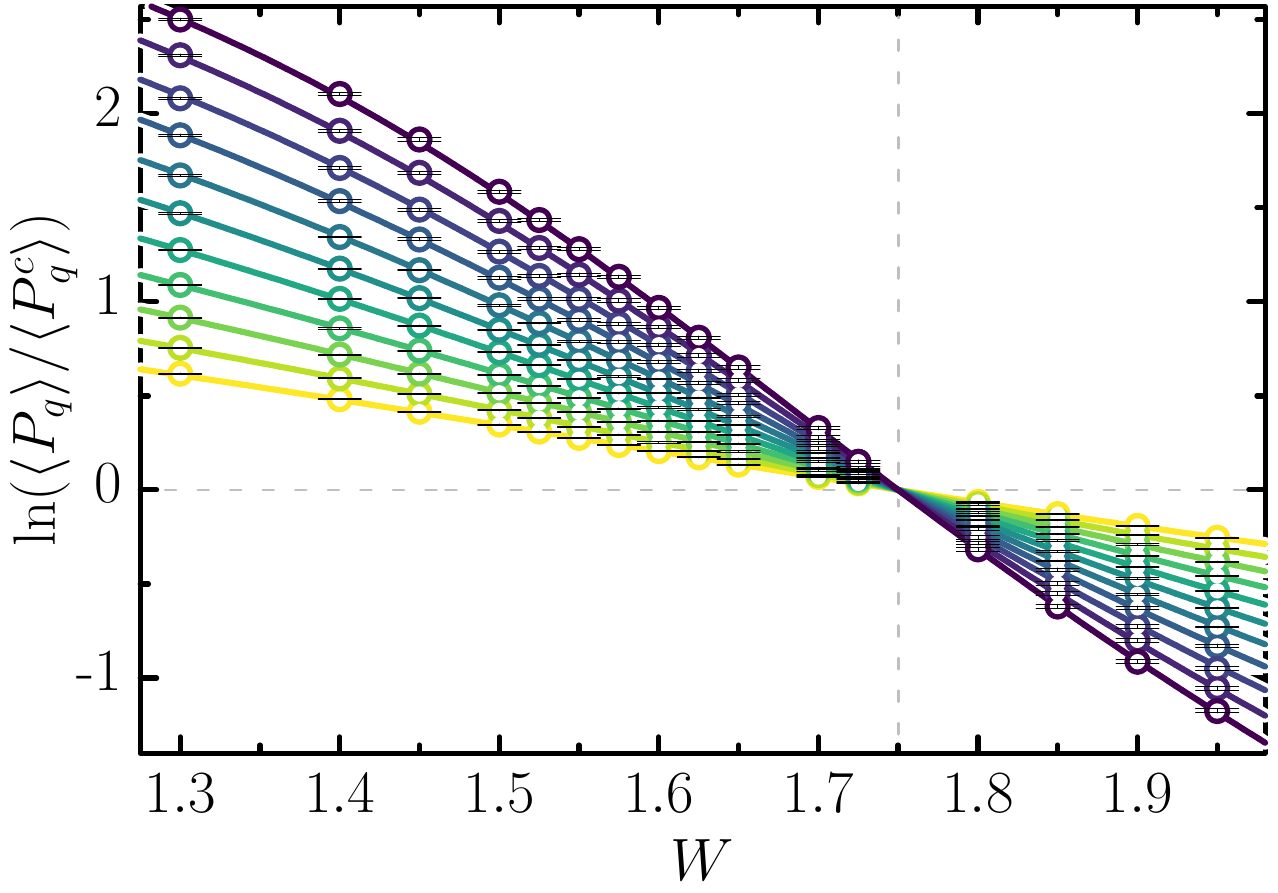}
	\caption{\label{fig:scaSlevin-lin-smallq} Finite-size scaling  analysis of $\langle P_q\rangle$ for $q=0.25<1/2$ using a Taylor expansion of the scaling function Eq.~\eqref{eq:scaTaylor} with a chosen value of critical disorder $W_c=1.75$. 
	The data for $ \ln (\langle P_q \rangle / \langle P_q^c \rangle)$ are represented as symbols with their error bar, while the lines represent the fit. The goodness of fit $Q=0.73$ confirms quantitatively that the fit goes through all data within error bars. The critical exponent obtained from this procedure is $\nu \approx 0.44$ close to $0.5$, as obtained using the standard finite-size scaling  approach, see Fig.~\ref{figFSS05p025}. The parameters are $p=0.06$, $10 \le \log_2 N  \le 20$ (from lighter shade $\log_2 N=10$ to darker shade $\log_2 N=20$) and $ 1.2\le W \le 2.2$.}
\end{figure} 

For $ q<0.5 $, a linear scaling behavior is observed in the vicinity of the transition with the same critical exponent on both sides of the transition.
In this case, it is possible to use the more controlled scaling approach which consists in a fit of the data by a Taylor expansion of the scaling function in the vicinity of the transition \cite{PhysRevLett.82.382, PhysRevB.84.134209}.

In order to put in operation this approach, we consider the quantity $ \langle P_q(W) \rangle / \langle P_q(W_c) \rangle $, where $ W_c $ is some fixed value of the critical disorder, chosen arbitrarily. We assume that this quantity follows the scaling law (corresponding to a linear scaling)
\begin{equation}\label{eq:scaTaylor}
 \ln\frac{\langle P_q(W) \rangle }{\langle P_q(W_c) \rangle} = F( w \; d_N^{1/\nu}) \; ,
\end{equation}
where $ w = (W-W_c) + A_2 (W-W_c) ^ 2 + A_3 (W-W_c) ^ 3 $ and $ F (X) = \sum_ {k = 1} ^ 5 B_k X ^ k $. In this analysis, the fitting parameters are the $ \{A_k\} $, $ \{B_k\} $ and $ \nu$, whereas $ W_c $ is fixed and all data for different $W$ and $N$ are fitted simultaneously.

This approach is illustrated in Fig.~\ref{fig:scaSlevin-lin-smallq} for $p=0.06$, $q=0.25$. Importantly, we quantify the quality of our fits by calculating the sum of squared residuals, which incorporates the error bars on the fitted data, and the associated goodness of fit $Q$. We obtain for $W_c=1.75$ and the range of $1.3\le W \le1.95$ and $2^{10}\le N\le2^{20}$ values considered, a fit whose goodness $Q \approx 0.7$ is perfectly acceptable. 
A critical exponent $ \nu \approx 0.5$ is obtained. 

We have checked the robustness of the scaling results with respect to the choice of $W_c$ in the vicinity of $1.75$ (data not shown). Moreover, we have repeated this procedure for different values of $q$ and find acceptable goodness of fit for $W_c\approx 1.75$ and consistent values of $\nu\approx 0.5$. However, the range of values of $W$  for which the quality
of fit is acceptable narrows down as $ q \rightarrow 0.5 $. This confirms the results obtained in Fig.~\ref{chi2dosp} by the standard finite-size scaling  method and discussed in Sec.~\ref{sec:volorlin}:
linear scaling is observed in a restricted range near the transition. For $q\simeq 0.5$ volumic and linear scalings compete, and a two-parameter scaling is in order. For $q\gtrsim 1$ the scaling is volumic in the delocalized phase and linear in the localized phase.

\section{Finite-size scaling  of spectral gap ratios for other parameter values}
\label{AppendixB}

\begin{figure}
\includegraphics[width=0.99\linewidth]{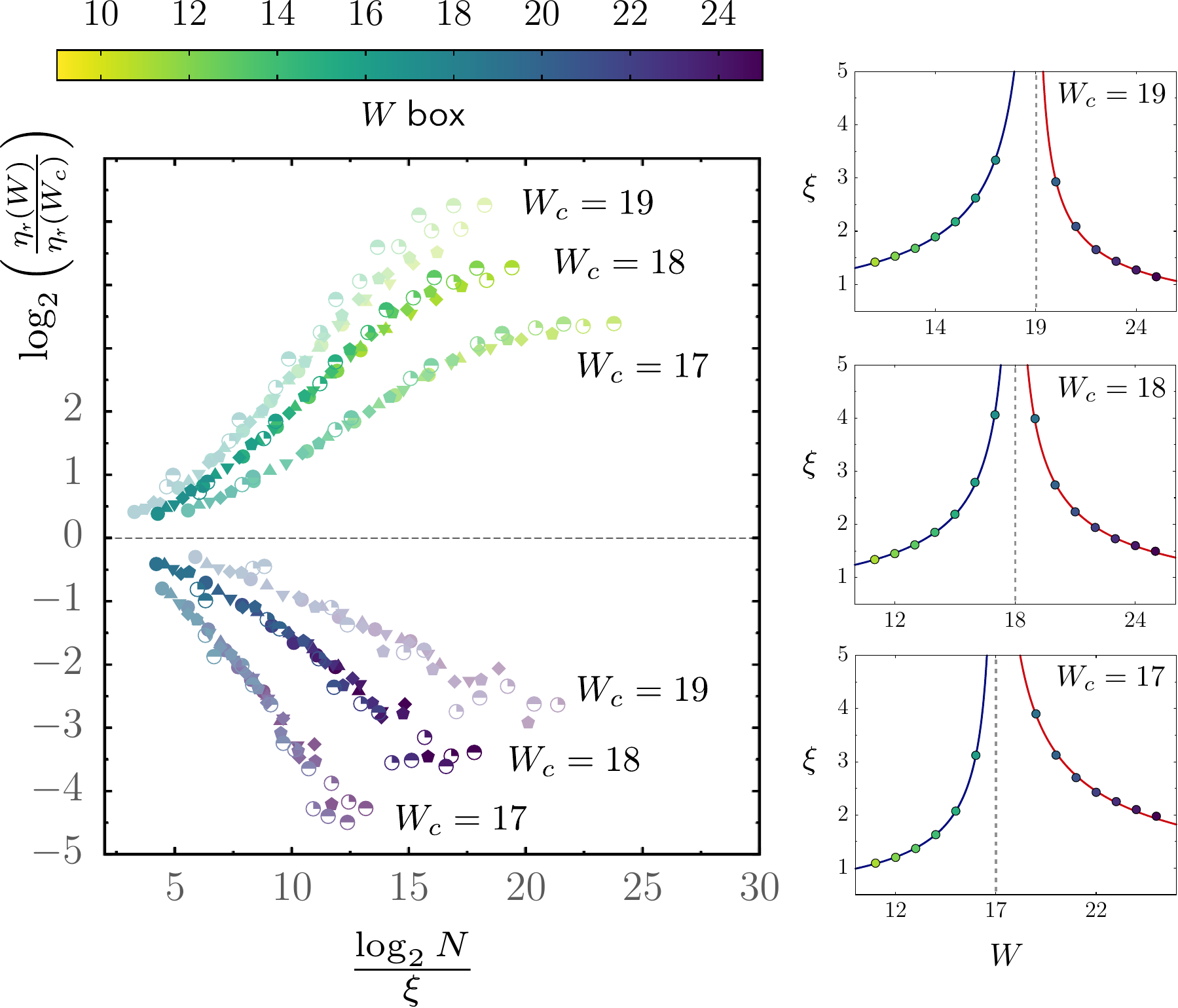}
\caption{Robustness of finite-size scaling of spectral data with respect to changes in the estimated value $W_c$ of the critical disorder strength for $p=0.49$ and box distributed disorder. The critical exponents extracted from these finite-size scalings are displayed in Fig.~\ref{fignuWcp0d0649}, top-right panel.}
\label{figvWcp0d49}
\end{figure}

\begin{figure}
\includegraphics[width=0.99\linewidth]{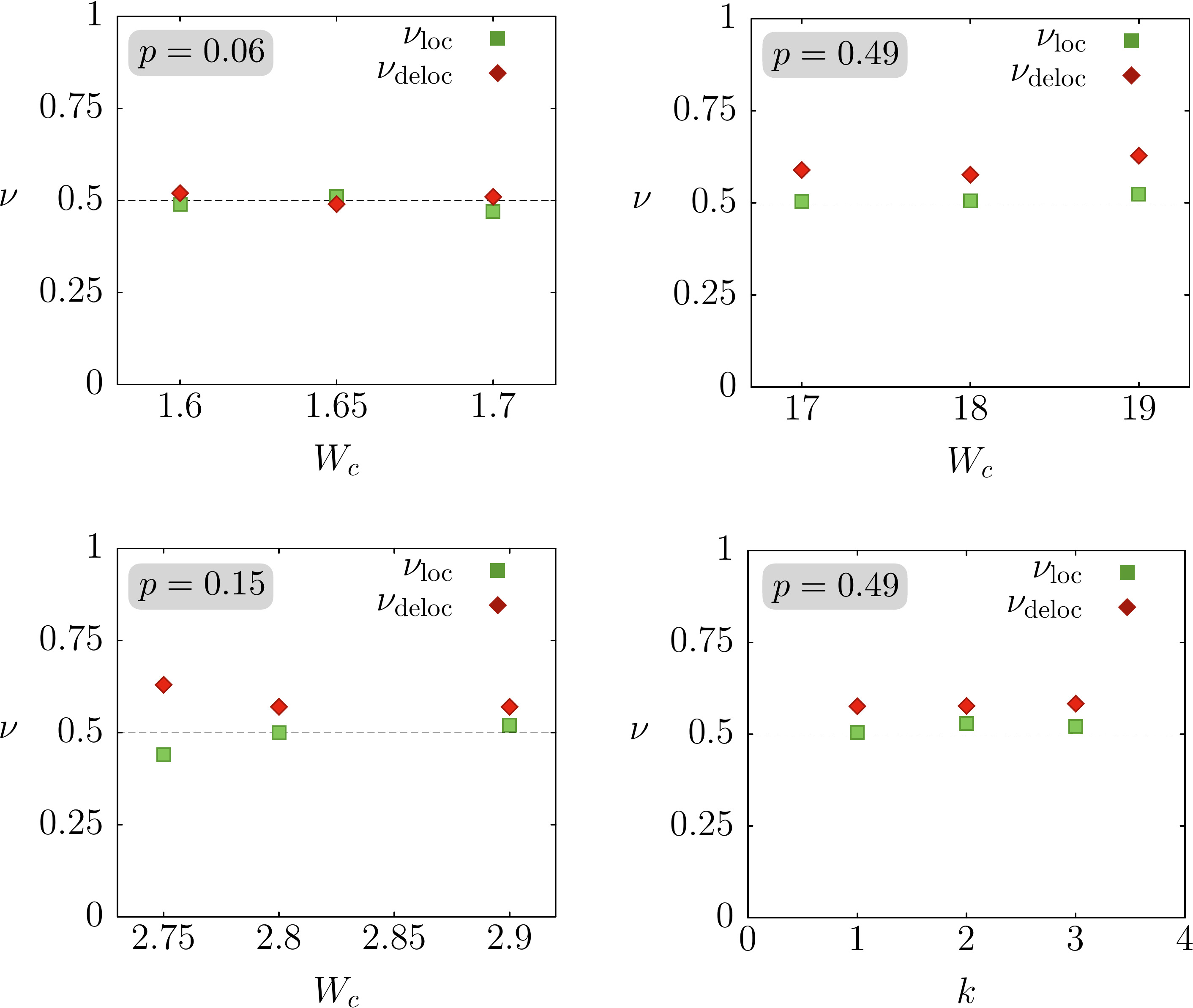}
\caption{Robustness of the critical exponents of spectral data for the localized and delocalized branches with respect to changes in the estimated value $W_c$ of the critical disorder strength for $p=0.06,0.15$ and Gaussian distributed disorder (left panels) and for $p=0.49$ and box distributed disorder (top-right panel). The bottom-right panel shows the robustness of the critical exponents extracted from the finite-size scaling  based on higher-order spacing ratios.}
\label{fignuWcp0d0649}
\end{figure}

In Fig.~\ref{figvWcp0d49} (left) we show the result of a finite-size scaling  analysis of $\eta_\raur$ for different values of the estimated critical disorder $W_c$ for $p=0.49$ and box distributed disorder. In all cases, after an appropriate rescaling $\eta_\raur(W)/\eta_\raur(W_c)=F_\mathrm{lin}(\log_2N/\xi(W))$, the data can be satisfactorily made to collapse. The rescaling lengths $\xi(W)$, which are shown on the panels on the right, are then fitted as $\xi(W)=A\,|W-W_c|^{-\nu}$. This yields approximately the same exponent $\nu\approx 0.5\pm 0.05$ on both (localized and delocalized) sides of the transition for the values of $W_c$ associated with the best collapse ($W_c=18$ for $p=0.49$). When $W_c$ is slightly different from these values the exponents associated with the divergence of $\xi(W)$ remain close to $1/2$, as shown in Fig.~\ref{fignuWcp0d0649} for different values of $p$. We also obtained similar plots for values $p=0.01,0.25$, data not shown.

\section{Finite-size scaling  of higher-order spacing ratios}
\label{higherorderapp}
\begin{figure*}
\includegraphics[width=0.9\linewidth]{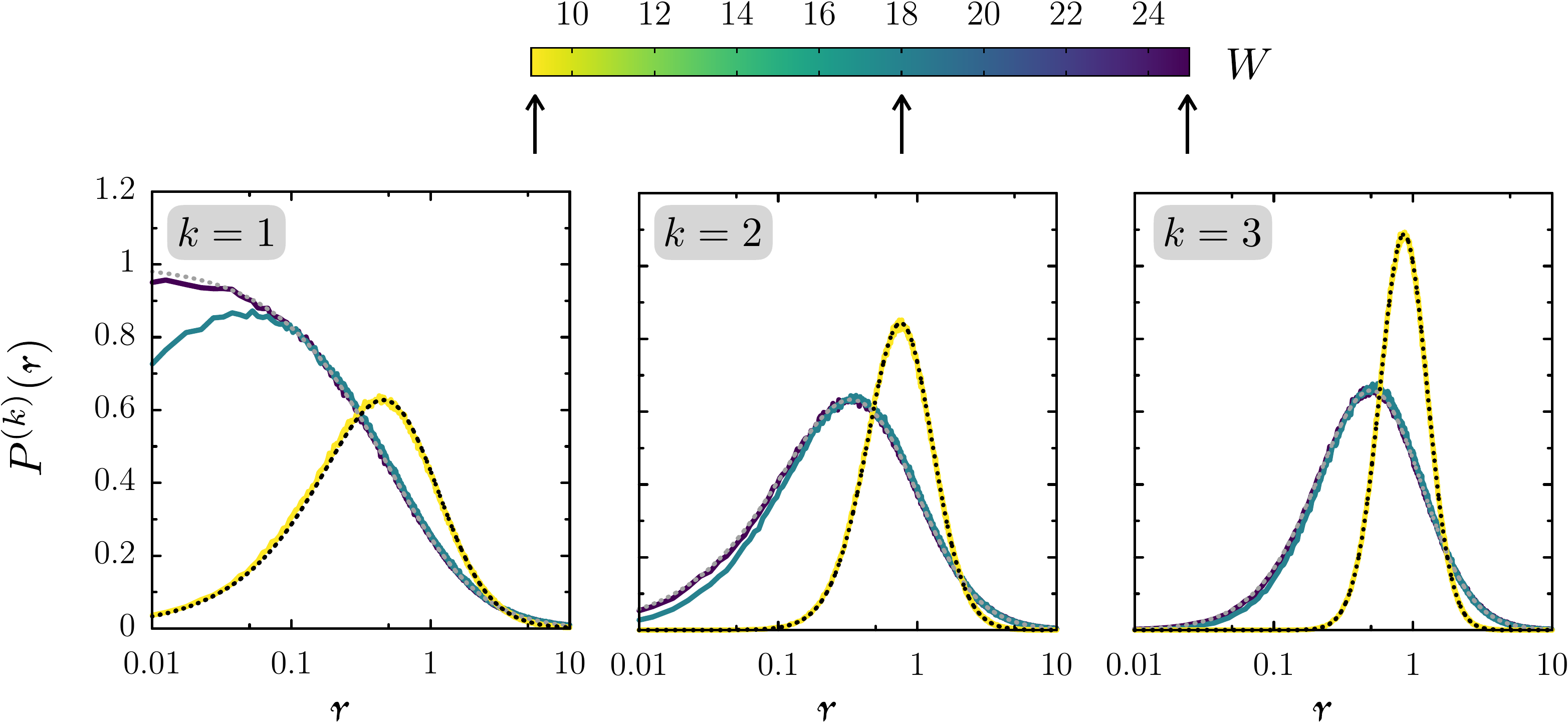} 
\caption{Probability density of the higher-order spacing ratios $\raur^{(k)}$ of the spectra for $k=1$ (left), $k=2$ (middle), and $k=3$ (right). Here, $p=0.49$, $N=2^{15}$, and the strength of the box distributed disorder is $W=9$ (yellow curves), $W=18$ (blue-green curves), and $W=25$ (dark-purple curves).}
\label{figPrk}
\end{figure*}

\begin{figure*}
\includegraphics[width=0.975\linewidth]{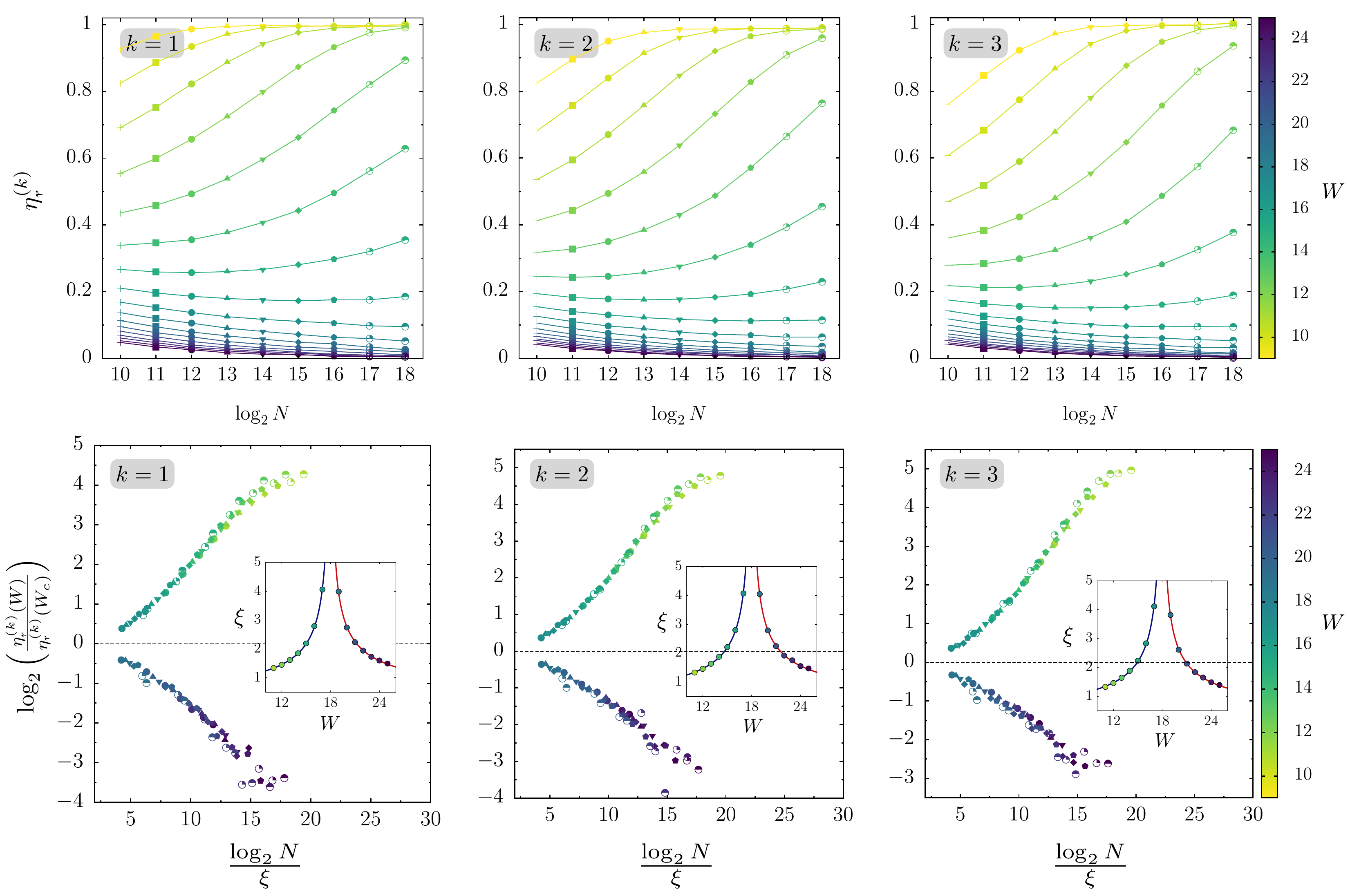} 
\caption{Top: Finite-size scaling of higher-order spectral function $\eta^{(k)}_\raur$ as function of system size for $k=1$ (left), $k=2$ (middle), and $k=3$ (right), same data for all values of $k$. Here, $p=0.49$ and the disorder is box distributed. Bottom: Finite-size scaling  of $\eta^{(k)}_r(W)$ for different values of $k$. For $W_c=18$, the critical exponents extracted from these finite-size scalings are $\nu_\mathrm{deloc}\approx 0.58$, $\nu_\mathrm{loc}\approx 0.51$ for $k=1$, $\nu_\mathrm{deloc}\approx 0.58$, $\nu_\mathrm{loc}\approx 0.53$ for $k=2$, and $\nu_\mathrm{deloc}\approx 0.58$, $\nu_\mathrm{deloc}\approx 0.52$ for $k=3$.}
\label{figetark}
\end{figure*}

Following Refs.~\cite{tekur2018higher,tekur2020symmetry,rao2020distribution}, we consider the distribution of non-overlapping higher order spacing ratios defined by
\begin{equation}
\raur_{i}^{(k)}=\frac{s_{i+k}^{(k)}}{s_{i}^{(k)}}=\frac{E_{i+2 k}-E_{i+k}}{E_{i+k}-E_{i}}, \quad i, k=1,2,3, \ldots
\end{equation}
This spectral quantity allows us to probe spectral fluctuations at a scale different from the one probed with nearest neighbor spacing ratios.

Let us denote by $P^{(k)}(\raur)$ the probability distribution of the $\raur_{i}^{(k)}$. For random matrices, it was conjectured and numerically checked that~\cite{tekur2018higher}
\begin{equation}
P^{(k)}_{\mathrm{WD}}(\raur) = C_{\alpha} \frac{\left(\raur+\raur^{2}\right)^{\alpha}}{\left(1+\raur+\raur^{2}\right)^{1+3 \alpha / 2}}
\end{equation}
where $C_{\alpha}$ is a normalization constant and
\begin{equation}
\alpha =\frac{(k+2)(k+1)}{2}-2, \quad k \geq 1
\end{equation}
In contrast, for randomly distributed energy levels (Poisson statistics), we have~\cite{tekur2018higher}
\begin{equation}
P_{\mathrm{P}}^{(k)}(\raur)=\frac{(2 k-1) !}{[(k-1) !]^{2}} \frac{\raur^{k-1}}{(1+\raur)^{2 k}}
\end{equation}
As an illustration, Fig.~\ref{figPrk} displays the distributions $P^{(k)}(\raur)$ for $1\leq k\leq 3$ in the localized, critical and delocalized regimes.

A set of parameters $\{\eta^{(k)}_\raur\}_{k\geq 1}$ can be defined to interpolate between Poisson and Wigner-Dyson level statistics as
\begin{equation}
\label{defetark}
\eta^{(k)}_\raur=\frac{\displaystyle \left\langle\min\left(\raur^{(k)},1/\raur^{(k)}\right)\right\rangle-I_{\mathrm{P}}}{\displaystyle I_{\mathrm{WD}}-I_{\mathrm{P}}}
\end{equation}
where $\langle . \rangle$ denotes an ensemble average, and $I_{P}$ and $I_{\mathrm{WD}}$ are the average of $\min\left(\raur^{(k)},1/\raur^{(k)}\right)$ when $\raur^{(k)}$ is distributed according to $P_{\mathrm{P}}^{(k)}(\raur)$ and $P_{\mathrm{WD}}^{(k)}(\raur)$, respectively. The parameter $\eta_\raur^{(k)}$ is equal to $0$ for Poisson statistics and to $1$ for Wigner-Dyson level statistics. 
Figure~\ref{figetark} shows the behavior of $\eta^{(k)}_\raur$ for $k=1,2,3$ for various disorder strengths as function of system size. The top row shows the raw data and the bottom row shows the result of the same linear finite-size scaling  procedure as described in the main text for $W_c=18$. These finite-size scalings consistently yield critical exponents close to $0.5$ for both localized and delocalized branches (see caption of Fig.~\ref{figetark} for details). \\

\begin{figure}[!ht]
  \includegraphics[width=.9\linewidth]{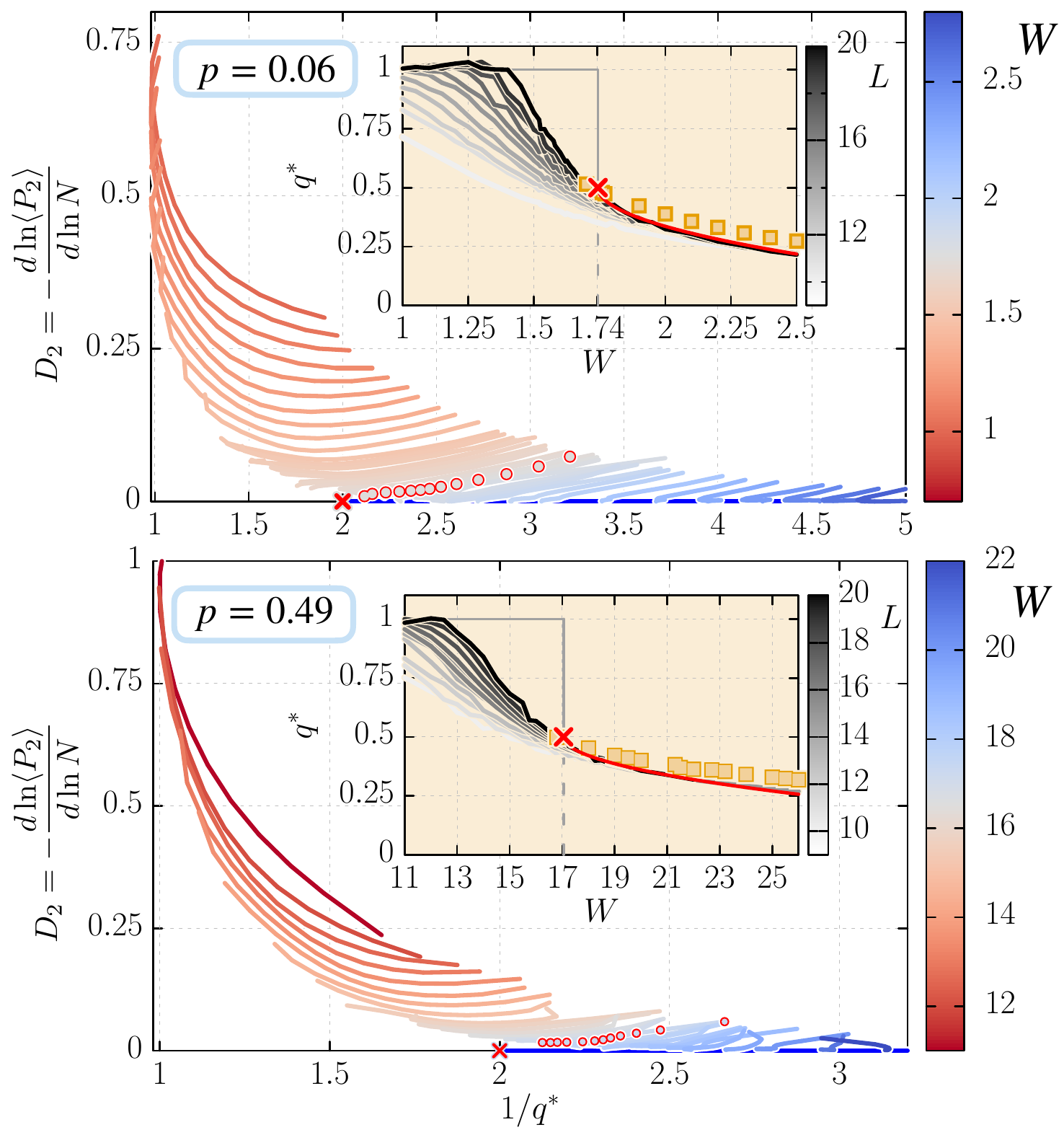}
\caption{The same two-parameter RG flow plot as in Fig.~\ref{figurasRGflow} for moments, with $p=0.06$ (top) and $p=0.49$, box disorder (bottom). The insets show $q^*$ as a function of $W$ for different system sizes (gray scale). The squares in the insets correspond to $\xi_\perp \ln K$ where $\xi_\perp$ was obtained from the exponential fit of $\Cttyp\sim e^{-d/\xi_\perp}$. For $p=0.06$ the correlation functions were computed for $N=2^{18}$ for $p=0.49$ we used $N=2^{17}$.}
\label{figurasRGflow2p}
\end{figure}

\section{Anderson transition and MBL transition: two-parameter RG flow for other parameter values}
\label{twoparamapp}
In Fig.~\ref{figurasRGflow} we showed that for $p=0.25$ the RG flow is identical to that of MBL. The same can be seen in Fig.~\ref{figurasRGflow2p} for other parameter values.

\section{Another observable: the radial probability distribution}
\label{radialprob}

\begin{figure*}[t]
  \includegraphics[width=0.85\linewidth]{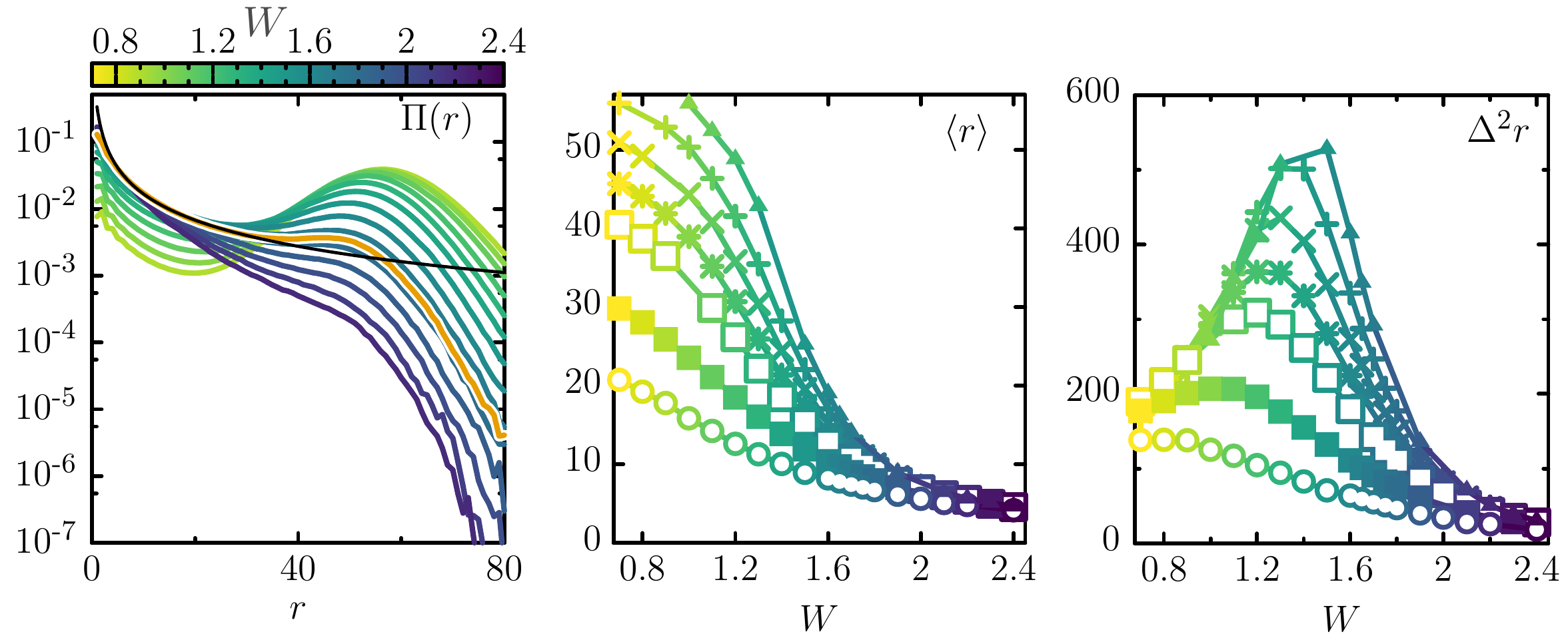}
\caption{Radial probability distribution of eigenstates. Left: $\Pi(r)$ for $W\in [0.7,2.4]$ and $N=2^{17}$. Its behavior can be inferred from that of $\Ctav$, Eq.~\eqref{chav}, shown in Fig.~\ref{fig:corr1}. It decays exponentially in the localized phase, $\Pi(r) \sim e^{-r/\xi_\parallel}/r^{\alpha}$, as a power law at the transition (where $\xi_\parallel$ diverges), $\Pi(r) \sim r^{-\alpha}$, and follows $N_r$ (see Fig.~\ref{fig1Remy}), thus is peaked at $r\approx d_N/2$, for $r$ sufficiently large, in the delocalized phase. The orange curve corresponds to $W=1.75\approx W_c$. The black line corresponds to the fit $\Pi(r) \sim r^{-\alpha}$, where we find $\alpha\approx 1.36$. This is compatible with $\alpha\approx 1.3$ which we obtain from what we find for $\Ctav(r)\sim K^{-r} e^{-r/\xi_\parallel}/r^\alpha$ for $p=0.06 $ (data not shown).
Center: Mean distance $\langle r\rangle$, calculated from the radial probability distribution of eigenstates $\Pi(r)$ defined by Eq.~\eqref{eq:radprob} as in \cite{warzel21} for the MBL problem, as a function of $W$. Different curves correspond to different system sizes $N=2^L$.
In the ergodic delocalized phase, $\langle r\rangle \sim d_N/2$ at large $N$, while $\langle r\rangle$ is finite (independent of $N$) in the localized phase. At the transition, $\langle r\rangle \sim {d_N}^{2-\alpha}$.  Right: $\Delta^2 r$ for the same values as the left panel. Importantly, $\Delta^2 r$ is finite in both localized and delocalized regimes, but diverges at the transition. The behaviors at criticality and in the delocalized phase are very similar to what is found in the MBL problem \cite{warzel21}. The localized behaviors are to be contrasted with the MBL case where $\langle r \rangle \approx p d_N$ with $p<1/2$, which reflects the multifractal property of this phase.
The parameters are as follows:
$p=0.06$ and system sizes $N=2^{10},2^{12},2^{14},2^{15},2^{16},2^{17}, 2^{18}$.}
\label{distmax}
\end{figure*}
Another way to compare the Anderson transition on random graphs with the MBL transition is to consider the radial probability distribution
of eigenstates introduced in \cite{warzel21}.
By denoting $i_{\rm max}$ the site where a given state reaches its maximum amplitude, we define the radial probability distribution
as:
\begin{equation}\label{eq:radprob}
    \Pi(r)=\left\langle{\sum_{d(i_{\rm max},i)=r}|\psi_i|^2}\right\rangle,
\end{equation}
where we average over the different eigenstates, disorder and graph configurations considered.
This distribution, plotted in Fig.~\ref{distmax}(b), allows us to assess the spatial properties of wave functions on the graph, in complement to the average $\Ctav$ [Eq.~\eqref{chav}] and typical $\Cttyp$ [Eq.~\eqref{cttyp}] correlation functions. 

The mean value $\langle r\rangle$ is displayed in Fig.~\ref{distmax}(a). In the localized regime, the curves of $\langle r \rangle$ as a function of $W$, for different $N$, collapse. This collapse is compatible with the exponential decay of the average correlation function 
$\Ctav $ [Eq.~\eqref{chav}]. Indeed, Eq.~\eqref{chav} implies that $\Pi(r) \sim e^{-r/\xi_\parallel}/r^{\alpha}$, thus (neglecting the $r^\alpha$ correction) $\langle r \rangle \approx e^{-1/\xi_\parallel}/(1-e^{-1/\xi_\parallel})$
and the variance $\Delta^2 r = \langle r^2 \rangle$ [shown in Fig.~\ref{distmax}c] are finite,
independent of $N$, in the localized phase $W>W_c$. This behavior is to be contrasted with that found in the MBL phase where $\langle r \rangle \sim p d_N$ with $p<1/2$ \cite{warzel21}, which reflects the fact that the MBL phase is multifractal. 

At the transition, $\xi_\parallel$ diverges and $\Pi_c(r) \sim r^{-\alpha}$ with $1<\alpha<2$. Therefore, $\langle r \rangle \sim d_N^{2-\alpha}$ and $\Delta^2 r \sim d_N^{3-\alpha}$ diverge at the transition. 
In the delocalized phase, $\Ctav$ tends to a constant (for $r\gg \xi$, $\xi$ the correlation length) such that $\Pi_c(r)$ is peaked at $\approx d_N/2$ ($\Pi_c(r)$ behaves like $N(r)$, shown in Fig.~\ref{fig1Remy}, for $r \gg \xi$), hence $\langle r \rangle \approx d_N/2$ and $\Delta^2 r$ is finite at large $N$. These behaviors are strikingly similar to what is found in the critical and delocalized regimes of the MBL transition \cite{warzel21}.

%

\end{document}